\documentclass[referee]{aa} 
\usepackage{xcolor}
\usepackage{pdflscape}
\usepackage{graphics}
\usepackage{txfonts}
\def\l{$\lambda$}

\def\mbh{$M_{\rm BH}$\/}

\def\lledd{$L/L_{\rm Edd}$}

\def\nc{$N_{\rm c}$\/}
\def\rfe{$R_{\rm FeII}$}

\def\feiiq{\rm Fe{\sc ii}$\lambda$4570\/}

\def\chm{$c(\frac{1}{2})$\/}
\def\cqm{$c(\frac{1}{4})$\/}
\def\ltsima{$\; \buildrel < \over \sim \;$}
\def\ltsim{\lower.5ex\hbox{\ltsima}}  
\def\gtsima{$\; \buildrel > \over \sim \;$}

\def\gtsim{\lower.5ex\hbox{\gtsima}}

\def\civ{{\sc{Civ}}$\lambda$1549\/}
\def\civonly{{\sc{Civ}}\/}
\def\civnc{{\sc{Civ}}$\lambda$1549$_{\rm NC}$\/}
\def\civbc{{\sc{Civ}}$\lambda$1549$_{\rm BC}$\/}
\def\civvbc{{\sc{Civ}}$\lambda$1549$_{\rm VBC}$\/}
\def\cm3{cm$^{-3}$\/}
\def\hb{{\sc{H}}$\beta$\/}

\def\hbbc{{\sc{H}}$\beta_{\rm BC}$\/}
\def\hbvbc{{\sc{H}}$\beta_{\rm VBC}$\/}
\def\hbnc{{\sc{H}}$\beta_{\rm NC}$\/}
\def\mgii{{Mg\sc{ii}}$\lambda$2800\/}

\def\ciii{{\sc{Ciii]}}$\lambda$1909\/}
\def\oiiiopt{{\sc{[Oiii]}}$\lambda\lambda$4959,5007\/}
\def\oii{{\sc{[Oii]}}$\lambda$3727\/}

\def\caii{{Ca{\sc ii}}}

\def\aliii{Al{\sc iii}$\lambda$1860\/}
\def\heiiuv{He{\sc{ii}}$\lambda$1640}

\def\feiiuv{{{\sc{Feii}}}$_{\rm UV}$\/}

\def\feii{{Fe\sc{ii}}\/}

\def\fe{{\sc{Fe}}\/}

\def\fe76087{{\sc [Fe vii]}$\lambda$6087\/}

\def\kms{km~s$^{-1}$}

\def\rk{$R_{\rm K}$\/}
\def\ergss{erg s$^{-1}$\/}

\def\rk{{$R{\rm _K}$}\/}
\def\heii{{{\sc H}e{\sc ii}}$\lambda$4686\/}

\def\siiv{Si{\sc iv}$\lambda$1397\/}

\def\cmp{$c(\frac{1}{2})$}
\def\cqp{$c(\frac{1}{4})$}
\usepackage{enumerate}

\definecolor{darkorange}{rgb}{1,0.612,0}
\definecolor{aquamarine}{rgb}{0.498,1,0.8314}
\newcommand\crule[3][black]{\textcolor{#1}{\rule{#2}{#3}}}

\begin{document}
\authorrunning{}
\titlerunning{The \civ\ and \hb\ emission lines as \mbh\ estimators in  quasars}
\title{Black hole mass estimates in quasars
}

\subtitle{A comparative analysis of high- and low-ionization lines}
\bibstyle{apj} 
\author{Paola Marziani\inst{1} \and  Ascensi\'on del Olmo\inst{2}  
\and M. A. Mart\'\i nez-Carballo\inst{3}
\and Mary Loli Mart\'\i nez-Aldama\inst{4} 
\and Giovanna M. Stirpe\inst{5}
\and  C. Alenka Negrete\inst{6,}\thanks{CONACyT research fellow, Instituto de Astronom\'{\i}a, UNAM} 
\and Deborah Dultzin\inst{6}  
\and Mauro D' Onofrio\inst{7,}\thanks{INAF associate, Os\-servatorio Astro\-nomico di Pado\-va}   
\and \hspace{0.4cm}Edi Bon\inst{8} 
\and Natasha Bon\inst{8} }
\institute{{INAF, Os\-servatorio Astronomico di Padova, vicolo dell' Osservatorio 5, IT 35122, Padova, Italy,  \email{paola.marziani@oapd.inaf.it}}
\and{Instituto de Astrofis\'{\i}ca de Andaluc\'{\i}a, IAA-CSIC, Glorieta  de la Astronomia s/n 18008 Granada, Spain, \email{chony@iaa.es}} 
\and {Departamento de  Matem\'atica Aplicada and IUMA, Universidad de Zaragoza, E-50009 Zaragoza, Spain
\email{gelimc@unizar.es}}
\and {Center for Theoretical Physics, Polish Academy of Science 02-668, Warsaw, Poland \email{mmary@cft.edu.pl}} 
\and {INAF, Os\-serva\-torio di Astro\-fisica e Scien\-za dello Spa\-zio, {IT-40129 Bologna},  Italy    \email{giovanna.stirpe@inaf.it}}
\and {Instituto de Astronom\'{\i}a, UNAM, Mexico D.F. 04510, Mexico, \email{alenka,deborah@astro.unam.mx}}
\and {Dipartimento di Fisica \& Astronomia ``Galileo Galilei'', Universit\`a\ di Padova, Vicolo dell'Osservatorio 3, IT 35122 Padova,  Italy  \email{mauro.donofrio@unipd.it} }
\and {Astronomical Observatory, Belgrade, Serbia  \email{ebon,nbon@aob.rs}}
}

\date{}
\abstract
{The inter-line comparison between high- and low-ionization emission lines has yielded a wealth of information on the quasar broad line region (BLR) structure and dynamics, including perhaps the earliest unambiguous evidence in favor of a disk + wind structure in radio-quiet quasars. }
{We carried out an analysis  of the \civ\ and \hb\  line profiles of 28 Hamburg-ESO high luminosity quasars and of 48 low-$z$, low luminosity sources in order  to test whether   the high-ionization line  \civ\ width could be correlated with \hb\ and be used as a virial broadening estimator. }
{We analyze   intermediate- to high-S/N, moderate resolution optical and NIR spectra covering the redshifted \civ\ and \hb\ over a broad range of luminosity $\log L \sim 44 - 48.5$ [\ergss] and redshift ($0 - 3$), following an approach based on the quasar main sequence.  }
{The present analysis indicates that the line width of \civ\ is not immediately offering a virial broadening estimator equivalent to \hb. At the same time a virialized part of the BLR appears to be preserved even at the highest luminosities. We suggest a  correction to FWHM(\civ)  for Eddington ratio  (using the \civ\ blueshift   as a proxy) and luminosity effects  that can be applied over more than four dex in luminosity.  
}
{ Great care should be used in estimating high-$L$\ black hole masses \mbh\ from \civ\ line width. However, once corrected FWHM \civ\ are used, a \civ-based  scaling law  can yield unbiased  \mbh\ values  with respect to the ones based on \hb\ with sample standard deviation $\approx$ 0.3 dex. 
 }  
\keywords{
                quasars: general --
                quasars: emission lines --
                quasars: supermassive black holes --
                ISM: jets and outflows --
                line: profiles   
               }
\maketitle
\defcitealias{sulenticetal06}{S06}
\defcitealias{sulenticetal14}{S14}
\defcitealias{sulenticetal07}{S07}
\defcitealias{sulenticetal17}{Paper~I}
\defcitealias{coatmanetal17}{C17}
\defcitealias{coatmanetal16}{C17}

\section{Introduction}
\label{intro}

Type-1 active galactic nuclei (AGN) and quasars  show the same broad optical-UV lines almost always accompanied by broad permitted \feii\ emission \citep[e.g.,][]{vandenberketal01}.  However, even among type-1 sources  we face a large  diversity in observational manifestations involving line profiles, internal line shifts as well as emission line intensity ratios \citep[e.g.,][and \citealt{sulenticmarziani15} for a recent review]{sulenticetal00a,bachevetal04,yipetal04,kuraszkiewiczetal09,zamfiretal10,shenho14}.  Broad line measurements involving \hb\ line width and  \feii\ strength are not randomly distributed but instead  define a quasar  ``main sequence'' (MS) \citep[e.g.,][]{borosongreen92,sulenticetal00a,shenho14}. The MS can be traced in an optical plane defined by \feii\ emission prominence and the Hydrogen \hb\ line width. The \feii\ strength is parametrized by the intensity ratio involving  the \feii\ blue blend at 4570 \AA\ and broad \hb\ i.e., \rfe = I(\feiiq)/I(\hb), and the Hydrogen \hb\ line width by   its FWHM.  Along the MS, sources  with higher \rfe\   show  narrower broad \hb\  (Population A, FWHM(\hb)$\lesssim 4000$ \kms, \citealt{sulenticetal00a}). Lower \rfe\ is associated  with sources with broader \hb\ profiles  (Pop. B with FWHM(\hb)$\gtrsim 4000$ \kms, \citealt{sulenticetal11}).  { A glossary of the MS-related terminology is provided in Appendix \ref{glossary}.}

Studies of the Balmer lines have played a prominent role for characterizing the MS and the properties  of the broad line emitting region (BLR) in low $z$ ($\lesssim$0.8) quasars with \hb\ providing information for the largest number of sources  \citep[e.g.,][for a variety of observational and  statistical approaches]{osterbrockshuder82,willsetal85,sulentic89,zamfiretal10,huetal12,steinhardtsilverman13,shen16}.   
A most important application of the FWHM \hb\ has been its use as a virial broadening estimator (VBE) to derive black hole masses (\mbh) from single-epoch observations of large samples of quasars \citep[e.g.,][and references therein]{mclurejarvis02,mcluredunlop04,vestergaardpeterson06,assefetal11,shen13,peterson14}. The underlying assumption is that the \hb\ line width provides the most reliable VBE,  which is likely to be the case, even if with some caveats \citep[e.g.,][see also \citealt{shen13}, \citealt{peterson14} for reviews]{trakhtenbrotnetzer12}.  

Balmer lines provide a reliable VBE up  $z \lesssim$ 2 \citep{matsuokaetal13,karouzosetal15}  at cosmic epochs less than a few Gyr. 
The importance to have a reliable VBE at even earlier cosmic epochs cannot be underemphasized. The entire scenario of early structure formation is  affected by inferences from estimates of quasar black hole masses. 
Overestimates of \mbh\ by  lines whose broadening is in excess to the virial one can have implications on the quasar mass function, and at high redshift ($z \gtrsim$ 6) when the Universe was less than 1 billion year of age,  on the formation and mass spectrum of   the seed black holes \citep{latifferrara16} that may have been responsible, along with Pop. III stars, of the reionization of the process at $z \sim 7 - 10$ \citep[e.g.,][for a review]{galleranietal17}. 

Strong and  relatively unblended \civ\ has been the best candidate for a VBE beyond $z \sim 1.5$, where  \hb\ is shifted into the IR domain.  \civ\ can be  observed up to redshift $z \approx 6$\, with optical spectrometers, and in the NIR bands up to redshift $z \approx 7.5 $\, \citep{banadosetal18}, and beyond. 
 Can \civ\ be used as an immediate surrogate for \hb\ when H$\beta$\ is invisible or hard to obtain?  Before attempting an answer to this question, two considerations are in order. 



Firstly, measures of the \civ\ line profiles remain  of uncertain interpretation without a precise determination of the quasar rest frame: an accurate $z$ measurement is not easy to obtain from broad lines, and redshift determinations at $z \gtrsim 1$\  from optical survey data suffer systematic biases as large as several hundreds \kms\   \citep{hewettwild10,shenetal16}.  Reliable studies tie \civonly\ measures to a rest frame derived from \hb\ narrow component (+ \oiiiopt\ whenever applicable \citealt[e.g.,][]{mejia-restrepoetal16}; for problems in the use of \oiiiopt,  see \citealt{zamanovetal02,huetal08}).  

Secondly, significant \civ\ blueshifts  are observed over a broad range in $z$\ and luminosity, from the nearest Seyfert 1 galaxies to the most powerful radio-quiet quasars \citep{willsetal93,sulenticetal07,richardsetal11,coatmanetal16,shen16,bischettietal17,bisognietal17,vietrietal17,sulenticetal17}.   Measures of the \civ\ profile velocity displacement provide an additional dimension to a 4D ``eigenvector 1'' (4DE1)  space  built on parameters that are  observationally independent (``orthogonal'')  and related to different physical aspects \citep[][]{sulenticmarziani15}.   Inclusion of \civ\ shift  as a 4DE1 parameter was motivated by the earlier discovery of internal redshift differences between low- and high-ionization lines \,  \citep{burbidgeburbidge67,gaskell82,tytlerfan92,brothertonetal94,corbinboroson96,marzianietal96}.   
 
The current interpretation of the BLR  in quasars sees the broad lines arising in a region that is  physically and dynamically composite \citep[e.g.,][]{collinsouffrinetal88,elvis00,ferlandetal09,kollatschnyzetzl13,grieretal13,duetal16}.  \civ\ is  a   doublet originating from an ionic species  of ionization potential (IP) four times larger than Hydrogen (54 eV vs. 13.6 eV), and is therefore a prototypical high-ionization line (HIL). The line is mainly produced by collisional excitation from the  ground state $^{2}${S}$_{0}$ to $^{2}${S}$_{\frac{1}{2},\frac{3}{2}}$\ at the temperature of photo-ionized BLR gas (T$\sim 10^{4}$K, \citealt{netzer90}), in the fully-ionized zone of the line emitting gas. Empirically, the line is relatively strong  (rest frame equivalent width $ W \sim$ 10 -- 100 \AA\ depending on the source location on the MS) and only moderately contaminated on the red side (red shelf) by \heiiuv\ and {\sc Oiii}] \l 1663 plus weak emission from \feiiuv\ multiplets \citep{fineetal10}. The  Balmer line \hb\ assumed to be representative of the low-ionization lines (LILs, from ionic species with IP $\lesssim 20$ eV) is instead enhanced in a partially-ionized zone due to the strong X-ray emission of quasars and to the large column density of the line emitting gas (\nc$\gtrsim $ 10$^{23}$ cm$^{-2}$; \citealt{kwankrolik81}).  Comparison of \hb\ and \civ\  profiles in the same sources tells us that  they provide independent inputs to BLR models  --- their  profiles can be dissimilar and several properties uncorrelated \citep[see, for instance, Fig. C2 of][]{mejia-restrepoetal16}. 

It is possible to interpret  \hb\ and \civ\ profiles  as associated with two sub-regions within the BLR \citep[e.g., ][]{baldwinetal96,halletal03,leighly04,sneddengaskell04,czernyhryniewicz11,plotkinetal15}: one emitting predominantly LILs (e.g., \citealt{dultzinhacyanetal99,matsuokaetal08}), and a second HILs,  associated with gas outflows and winds \citep[e.g.,][]{richardsetal11,yongetal18}. This view is in accordance with early  models of the BLR structure involving a disk and outflow or wind component \citep{collinsouffrinetal88,elvis00}.  Intercomparison of  \civ\ and \hb\  at low $z$\ and moderate luminosity provided the most direct observational evidence that this is the case at least for radio-quiet (RQ) quasars \citep{corbinboroson96,sulenticetal07,wangetal11,coatmanetal16}. Modeling  involves a  disk + wind system \citep[e.g.,][for different perspectives]{progaetal00,progakallman04,flohicetal12,sadowskietal14,vollmeretal18}, although the connection between disk structure and BLR (and hence the \hb\ and \civ\ emitting regions) is still unclear.   

There are additional caveats, as the \civ\ blueshifts are not universally detected. Their amplitude is a strong function of the location along the MS \citep{sulenticetal00b,sulenticetal07,sunetal18}. { Large blueshifts are clearly detected in Population A, with sources accreting at relatively high rate, and reach extreme values for  quasars  at the high \rfe\ end along the MS.  } In Pop. B, the wind component is not dominating the line broadening of \civ\ at moderate luminosity; on the converse, the \civ\ and \hb\ line profile intercomparison indicates that the dynamical relevance of the \civ\ blueshift is small i.e., that the ratio between the centroid at half-maximum \cmp\ and the FWHM is $\ll 1$ \citep{sulenticetal07}. Reverberation mapping studies indicate that the velocity field is predominantly Keplerian (\citealt{peietal17} and references therein for the protypical source NGC 5548, \citealt{denneyetal10,grieretal13}), and that the \civ\ emitting region is closer to continuum source than the one of \hb\ \citep[e.g.,][]{petersonwandel00,kaspietal07,treveseetal14}.  The issue is complicated by luminosity effects on the \civ\ shifts that may have gone undetected at low-$z$. { Both Pop. A and B sources} at $\log L \gtrsim 47$ \ergss\  show large amplitude blueshifts in \civ\  \citep{sulenticetal17,bisognietal17,vietrietal18}. { The  present work  considers the trends associated with the MS as well as the luminosity effects that may appear second-order in low-luminosity samples to provide corrections to the FWHM of \hb\ and ultimately a scaling law based on \civ\ FWHM and UV continuum luminosity that may be unbiased with respect to \hb\ and with a reasonable scatter.}

The occurrence of \civ\  large shifts  challenges  the suitability of  the \civ\ profile broadening as a VBE for \mbh\ estimates \citep[see e.g.,][for  a review]{shen13}. Results at low-redshift suggest that the \civ\ line is fully  unsuitable for part of Pop. A sources \citep{sulenticetal07}. A similar conclusion was reached at $z \approx 2$ on a  sample of 15 high-luminosity quasars \citep{netzeretal07}.  More recent work tends to confirm that the \civ\ line width is not straightforwardly related to virial broadening \citep[e.g.,][]{mejia-restrepoetal16}. However, the \civ\ line is strong and observable up to $z \approx 6$\ with optical spectrometers. It is so highly desirable to have a consistent VBE up to the highest redshifts that various attempts \citep [e.g.,][]{brothertonetal15} have been done at rescaling \civ\ line width estimators to the width of LILs such as \hb\ and \mgii.  Several conflicting claims have been recently made on the valid use of \civ\ width in high-$z$\ quasars  \citep[e.g., ][]{assefetal11,shenliu12,denneyetal12,karouzosetal15,coatmanetal17,mejia-restrepoetal18}.


From the previous outline we infer that a proper approach  to testing the suitability of the \civ\ line width as a VBE is to compare \civ\ and \hb\ profiles along the quasar MS, and to extend the luminosity range including intermediate-to-high $z$\ ($\gtrsim 1.4$)\ sources when \hb\ is usually not covered by optical observations.  { A goal of this paper is to analyse the factors yielding to large discrepancies between the \mbh\ estimates from \hb\ and \civ, with a focus on the aspect and physical factors affecting the broadening of the two lines.}  

The quasar sample used in the present paper joins { two samples with both \hb\ and \civ\ data}, one at low luminosity and $z$\ ($\lesssim 0.7$, \citealt{sulenticetal07}), and one at high luminosity, in the range $  1.5 \lesssim z \lesssim 3$ presented and analyzed by \citet[][hereafter \citetalias{sulenticetal17}]{sulenticetal17}.  The sample provides   a wide coverage in luminosity, and Eddington ratio (Sec. \ref{sec:sample}); \hb\ line coverage for each \civ\ observation; consistent analysis of the line profiles of both \civ\ and \hb\ (Sec. \ref{anal}). Our approach  is intended to overcome some of the sample-dependent difficulties encountered by past studies. Results involve the reduction of the measured \civ\ line width to a VBE (Sec. \ref{virial}) { with a correction factor dependent on both shift amplitude and luminosity}.   They are discussed in terms of BLR structure (Sec. \ref{wind}), { and specifically of the interplay between broadening associated with the outflow (very relevant for \civ) and with orientation effects (which are dominating for \hb).} Finally, a new \mbh\ scaling law with line width and luminosity (\S \ref{mass}) is presented. The new \civ\ scaling law, which considers different corrections for Pop. A and B separately, may provide an unbiased estimator of black hole masses derived from \hb\ over a wide range in luminosity (Sect. \ref{large}).

\section{Sample}
\label{sec:sample}

\subsection{High-luminosity VLT data for Hamburg-ESO quasars}

The high-$L$\ quasars considered in the present study are 28 sources  identified in the HE survey \citep[][hereafter the HE sample]{wisotzkietal00}, in the redshift range $1.4 \lesssim   z  \lesssim 3.1$.  All satisfy the conditions on the absolute B magnitude $M_\mathrm{B} \lesssim -27.5$\ and on the bolometric luminosity $\log L \gtrsim 10^{47.5}$ \ergss. They are therefore among the most luminous quasars ever discovered in the Universe, and a relatively rare population even at $z \approx 1 - 2$ when luminous quasars were more frequent than at low-$z$ (the luminosity function at $M_\mathrm{B} \approx -27.5$\ is $\Phi(M_\mathrm{B}) \sim 10^{-8}$ Mpc$^{-3}$ mag$^{-1}$ compared to $\sim 10^{-6}$ Mpc$^{-3}$ mag$^{-1}$ at $M_\mathrm{B} \approx -25$, corresponding to the ``knee'' of the \citealt{boyleetal00} luminosity function).  

The \civ\ data were obtained with the FORS1 spectrograph at VLT and Dolores at TNG; the matching \hb\ observations with the ISAAC spectrometer were analyzed in detail in \citet{sulenticetal06}.  The resolution at FWHM of the \civ\ data is $\lesssim 300$ \kms\ and $\lesssim$\ 600 \kms for FORS1 and Dolores, respectively; the H$\beta$\ resolution is $\approx$ 300 \kms\ \citep{sulenticetal04}. Typical S/N values are $\gtrsim 50$. 

Resolution and S/N are adequate for a multicomponent nonlinear fitting analysis using the IRAF routine {\tt specfit} \citep{kriss94}, involving an accurate deconvolution of \hb, \oiiiopt, \feii, \heii\  in the optical, and of \civ\ and \heiiuv\ in the UV. The  \civ\ and \hb\ data and the immediate results of the {\tt specfit}   analysis were reported in \citetalias{sulenticetal17}. 

\subsection{Low-luminosity \civ\ and \hb\ data}

We considered a Faint Object Spectrograph (FOS) sample from \citet{sulenticetal07} as a complementary sample at low-$L$ and low-$z$. For the sake of the present paper, we restrict the  FOS sample to 29 Pop. A and 19 Pop. B RQ (48 in total) sources covering the \civ\ blend spectral range and with previous measures for the \hb\ profile and \rfe\ \citep{marzianietal03a}. The list of sources can be obtained by the cross-correlation of the \citet{sulenticetal07} RQ sources (Kellermann's ratio $\log$\rk $< 1.8$) and the \citet{marzianietal03a} catalog on Vizier. We excluded NGC 4395 and NGC 4253 whose luminosities are $\log L \approx$ 40.4 and 41.7\ [\ergss] respectively,  outlying with respect to the $L$\ distribution of the FOS sample. The FOS high-resolution grisms yielded an inverse resolution $\lambda/\delta \lambda \sim 1000$, equivalent to typical resolution of the \citet{marzianietal03a}'s data. The S/N is above $\gtrsim 20$\ for both the  optical and UV  low-$z$\ data.   The FOS sample has a typical bolometric luminosity $\log L \sim 45.2$ [\ergss] and a redshift $z \lesssim  0.5$.  


\subsection{Joint HE+FOS sample}

\begin{figure}[t!]
\centering
\includegraphics[width=0.9\columnwidth]{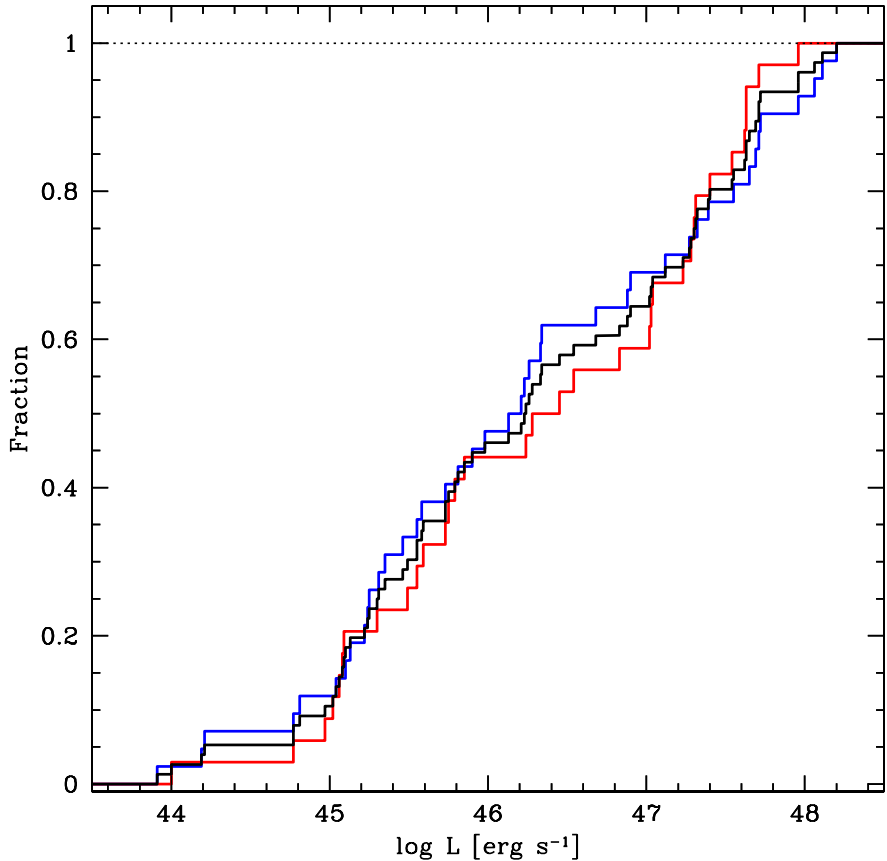}
\caption{Cumulative distribution of bolometric luminosity $L$ of the FOS+HE sample (black), Pop. A (blue) and Pop. B (red). }
\label{fig:lumab}
\end{figure}

The HE+FOS sample has therefore 76 sources, of which 43 are Pop. A and 33 Pop. B.   The distribution of $\log L$ for the 76 sources of the joint sample (derived from the rest-frame luminosity at 1450 \AA, assuming a constant bolometric correction equal to 3.5) uniformly covers the range 44 -- 48.5,  with similar distributions for Pop. A and B (lower panel of Fig. \ref{fig:lumab}; a K-S test confirms that the two distributions are not significantly different). The Eddington ratio (\lledd) covers the range 0.01 -- 1 which means complete coverage of \lledd\ range where most sources in optically-selected samples are found.  

 \section{Emission line profile analysis for the  FOS+HE sample}
\label{anal}

\subsection{Line modeling and measured parameters} 

In the following we consider the merit of \hb\ and \civ\ as VBEs. Previous work has shown that the \hb\ and \mgii\ profiles are almost equally reliable estimators of the ``virial'' broadening in samples of moderate-to-high luminosity \citep[e.g.,][excluding the \mgii\ extreme Population A that is significantly broadened by a blueshifted component, \citealt{marzianietal13}]{wangetal09,trakhtenbrotnetzer12,shenliu12,marzianietal13}. However, the { broad} \hb\ line full profile is often affected by asymmetries toward the line base and by significant line centroid shifts. Typically, the \hb\ line profiles are characterized  by two main asymmetries, differently affecting  sources in  spectral types along the MS { (Appendix \ref{glossary} provides the definition of spectral types)}:
\begin{itemize}
\item Pop. A: a blueshifted excess, often modeled with a blueward asymmetric Gaussian component (BLUE) related to the outflows strongly affecting the \civ\ and \oiiiopt\ line profiles \citep[e.g.][\citetalias{sulenticetal17}, and references therein]{negreteetal18}. 
\item Pop. B: a redward asymmetry modeled with a broader redshifted (FWHM $\sim 10000$ \kms, \cmp $\sim 2000$ \kms) Gaussian. 
The very broad Gaussian is meant to represent the innermost part of the BLR, providing  a simple representation of the radial stratification of the BLR in Pop. B suggested by reverberation mapping \citep[e.g.,][]{sneddengaskell07}. This component (hereafter the very broad component, VBC) has been associated with a physical region of high-ionization virialized and closest to the continuum source   \citep{petersonferland86,brothertonetal94a,sulenticetal00c,sneddengaskell07,wangli11}. While the properties of the Very Broad Line Region (VBLR) remains debatable, a decomposition of the full \hb\ profile into a symmetric, unshifted \hb\ component (\hbbc) and a \hbvbc\ provides an excellent fit to most \hb\ Pop. B profiles \citep{sulenticetal02,zamfiretal10}.  
\end{itemize}

Figs. 4 and 5 of \citetalias{sulenticetal17} show the \hb\ and \civ\ profiles of the HE sample, and their multicomponent interpretation. To extract a symmetric, unshifted component that excluded the blueshifted excess and the VBC, we considered a model of the { broad} \hb\  and \civ\ line with the following components (see also Appendix \ref{glossary}): 
\begin{itemize}
\item Pop. A \hb\ and \civ: an unshifted Lorentzian profile (\hbbc)  + one or more asymmetric Gaussians to model the blueward excess (BLUE). 
\item Pop. B \hb\ and \civ: an unshifted Gaussian (\hbbc) + a redshifted VBC for \hb\ (\hbvbc). In the \hb\ case, there is no evidence of a blueward excess even at the highest luminosity. However, among Pop. B sources of the HE sample, a prominent \civ\ BLUE appears, implying an intensity ratio \civ/\hb$\gg1$ in the BLUE component. The \civ\ BLUE is usually  fainter in the low-luminosity FOS sample \citepalias{sulenticetal17}.  
\end{itemize}
{ In the fits, narrow components of both \hb\ (\hbnc) and \civ\ (\civnc) were included. In the case of \civ, separation of the broad and narrow component is subject to significant uncertainty, so that the effect of the \civnc\ needs to be carefully considered (see discussion in Sect. \ref{civnc}).  }

{ The decomposition approach summarized above has a heuristic value, as the various components are not defined on the basis of a physical model, even if the assumptions on line shapes follows from  MS trends. The distinction between BC and VBC might be physically motivated (the emitting region associated with the BC is the one emitting most of all \feii), but the decomposition into two symmetric Gaussians is  a crude approximation at best. Full profile measures are added to avoid any exclusive dependence of the results on the profile decomposition.}
The full profiles of \hb\ and \civ\ are parameterized by the FWHM, an asymmetry index (AI) and centroid at  fractional intensity at 1/2 and 1/4 of peak, \cmp\ and \cqp. The definition of centroids and A.I.  follows \citet{zamfiretal10}: 

\begin{equation}
c(\frac{i}{4}) = \frac{\lambda_\mathrm{B}(\frac{i}{4})+\lambda_\mathrm{R}(\frac{i}{4})}{2 \lambda_{0}} c, \, i=1,2,3; \frac{i}{4} = 0.9 
\end{equation}

\begin{equation}
A.I. = \frac{\lambda_\mathrm{B}(\frac{1}{4})+\lambda_\mathrm{R}(\frac{1}{4})- 2\lambda_\mathrm{P}}{\lambda_\mathrm{B}(\frac{1}{4})+\lambda_\mathrm{B}(\frac{1}{4})}
\end{equation}

where $\lambda_\mathrm{P}$\ is the peak wavelength, and $\lambda_\mathrm{B}$\ and $\lambda_\mathrm{R}$ are the wavelengths on the blue and red side of the line at the $i/4$ fractional intensities. The centroids are referred to the quasar rest frame, while the AI is referred to the peak of the line that may be shifted with respect to rest-frame.  A proxy to $\lambda_\mathrm{P}$\ which will be used in this paper is $ \tilde{\lambda_\mathrm{P}} \approx \lambda_{0}(1+c(0.9)/c)$.  

We assume that the symmetric and unshifted \hbbc\ and \civbc\ are the representative line components of the virialized part of the BLR. It is expedient to define a parameter $\xi$\ as follows:
\begin{equation}
\xi_\mathrm{line} = \frac{{\rm FWHM}_\mathrm{vir}}{\rm FWHM}
\label{eq:xi}
\end{equation}

where the FWHM$_\mathrm{vir}$\ is the FWHM of the ``virialized'' component, in the following assumed to be \hbbc, and the FWHM\ is the FWHM measured on the full profile (i.e., without correction for asymmetry and shifts). The $\xi$\ parameter is a correction  factor that can be defined also using  components of different lines, for instance \civ\ full profile FWHM and \hbbc, where \hbbc\ is assumed to be a reference VBE.  
 
\subsection{The \civ\ narrow component in the HE sample and its role in FWHM \civbc\ estimates}
\label{civnc}
 

In only two cases does \civnc\  contribute to the total \civ\ flux  of the HE Pop. B sources by more than 10\%:  \object{HE2202-2557} and \object{HE2355-4621} (Pop. B, Fig. 5 of \citetalias{sulenticetal17}).  There is no evidence for a strong NC in the HE Pop. A sources except  for  \object{HE0109-3518} where I(\civnc)$\lesssim$ 0.09 of the total line flux and whose \civ\ profile resembles the ones of low-$z$\ sources that are 2-3 dex less luminous (the \object{HE0109-3518} \civ\ profile is shown in Fig. 4 of \citetalias{sulenticetal17}). 

In general, considering \hbbc\ as a reference for Pop. B sources, and comparing  FWHM \hbbc\ to FWHM \civ\  with and without removing the \civnc (i.e., to FWHM \civbc\ and FWHM \civbc + \civnc), the  \civnc\ removal improves the agreement with FWHM \hbbc\  in 5 cases out of 6 when \civnc\ has an appreciable effect on the line width (in the other eight cases there is no effect because \civnc\ is too weak).  The  FWHM measured on the \civ\ profiles without removing the \civnc\ (i.e., FWHM \civbc +\civnc) are, on average, $\approx$ -4 \%\ and  -11 \%\  of the  FWHM \civ, for Pop. A and B respectively.  Therefore, (1) subtracting the \civnc\ improves the agreement between \hb\ and \civ\ FWHM; (2) the average effect is too small to   affect our inferences concerning on the \civ\ line width as a VBE in the HE sample.  The \civnc\ has been always included as an independent component in the line profile fitting of \citetalias{sulenticetal17}, following an approach consistently applied for the low-$z$ FOS sample and described by \citet{sulenticetal07}.
 
\section{Results}
\label{virial}

\subsection{\hb\ in the HE sample}


We considered  several different measures of the \hb\ width following 
empirical corrections derived from previous work on low-$z$ samples:
 
 \begin{itemize}
\item substitution of the \hbbc\ extracted through the {\tt specfit} analysis  in place of the full \hb\ profile.  \end{itemize}

In principle, extraction of the \hbbc\ should be the preferred approach, and the FWHM \hbbc\ the preferred VBE.  To test  the reliability of the FWHM values, we performed  Monte-Carlo repetitions of the \hb\ fit for Pop. B sources with the broadest lines (FWHM \hbbc $\sim $ 7000 \kms, and FWHM \hbvbc $\sim$ 11000 \kms), under the assumption of S/N $\approx$20,\footnote{{ S/N is measured per pixel on the continuum.}} weak and relatively broad \oiiiopt, changing noise pattern and initial values of the fitting. The values of FWHM \hbbc\ and \hbvbc\ were chosen to represent  the broadest lines, where FWHM \hbbc\ measures  might be affected by a degeneracy in the BC+VBC decomposition. The dispersion of the Monte Carlo FWHM distribution is almost symmetric, and implies typical FWHM \hbbc\ uncertainties $\approx 10$\%\ at 1$\sigma$\ confidence level. Therefore, the blending should not be a source of strong bias or of large uncertainties in the \hbbc\ and \hbvbc\ FWHM.\footnote{If S/N\ is relatively high $(\gtrsim 20$) only in some peculiar cases the uncertainty might be significantly larger. For example, if the \hb\ profile is composed for a narrower core and a broader base, the FWHM measure is unstable, and may abruptly change depending on continuum placement.}  However, we still expect that in the case of very broad profiles, and low S/N or low dispersion, the decomposition of the \hb\ profile into \hbbc\ and \hbvbc\ is subject to large uncertainties difficult to quantify. To retrieve information on the \hbbc\ we introduce several corrections that can be applied to the full \hb\ profile without any multicomponent fitting (which makes the results also model-dependent).

\begin{figure}[htp!]
\centering
\includegraphics[width=0.4\columnwidth]{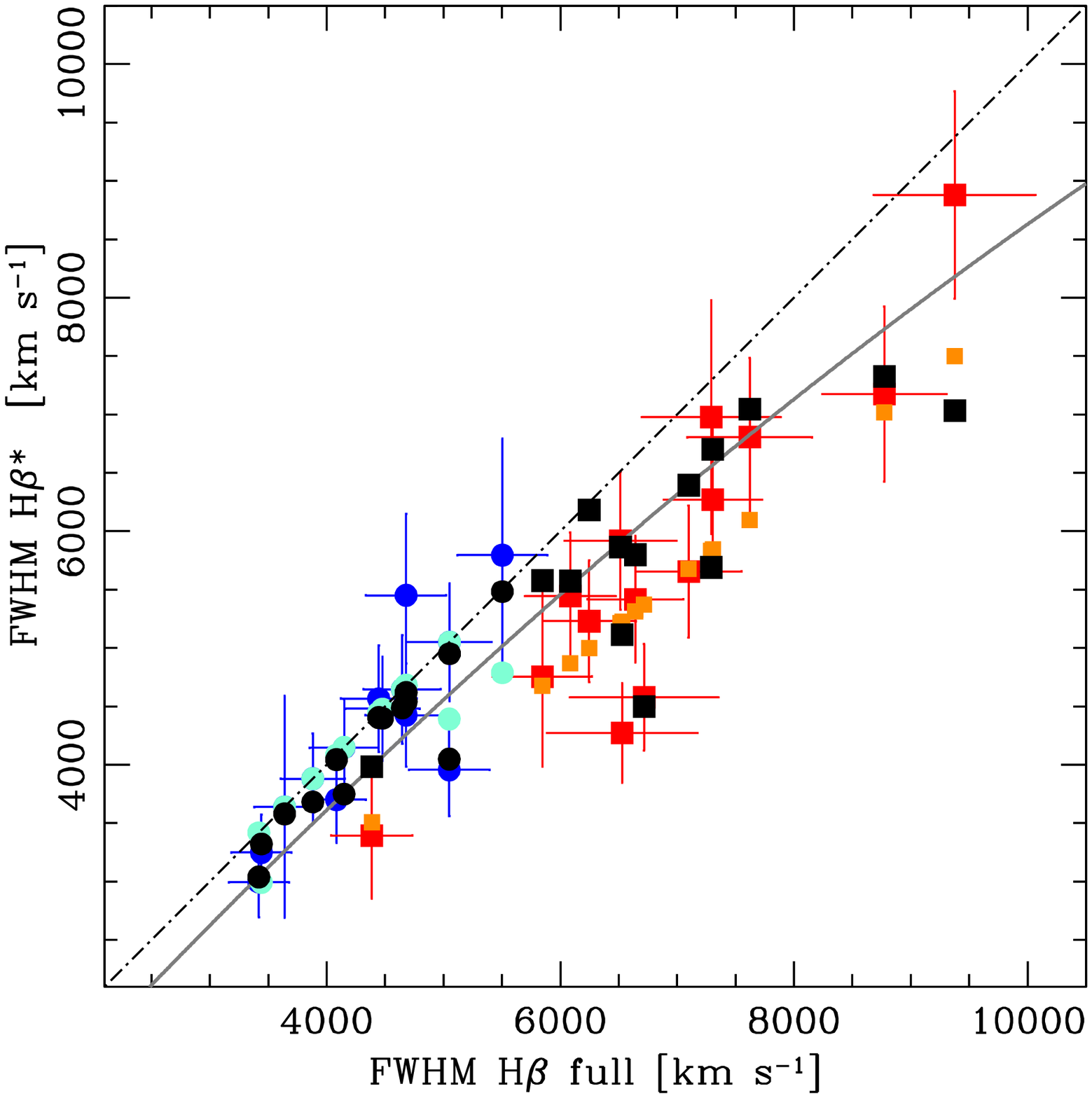}\\
\includegraphics[width=0.4\columnwidth]{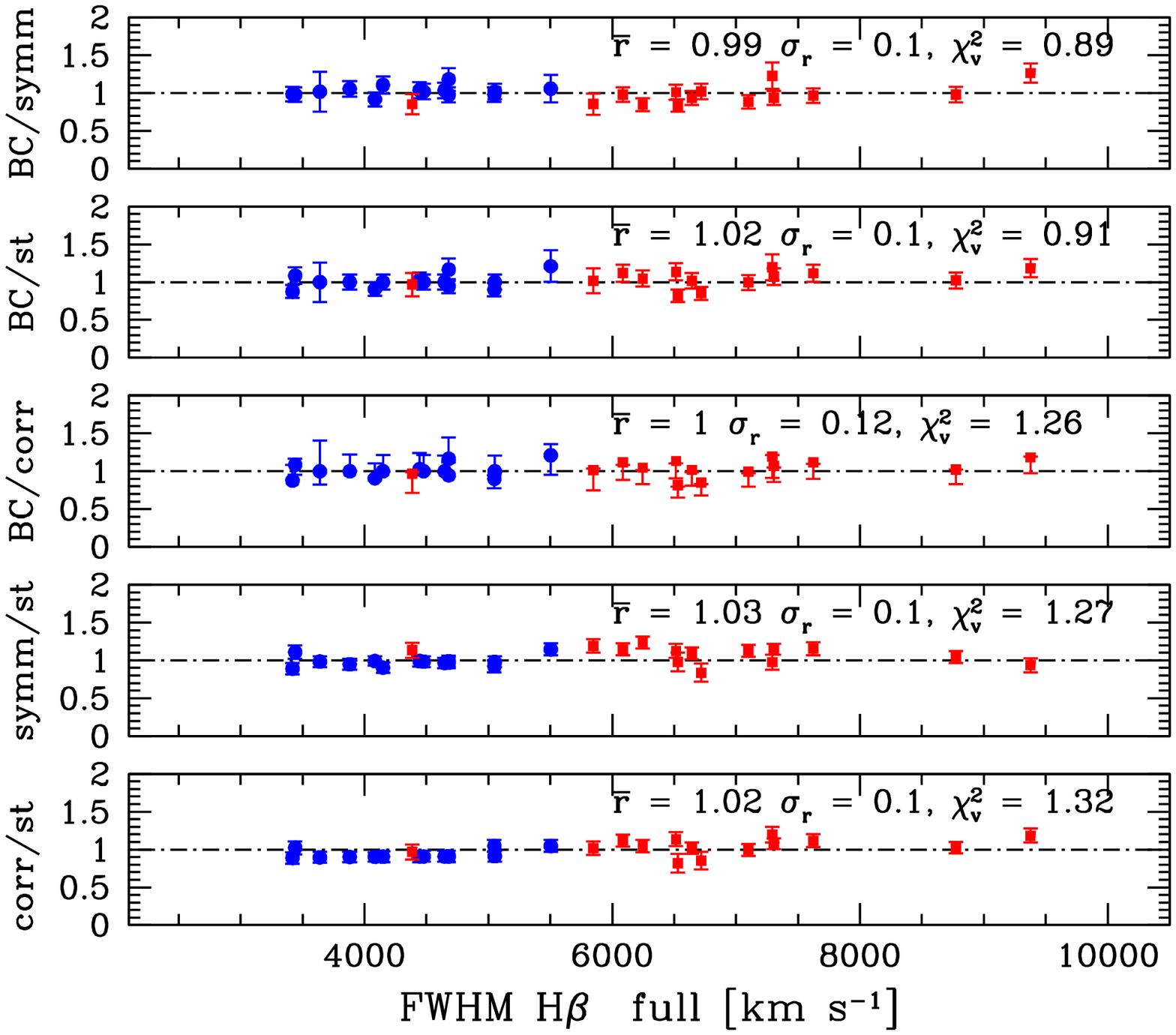}\\
\vspace{-1cm}
\includegraphics[width=0.4\columnwidth]{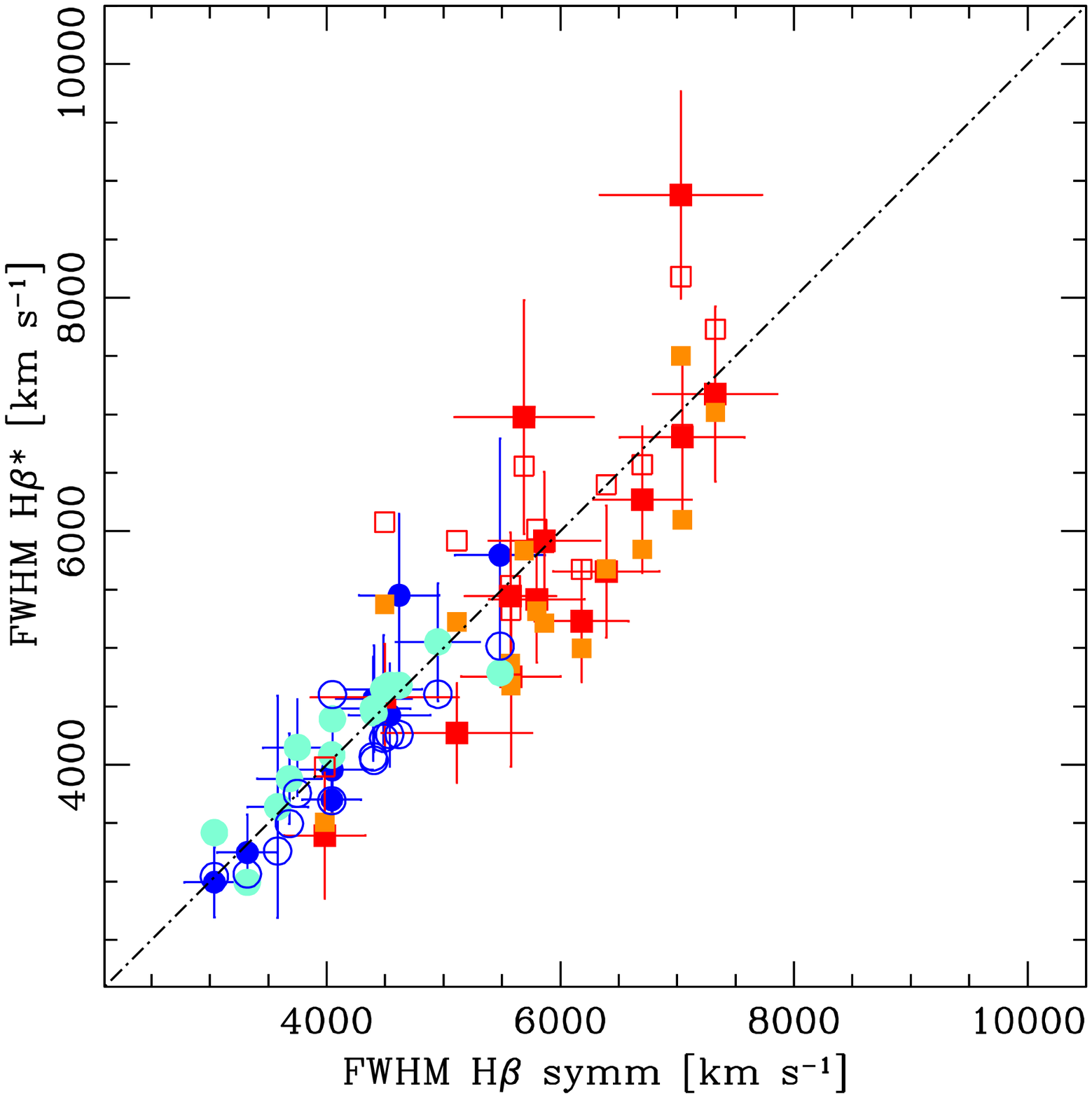}\\
\caption{Virial broadening estimators based on \hb, with several corrections  applied to the HE sample. Top square panel:  FWHM \hbbc\  (blue (Pop. A) and red (Pop. B)) with error bars, symmetrized FWHM \hb\  ({\tt symm},  black), and FWHM \hb\ corrected according to spectral type   ({\tt st}, aquamarine and dark orange for Pop. A and B respectively) versus  FWHM of the full \hb\ profile. The grey line traces   the correction ({\tt corr}) following the relation of \citet{sulenticetal06} reported in \S \ref{virial}. Middle panels: ratios of FWHM after various corrections vs. full profile FWHM. First panel from top:  BC/{\tt symm},  BC/{\tt st},  BC/{\tt corr},  {\tt symm}/{ \tt st}, {\tt corr}/{\tt st}. Average values, standard deviation and normalized $\chi^{2}_{\nu}$\ are reported in the upper right corner of the panels. Bottom square panel:  BC,  spectral type {\tt st} and reverberation corrected ({\tt corr}, open symbols) FWHM values vs. symmetrized FWHM. 
}
\label{fig:virialhb}
\end{figure}

\begin{itemize}
\item symmetrization of  the full profile: FWHM$_\mathrm{symm} $ =  FWHM -- 2 \cmp\  ({\tt symm} in Fig. \ref{fig:virialhb}). The physical explanation behind the symmetrization approach involves  an excess radial velocity on the red side that may be due to gas with a radial infall velocity component, with velocity increasing toward the central black hole \citep[e.g.,][and references therein]{wangetal17}. Generally speaking, redward displacements of line profiles have been explained by invoking a radial infall component plus obscuration \citep{huetal08,ferlandetal09};
\item { Substitution of  the  FWHM \hbbc\ with the FWHM measured  on the full broad profile of \hb, corrected according to its  spectral type.} The spectral types have been assigned following \citet[][for a conceptually equivalent approach see \citealt{shenho14}]{sulenticetal02}. The correction are as defined from the analysis of the \hb\ profile in a large SDSS-based sample at $0.4 \ltsim z \ltsim 0.7$\ (labeled as {\tt st} in Fig. \ref{fig:virialhb}). In practice, this means to correct \hb\ for Pop. B sources by a factor $\xi_\mathrm{H\beta} \approx 0.8$\ \citep{marzianietal13a} and extreme population A sources (\rfe $\ge 1$) by a factor $\xi_\mathrm{H\beta} \approx 0.9$. On average, spectral types A1 and A2 show symmetric profiles for which $\xi_\mathrm{H\beta} \approx 1$.   Recent work confirmed that the effect of a blueshifted excess on the full profile of \hb\ is small at half-maximum, $0.9 \lesssim \xi_\mathrm{H\beta} \lesssim 1.0$ \citep{negreteetal18}. We assume $\xi_\mathrm{H\beta} = 0.9$\ as an average correction. The  ratio we derive between BC and full profile FWHM of HE Pop. B \hb\ is $\approx 0.82 \pm 0.09$, consistent with the same ratio estimates at moderate luminosity \citep{marzianietal13a}. The {\tt st} correction can be summarized as follows:

\begin{center}
\begin{tabular}{lc}\hline
ST & $\xi_\mathrm{H\beta}$ \\
\hline
A3-A4 & 0.9\\
A1-A2 & 1.0\\
B1-B1+& 0.8\\
\hline
\end{tabular}
\end{center}

\item correction of the width of the full broad \hb\ profile based on the one derived at low $z$\ by pairing the observed  full broad \hb\ FWHM to the best width estimator from reverberation mapping, following the relation  FWHM$_\mathrm{c}\approx 1.14$  FWHM$ - 601- 0.0000217$FWHM$^{2}$\ derived by \citet[][{labeled \tt corr}]{sulenticetal06};
\end{itemize}

Fig. \ref{fig:virialhb} shows that these corrections all provide similar results if applied to the HE sample FWHM \hb. Error bars  of Fig. \ref{fig:virialhb} were estimated propagating the uncertainty values reported in \citetalias{sulenticetal17} for the full profiles, and the ones derived from {\tt specfit} for the line components (assuming a minimum error of 10\%).  


The middle  panels of  Fig. \ref{fig:virialhb} show the ratios of corrected FWHM measures as a function of the  FWHM of the full \hb\ profile. The low $\chi^{2}_{\nu}$ indicates that the $\chi_{\nu}^{2}$  associated with the ratios between {\tt BC} and {\tt symm}, and  {\tt BC} and {\tt st} not significantly different from unity.  In the case of {\tt BC} and {\tt corr} the two measurements are different but only at 1$\sigma$\ confidence level.  $F$ tests do not  exclude that BC, symmetrization, {\tt st}, and reverberation corrections can be  equivalent at a minimum confidence level of $2\sigma$.  The bottom panel of Fig. \ref{fig:virialhb} shows the behavior of the full and corrected FWHM versus the ``symmetrized'' \hb\ FWHM.  We consider the symmetrization, as it is relatively easy to apply (once the quasar rest frame is known), and the {\tt st} correction (that does not even require the knowledge of the rest frame) as reference corrections.   We remark again that these corrections are relatively minor { but still significant: a 20\%\ correction translates into a factor 1.44 correction in \mbh}. They do not undermine the value of the full line width of \hb\ as a useful VBE (with the caveats discussed in Sec. \ref{wind}), { since the \hb\ full line width remains preferable to the uncorrected \civ\ width for most objects}.



\subsection{\civ\ in the full HE+FOS sample}

The results on the HE \civ\ profiles  do not bode well for  the use of \civ\ FWHM as a VBE, as also found by \citet{sulenticetal07} and other workers (Sect. \ref{disc:virial} for a  brief critical review). The presence of very large blueshifts in both Pop. A and B makes the situation even more critical than at low $L$.     Fig. \ref{fig:virialciv} (top panel) shows that there is no obvious relation between the FWHM of \civ\ and the FWHM of \hb\ if  FOS+HE data are considered together. 


For the Pop. A sources in the HE+FOS sample, \civ\ is broader than \hb\ save in  two  cases in the HE sample, and     FWHM(\civ) shows a broad range of values for similar FWHM \hb\ i.e., FWHM(\civ) is almost degenerate with respect to \hb.  The \civ\ line FWHM values are so much larger than the ones of \hb\  to make it possible that the   \mbh\ derived from FWHM \civ\ might be higher by  even more than one order of magnitude.   Formally, the Pearson's correlation coefficient  $r \approx$ 0.52 is highly significant for a sample of $n = 43$, with significance at an $\approx 4.5 \sigma$ confidence level. 
A weighted least square fits yields FWHM(\civ) = (1.822 $\pm$ 0.204) FWHM(\hb) +( -624 $\pm$ 677) \kms, with a significant scatter, rms $\approx$1959 \kms. Unfortunately it is not possible to   apply a simple \civ\ symmetrization as done for \hb: subtracting 2 $\cdot$ \cmp\ to FWHM \civ\ leads to corrections that are unrealistically large.


If we combine the Pop. B FOS and HE samples,  FWHM \civ\ and \hb\ become loosely correlated (the Pearson's correlation coefficient is $\approx 0.4$, significant at $P \approx 98$\% for a sample of 33 objects). A weighted least-square fit yields FWHM(\civ) $\approx (0.764 \pm 0.165)$ FWHM(\hb) $+ (810 \pm 1030)$\ \kms, and rms $\approx 1090$ \kms, with a significant deviation from the 1:1 relation. In the case of Pop. B sources, the  trend  implies FWHM \civ $\sim$ FWHM \hb, and even a {\em slightly narrower}  FWHM \civ\ with respect to \hb. 

The large scatter induced by   using uncorrected \civ\ line FWHM may have contributed to the statement that line width does not contribute much to \mbh\ determinations \citep{croom11}.  

\subsection{Practical usability of \civbc}

The fitting procedure  scaled the \hb\ profile to model the red side of \civ\ so that the FWHM \civbc\ estimate is not independent from FWHM \hbbc. The  FWHM values of the two BCs are in agreement because of this enforced condition.  


The \civbc\ extraction is very sensitive to the assumed rest frame, and also requires that the \civ\ line is cleaned from contaminant such as \feii\ (weak) and \heiiuv\ (moderate, but flat topped and gently merging with the \civ\ red wing; \citealt{marzianietal10,fineetal10,sunetal18}). Without performing a line profile decomposition, one can consider the width of the red side with respect to rest frame as the half-width half maximum (HWHM) of the virial component. Again this requires (1) an accurate redshift that can be set, in the context of high $z$\ quasars,  either by using the \hb\ narrow component or by the \oii\ doublet \citep{eracleoushalpern04,huetal08}, and (2) the decomposition from \heiiuv\ emission blended on the \civ\ red side.  If \oii\ is covered, then \mgii\ is also likely to be covered. As mentioned in Sect. \ref{anal}, the \mgii\ line width is a  reliable VBE for the wide-majority of type-1 AGN. The same is not true for \civ. For spectra where \civ\ is conveniently placed at $z \gtrsim 1.45$, the \oii\ line is shifted beyond 9000 \AA, a domain where intense sky emission makes it difficult to analyze a relatively faint narrow line. The extraction of \civbc\ is therefore not a viable solution if single-epoch  \civ\ observations are available without the support of at least a narrow LIL that may set a reliable rest frame. This is unlikely to occur on the same optical spectra. An alternative strategy for \mbh\ estimation using \civ\ FWHM should consider the origin of the \civ\ non-virial broadening. 

\subsection{Reducing \civ\ to a VBE estimator consistent with \hb}
\label{redu}

The main results of \citetalias{sulenticetal17} suggest a strong dependence of the \civ\ blueshift on \lledd, especially above a threshold value \lledd $\approx 0.2 \pm 0.1$\ \citep[][and references therein]{sulenticetal14}. A correlation between Eddington ratio  and the FWHM(\civ) to FWHM(\hb) ratio   \citep[i.e., $1/\xi_\mathrm{CIV}$, c.f.,][]{saitoetal16}  is detected at a high significance level (Pearson's correlation coefficient $r \approx 0.55$) joining all FOS RQ sources of \citet[][Fig. \ref{fig:virialcivbr}]{sulenticetal07}.  In this context,  \lledd\  was computed from the \mbh\ scaling law of \citet{vestergaardpeterson06}, using the FWHM of \hb\ and  $\lambda L_{\lambda}$(5100). A bisector best fit with {\tt SLOPES} \citep{feigelsonbabu92} yields

\begin{equation}
\log \frac{1}{\xi_\mathrm{CIV}} \approx (0.426 \pm 0.043) \frac{L}{L_\mathrm{Edd}} + (0.401 \pm 0.035).
\end{equation}

An \lledd\ -- dependent correction is in principle a valid approach. However, it is not obvious how to calculate \lledd\ from UV spectra without resorting to \hb\ observations. In addition  FWHM  \hb\ is strongly affected by orientation and yields  biased values of \lledd\ (Sect. \ref{orien}).    Both FWHM(\hbbc)/FWHM(\civ) and \cmp\ are both correlated with  Eddington ratio. Consistently, the \civ\ blueshift is correlated with  FWHM \civ\ \citep[\citetalias{sulenticetal17}, ][]{coatmanetal16}, and accounts for the broadening excess in the \civ\ FWHM.  Measures of the \civ\ blueshift or  the FWHM(\hbbc)/FWHM(\civ) can be used as  proxies for \lledd.   At the same time,  \citetalias{sulenticetal17} reveals a weaker correlation with $L$, which is expected in the case of a radiation driven wind.   If the correction factor is  $\xi_\mathrm{CIV} = $FWHM(\hbbc)/FWHM(\civ), then it should include a term in the form $1/\zeta(L, $ \lledd).

\begin{figure}[htp!]
\centering
\includegraphics[width=0.87\columnwidth]{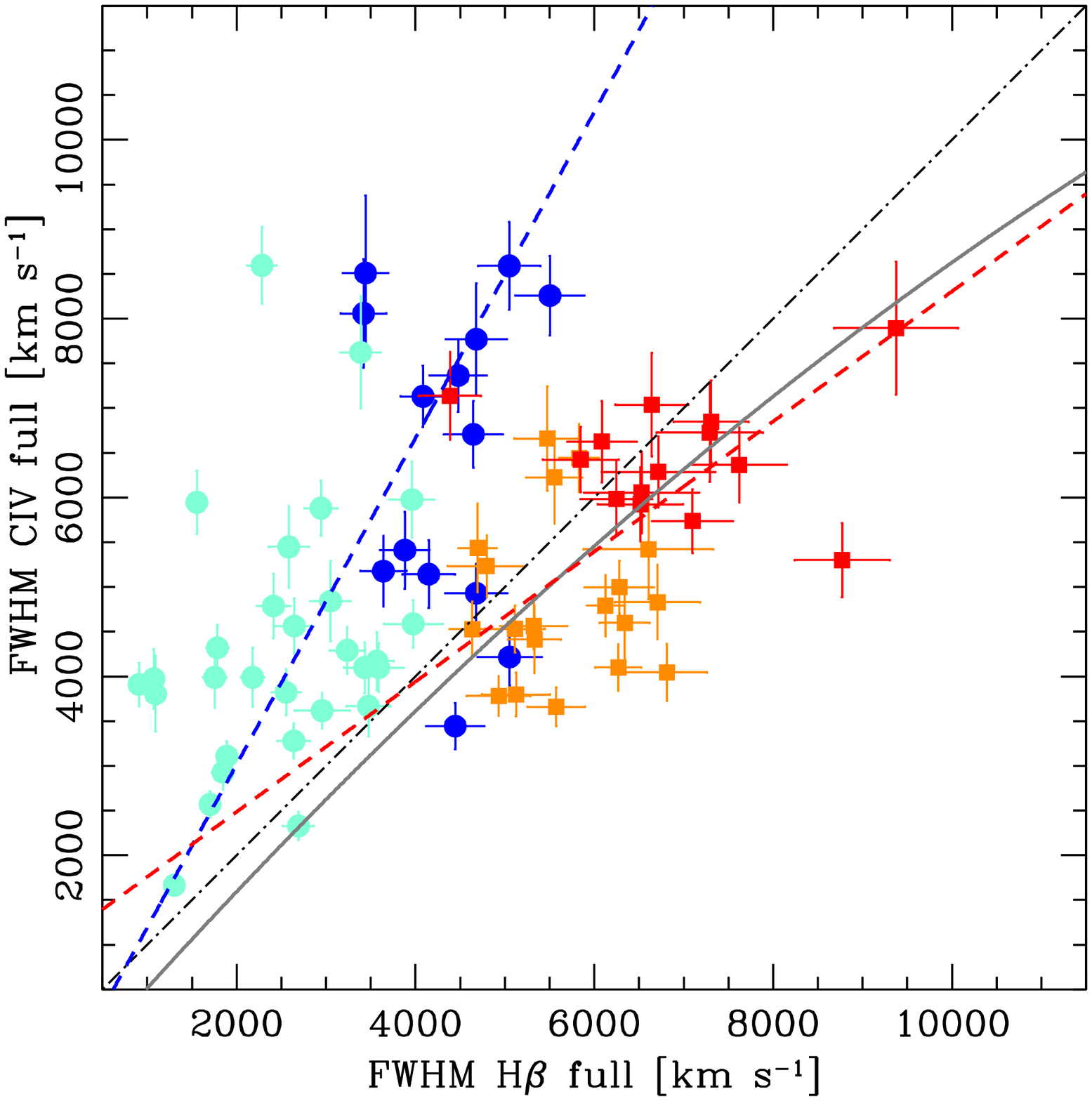}
\includegraphics[width=0.87\columnwidth]{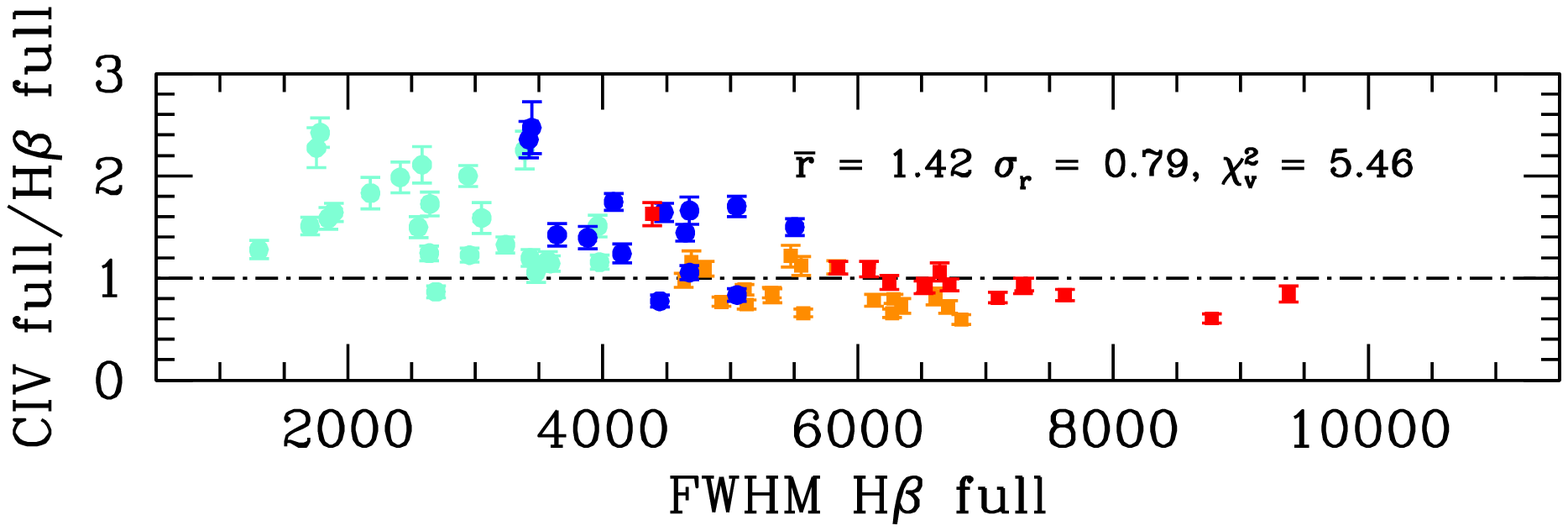}
\vspace{-5.cm}
\caption{Top panel: FWHM(\civ) vs FWHM(\hb) (full profiles) for the FOS+HE sample. { Data points are color-coded according to sample and population. HE Pop. A: blue circles   (\textcolor[rgb]{0,0,1}{$\bullet$}), HE Pop. B: red squares   (\crule[red]{0.15cm}{0.15cm}),  FOS Pop. A: aquamarine circles (\textcolor{aquamarine}{$\bullet$}), FOS Pop. B: golden squares (\crule[darkorange]{0.15cm}{0.15cm})}.  Best fitting lsq lines (dashed) are shown in the blue  for all Pop. A and red for all Pop. B. The black dot dashed line is the equality line. The continuous grey line is the expected FWHM following the correction of \citet[][{\tt corr}]{sulenticetal06}. 
Lower panel: ratio between FWHM \civ\ and FWHM \hb\ as a function of FWHM (\hb). }
\label{fig:virialciv}
\end{figure}

\begin{figure}[htp!]
\centering
\includegraphics[width=0.9\columnwidth]{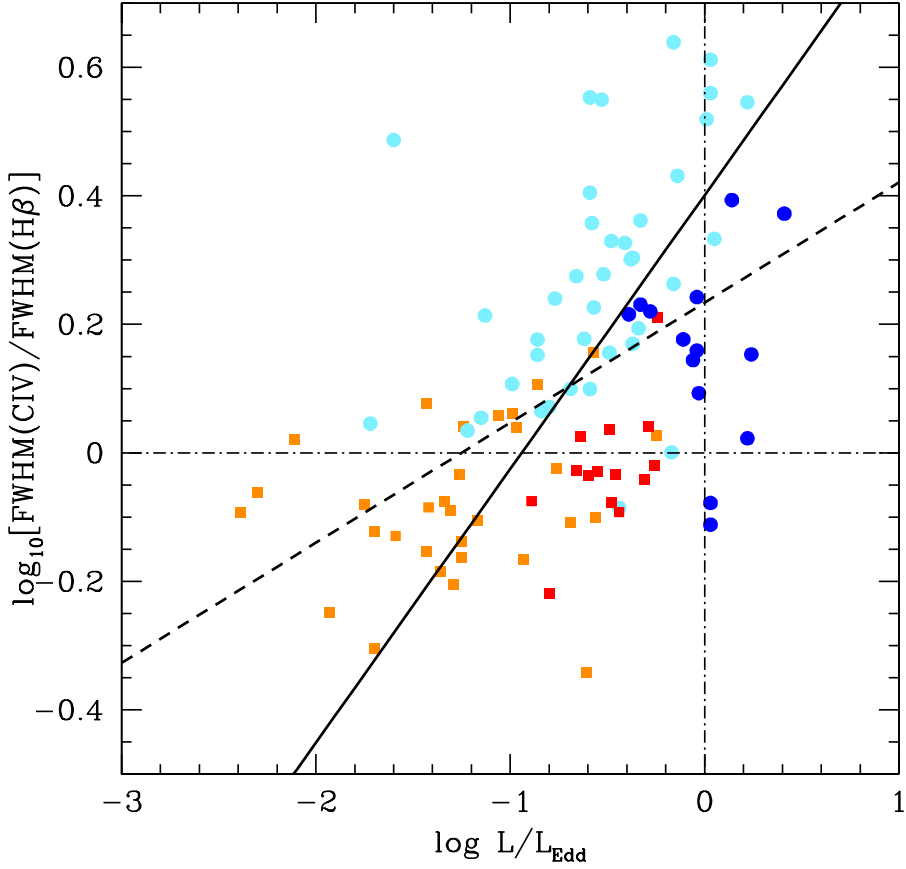}
\caption{Relation between the logarithm of the FWHM ratio of \civ\ to \hb\ and the logarithm of the Eddington ratio \lledd. The vertical dot-dashed line traces the Eddington limit.  The colors and shape of symbols are as in Fig. \ref{fig:virialciv}. The dashed line is an unweighted least squares fit, the filled line was obtained with the bisector method \citep{feigelsonetal92}.  }
\label{fig:virialcivbr}
\end{figure}

\begin{figure}[htp!]
\centering
\includegraphics[width=0.95\columnwidth]{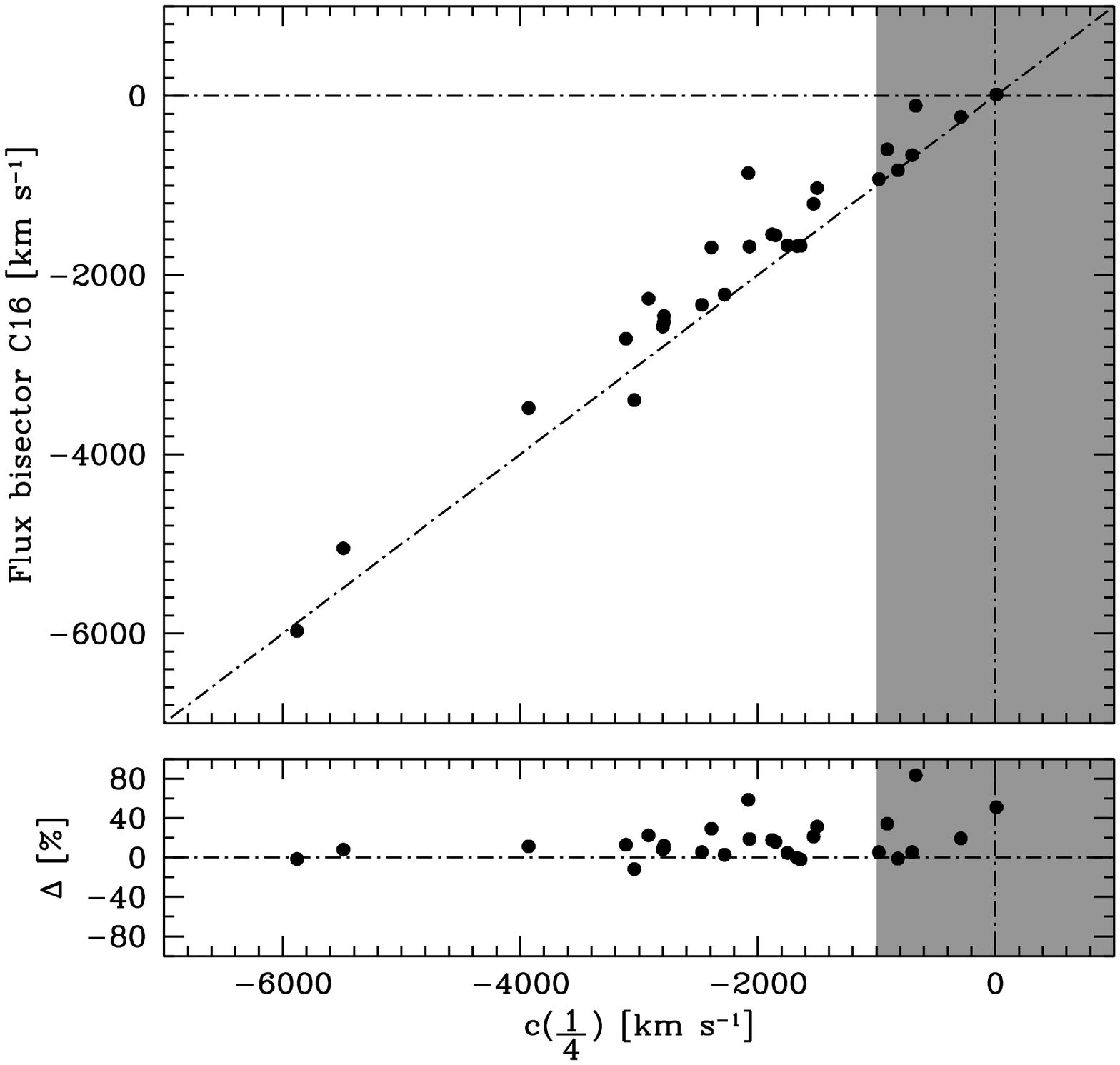}
\caption{Top panel: bisector flux estimator of \citet{coatmanetal17} vs \cqm, in \kms. The dot-dashed line is the equality line.  Bottom panel: Percentage residuals. The shaded areas indicate the average error of measurement at a $2\sigma$\ confidence level computed from Table 4 of \citetalias{sulenticetal17}. }
\label{fig:coatc}
\end{figure}

\subsection{Calibrating empirical corrections on FWHM \civ}
\label{empcorr}

\begin{figure}[htp!]
\centering
\includegraphics[width=0.9\columnwidth]{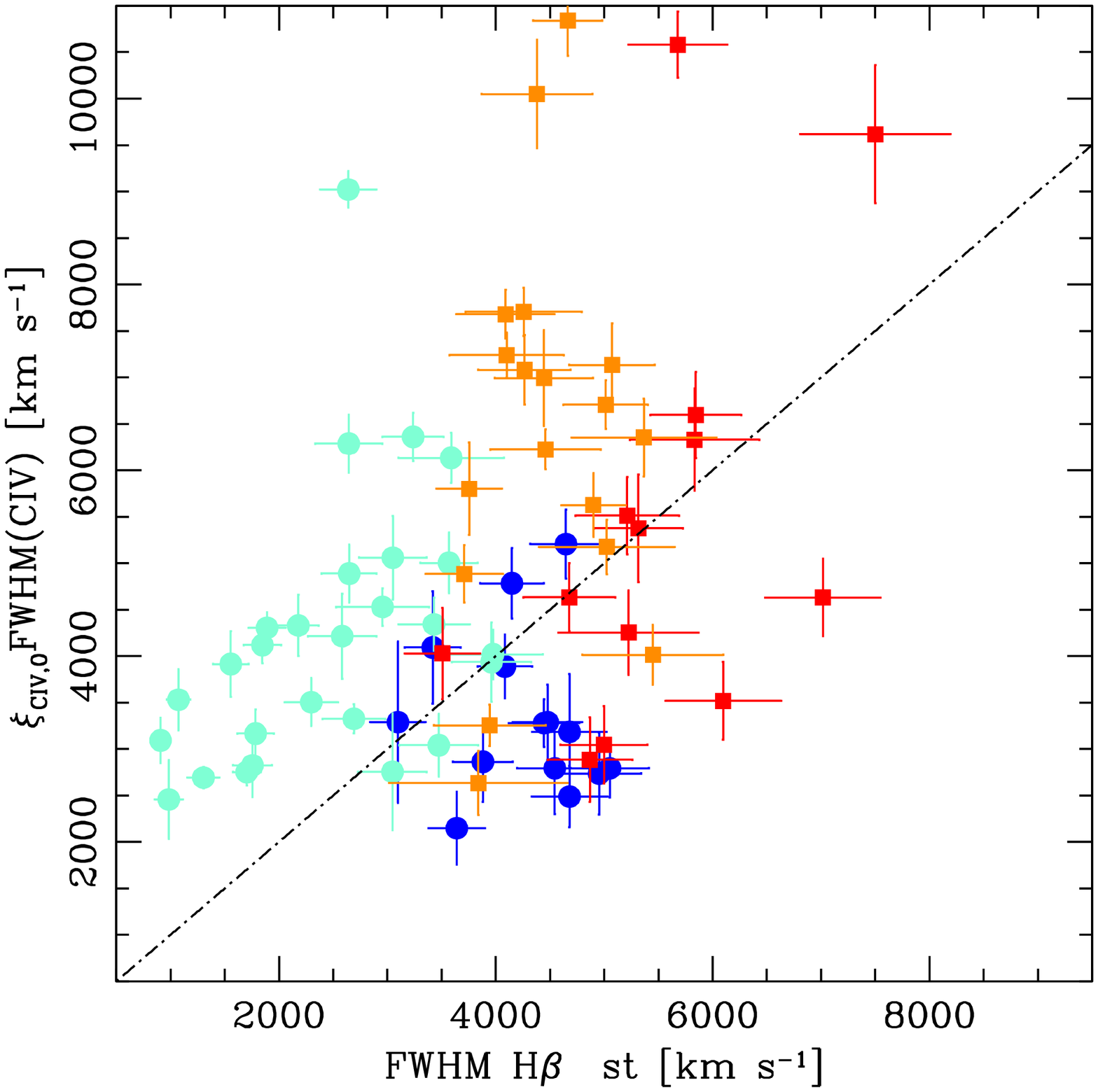}\\
\vspace{-1.25cm}
\includegraphics[width=0.9\columnwidth]{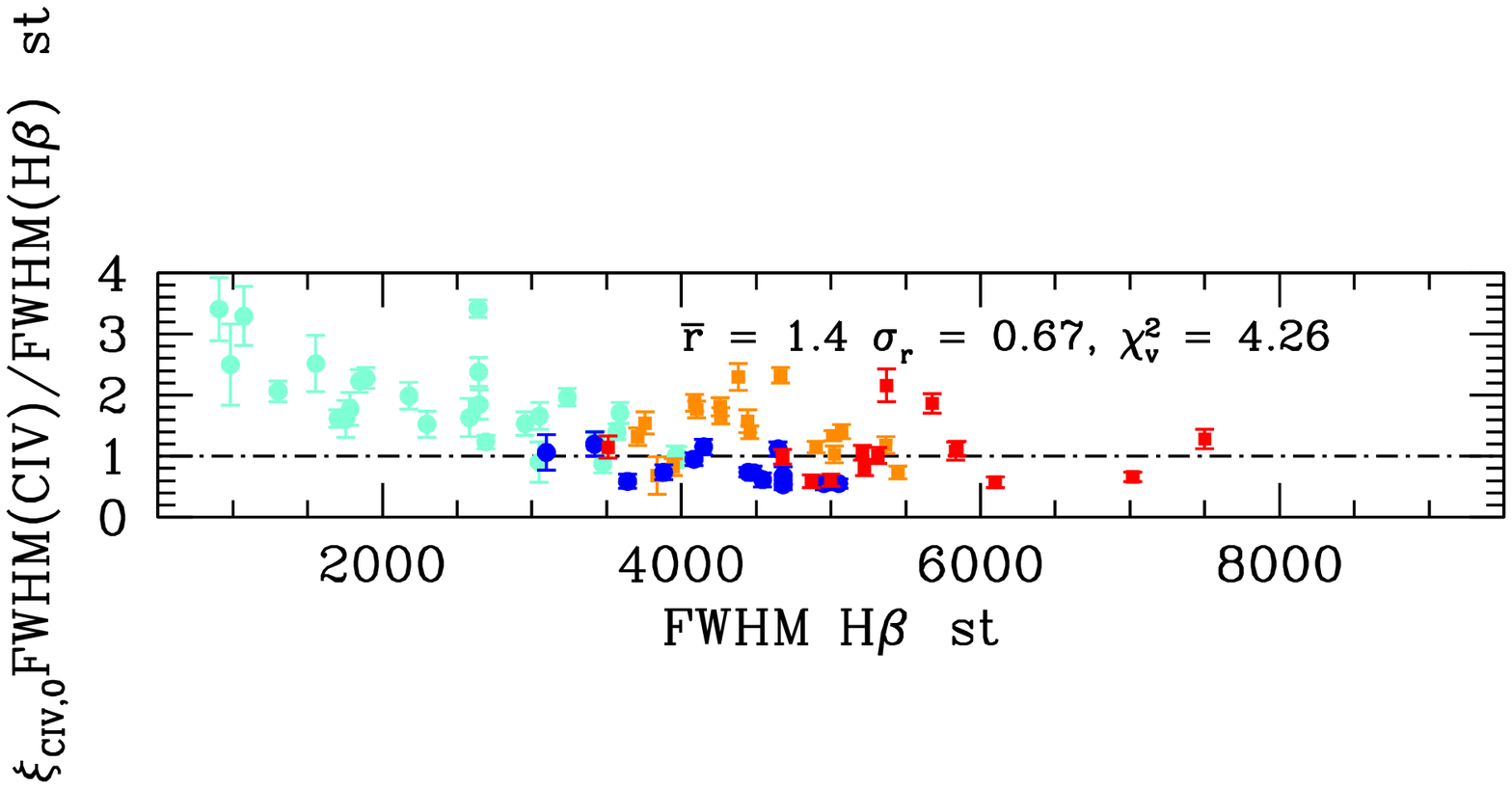}\\
\vspace{-4cm}
\caption{Top panel: FWHM(\civ) {\tt C16} i.e., corrected following \citet{coatmanetal16} vs  FWHM(\hb) {\tt st} for the  FOS+HE sample. Meaning of symbols is the same as for Fig. \ref{fig:virialciv}.    The black dot dashed line is the equality line. The bottom panel shows the residuals. Average ratio, dispersion and $\chi^{2}_{\nu}$\ refer to all sources. }
\label{fig:wcoat}
\end{figure}

\citet{coatmanetal17} introduced a non-parametric measure of the \civ\ blueshift associated with the wavelength that splits the line flux in equal parts on its blue and red side (flux bisector). The flux bisector is strongly correlated  with \cmp\ and \cqp, and \cmp\ and \cqp\ are correlated among themselves in the  FOS+HE sample (Pearson's $r \approx 0.95$):
$c(\frac{1}{2}) = (0.773 \pm 0.307) c(\frac{1}{4}) -(58 \pm 65)\, \, \mathrm{km \, s^{-1}}$. The bisector correlation is stronger with  \cqp\  (Pearson's correlation coefficient $r \approx 0.98$),  with flux bisector  $ \sim (0.98 \pm 0.04) c(\frac{1}{4}) +(220 \pm 110)\, \, \mathrm{km \, s^{-1}}$  (Fig. \ref{fig:coatc}).  The lower panel of Fig. \ref{fig:coatc} shows a few objects with  difference between \cqp\ and flux bisector $\gtrsim20$\%; these sources are either with small shifts (within the measurement uncertainties; shaded area of Fig. \ref{fig:coatc}), or sources strongly affected by broad absorptions, for which a measure of blueshift is tricky regardless of the method employed.  Therefore it is possible to apply Eq.  4 of \citet{coatmanetal17} substituting the \cqp\ to the flux-bisector blueshift measurements:
\begin{equation}
\xi_\mathrm{CIV,0} = \frac{1}{a \left(- \frac{c(\frac{1}{4})}{1000} \right) + b}
\label{eq:xic}
 \end{equation}

with $a =0.41 \pm 0.02$ \ and $b \approx 0.62 \pm 0.04$\ (the minus sign is because \citet{coatmanetal17} assumed blueshifts to be positive),  to correct the FWHM \civ\ of the FOS+HE sample.  The resulting trend is shown in Fig. \ref{fig:wcoat}. The Eq. 4 of \citet{coatmanetal17} undercorrects both Pop. A and B sources at low $L$\ (the FOS sample) and provides a slight overcorrection  for the HE sources.   {  The correction of \citet{coatmanetal17} does not yield FWHM (\civ) in agreement with the observed values of FWHM(\hb).  This does not necessarily mean that the FWHM(\civ) values are incorrect, as FWHM(\hb) is likely more strongly affected by orientation effects than FWHM(\civ) (see the discussion in Sect. \ref{wind}). }

\begin{figure*}[htp!]
\centering
\includegraphics[width=0.45\columnwidth]{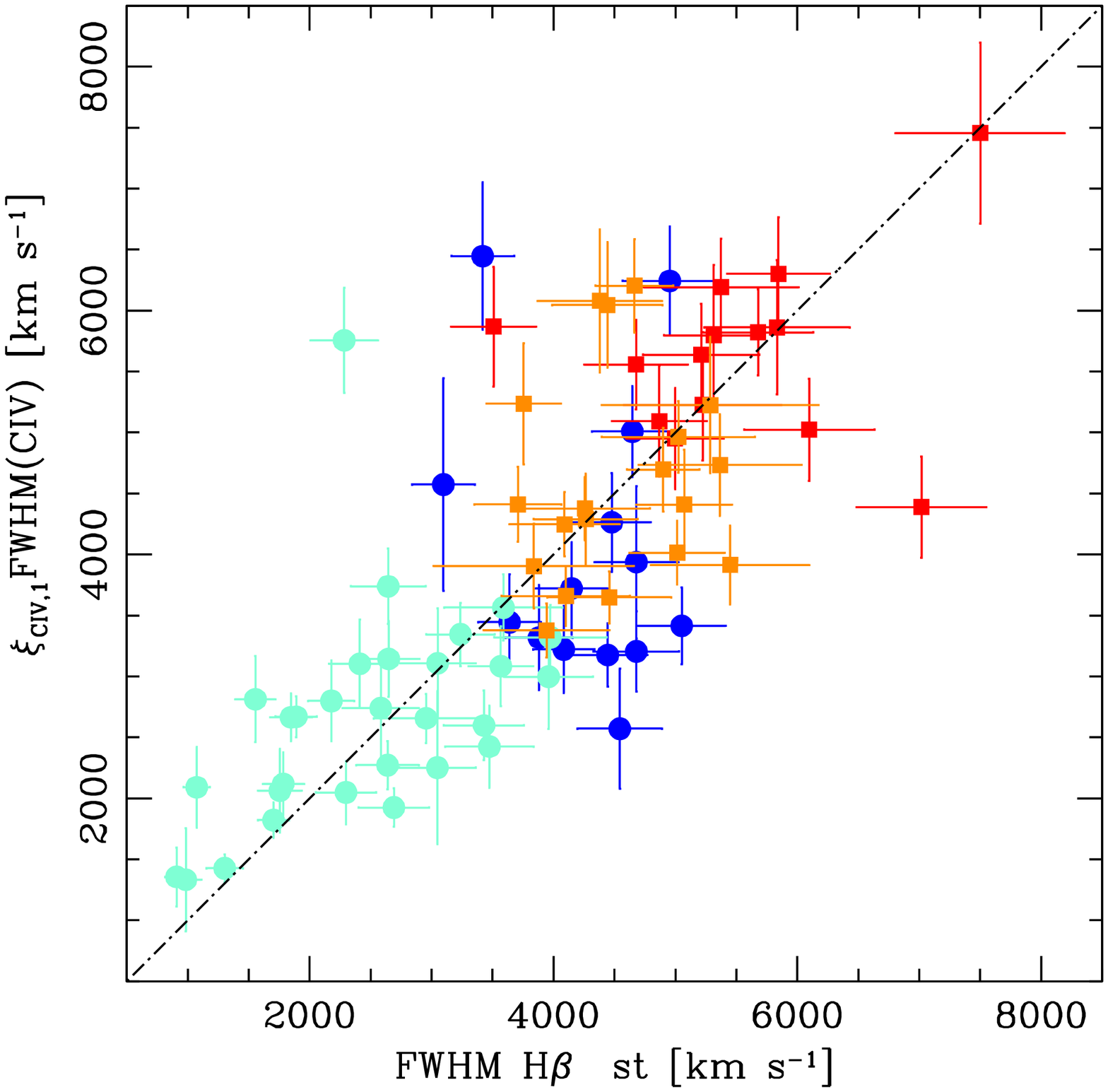}
\includegraphics[width=0.45\columnwidth]{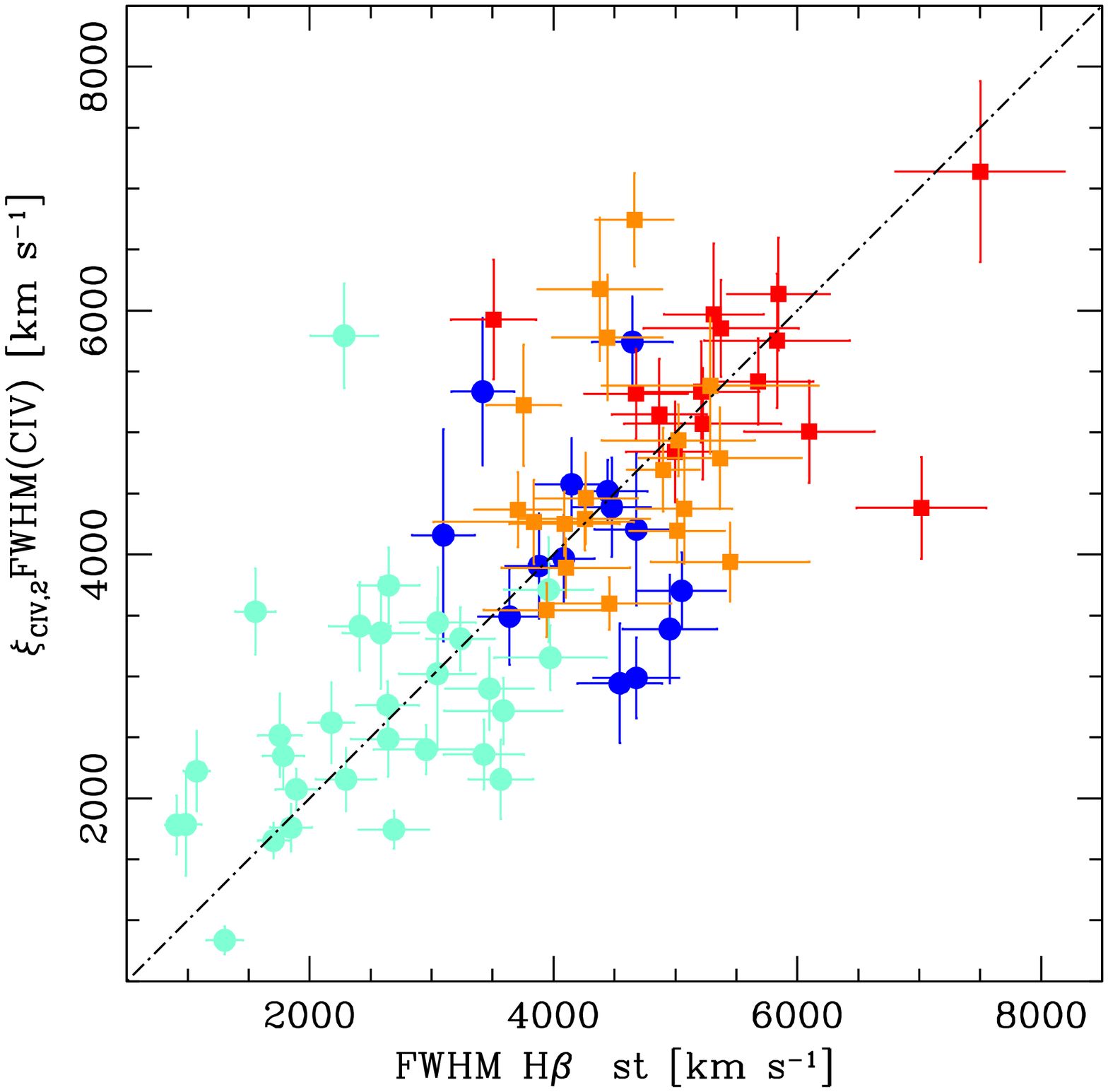}\\
\vspace{-0.75cm}
\includegraphics[width=0.45\columnwidth]{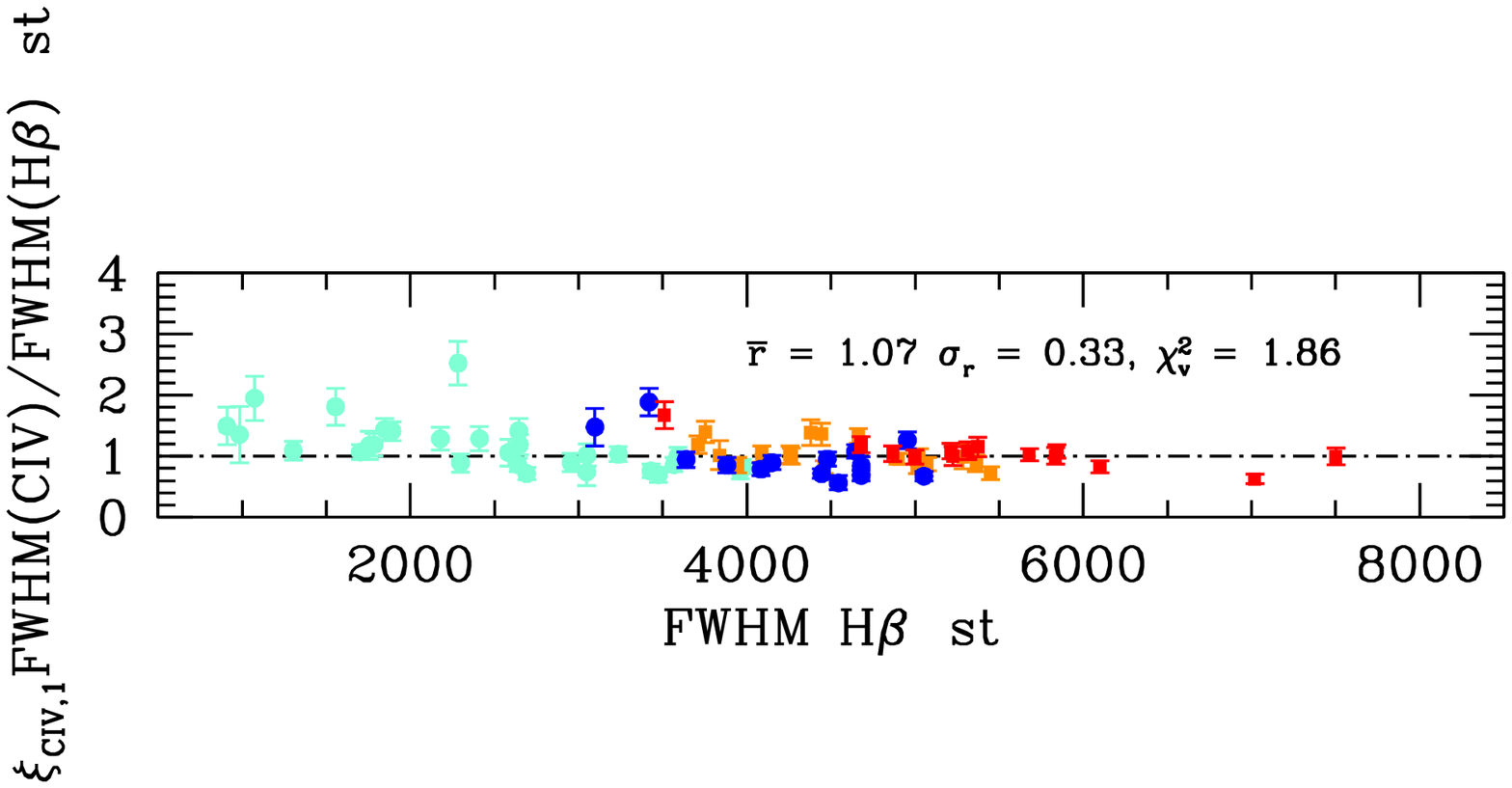}
\vspace{-0.75cm}
\includegraphics[width=0.45\columnwidth]{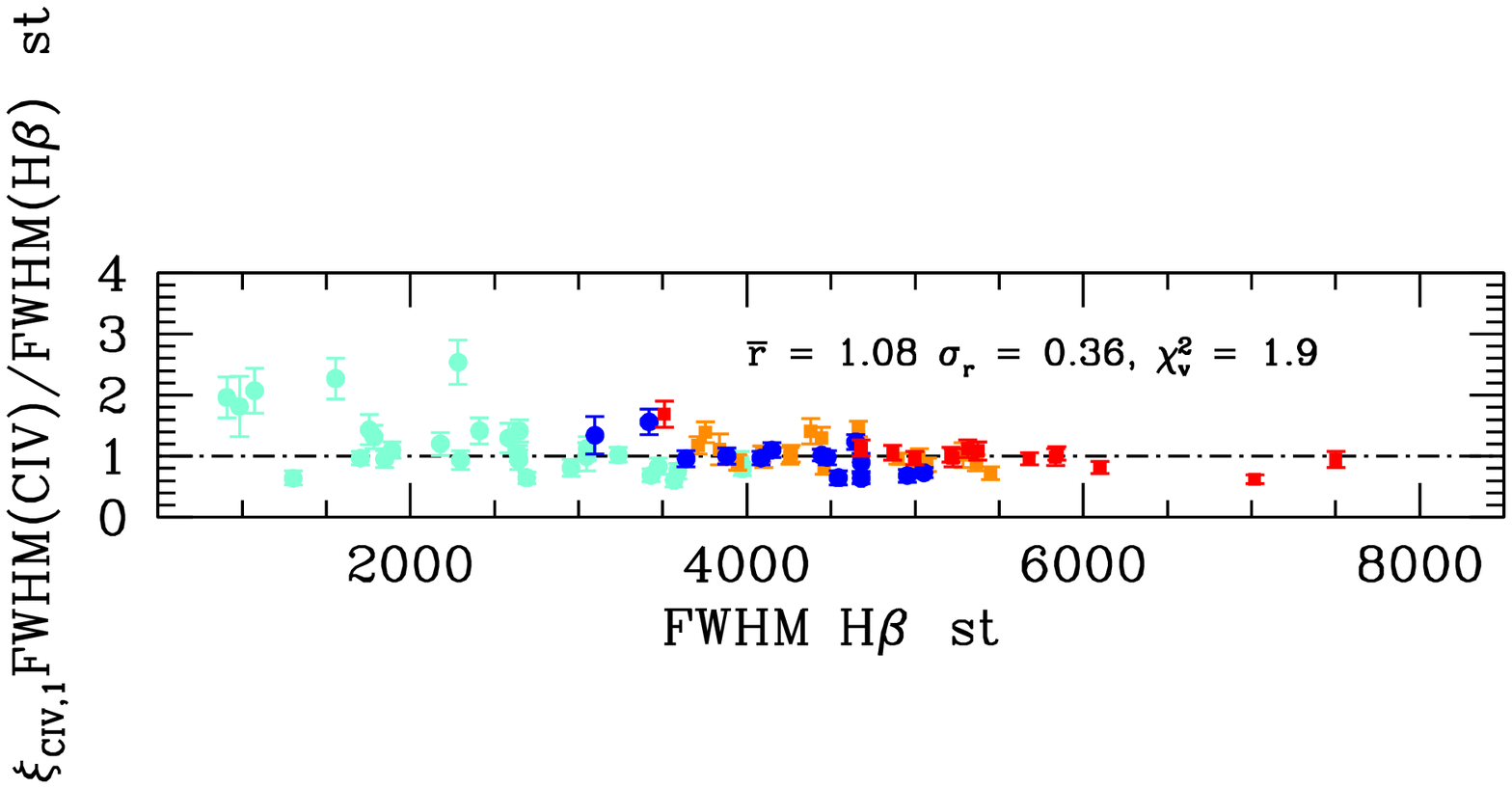}\\
\vspace{-1.5cm}
\caption{Top left panel: $\xi_\mathrm{CIV}$FWHM(\civ) i.e., FWHM(\civ) after correction for blueshift and luminosity dependence following Eq. \ref{eq:xic1}  vs FWHM(\hb) {\tt cm} for the FOS+HE  sample.   The black dot dashed line is the equality line, meaning of color code is the same as in Fig. \ref{fig:virialciv}. Top right: same, but with $\xi_\mathrm{CIV}$ computed from Eq. \ref{eq:xilinear}. 
  }
\label{fig:wcoatcorr}
\end{figure*}

A correction dependent on luminosity  reduces the systematic differences between the various samples in the present work but it  has  to be separately  defined for Pop. A and B (Fig. \ref{fig:wcoatcorr}). The following expression: 

\begin{equation}
\xi_\mathrm{CIV,1} = \frac{1}{b \left(a - \log \lambda L_{\lambda}(1450)\right) \cdot \left(|\frac{c(\frac{1}{2})}{1000} |\right) + c}
\label{eq:xic1}
 \end{equation}
 
provides a suitable fitting law, with $a$, $b$, $c$\ different for Pop. A and B.  Here we consider the \cmp\ because of its immediate connection with the FWHM, and because it is highly correlated with \cqp\ (Sec. \ref{large}). Eq. \ref{eq:xic1} is empirical: it entails a term proportional to shift and one to the product of $\log L_{1450}$ and shift. 
  Multivariate, nonlinear lsq results for Eq. \ref{eq:xic1} are reported in the first rows of Table \ref{tab:r}. For Pop. A the correction is rather similar to the one of \citet{coatmanetal17}, and is driven by the large blueshifts observed at high \lledd. The luminosity-dependent factor accounts for low-luminosity sources that are not present in  \citet{coatmanetal17} sample. The use of the absolute value operator provides an improvement with respect to the case in which blueshifts are left negative. There are only three objects for which \cmp\ is positive. The improvement is understandable if one consider any \civ\ shift as affecting the difference between the FWHM of \civ\ and \hb.   An A(+) sample was defined from the Pop. A  sample minus three objects with positive \chm, i.e., all A(+) sample sources show blueshifts. No significant improvement was found with respect to Eq. \ref{eq:xic1}.   

The correction for Pop. B is less well defined, considering the uncertainty in $a$, and the low value of $b$\ (Table \ref{tab:r}).  The corrections for Pop. B still  offer an improvement because { they remove a significant bias (as evident by comparing Fig. \ref{fig:virialciv} and Fig. \ref{fig:wcoatcorr}). }  In practice, for Pop. B, at low $L$\  the $\xi_\mathrm{CIV}$\ could be considered constant to a zero-order approximation, with $\xi_\mathrm{CIV} \sim 1$.  In other words, when the velocity field is predominantly virial, and no prominent blueshifted component affects the line width, \civ\ may be somewhat broader than \hb\ as expected for the stratification revealed by reverberation mapping of lines from different ionic species \citep{petersonwandel99,petersonwandel00}.  The fit is consistent with   $\xi_\mathrm{CIV}$\ depending on shift but only weakly on luminosity. The Pop. B correction is so ill-defined that larger samples are needed for a better determination of its coefficients.

Slightly different fitting laws    

\begin{equation}
\xi_\mathrm{CIV,2} = \frac{1}{a + b  \log \lambda L_{\lambda}(1450) + c \frac{c(\frac{1}{2})}{1000}}
\label{eq:xilinear}
 \end{equation}

which considers a linear combination $\log \lambda L_{\lambda}(1450)$ and shift, and

\begin{equation}
\xi_\mathrm{CIV,3} = \frac{1}{a - b  \log \lambda L_{\lambda}(1450) \cdot \frac{c(\frac{1}{2})}{1000}}
\label{eq:xicl}
 \end{equation}

which assume a dependence from the product  $\log \lambda L_{\lambda}(1450)$ and shift, provide consistent results, with fitting parameters, their $1\sigma$\ confidence level associated uncertainty, and rms residuals of $\xi_\mathrm{CIV}$ reported in Table \ref{tab:r}.  The fitting relations  yield a lower residual scatter in $\xi_\mathrm{CIV} $ than assuming no luminosity dependence. For instance, using Eq. \ref{eq:xilinear} we obtain a scatter in $\xi_\mathrm{CIV,2}$ that is a factor  1.86  lower than if Eq. \ref{eq:xic} is used.  We also considered the A.I. in place of \cmp\ in Eq. \ref{eq:xic1} (without the absolute value operator; bottom rows of Table \ref{tab:r}). The A.I. has a non-negligible advantage to be independent from the choice of the rest 
frame. The A.I. is correlated with both \cqp\ and \cmp, and shows higher correlation with \cqp\ (Pearson's $r\approx 0.66$).  However, the scatter in $\xi_\mathrm{CIV,A.I.}$ is unfortunately large, and would imply a scatter $\approx 1.5$\ higher in \mbh\ estimates than in the case Eq. \ref{eq:xic1} is considered for Pop. A. 

If Pop. A and B are considered together, the final scatter in $\xi_\mathrm{CIV}$\ is close for the different fitting function (Eq. \ref{eq:xic1} yields a slightly better result) but much higher than if Pop. A and B are kept separated.   It is therefore necessary to distinguish between Pop. A and B as the intrinsic structure of their BLR may be different \citep[e.g.,][]{goadkorista14,wangetal14a}. In Pop. B, at low \lledd, the  lines are mainly broadened following a virial velocity field  \citep{petersonwandel00}. The relative prominence of the blueshifted to the virialized component (ratio BLUE over BC), a consequence of the low \lledd\ for Pop. B sources. Both properties are expected to contribute to the overall consistency between \hb\ and \civ\ profiles in Pop. B sources. At any rate, $\xi_\mathrm{CIV}$\ should always be $\lesssim 1$, with $\xi_\mathrm{CIV} \approx 1 $\ for Pop. B at low-$L$, and $\xi_\mathrm{CIV} \ll 1$\ in case of very large shifts, as in Pop. A at high-$L$. 

It is possible, in most cases, to distinguishing between Pop. A and B from the UV spectrum  emission blend, making the correction applicable at least to a fraction of all quasars in large samples.  Several criteria were laid out by \citet{negreteetal14}:  (1) broad line width; (2) evidence of a prominent red wing indicative of a VBC; (3) prominence of \ciii. Population B sources show a \civ\ red wing and strong \ciii\ in the 1900 \AA.   Extreme Population A (xA) sources are easy to recognize; they show strong \aliii\ in 1900 \AA\ blend and low $W$(\civ).  A prototypical composite spectrum of xA sources is shown by \citet{martinez-aldamaetal18}. However, some intermediate cases along the MS (i.e., spectral type A1) may be easier to misclassify. Also, with only the UV spectral range available the redshift estimate may be subject to large errors. \citet{negreteetal14} provide a helpful recipe; however, their recipe applied to three of their 8 sources  allowed for a precision  $\sim 100$ \kms\ in the rest frame, but the remaining 5 had an uncertainty on average $\gtrsim$ 500 \kms. 


\begin{table*} 
\begin{center}
 \caption{Fits of $\xi_\mathrm{CIV}$ \label{tab:r}}
\tabcolsep=3pt
\begin{tabular}{lccccc}
 \hline \hline
Sample &  $a \pm \delta a$  & $b \pm \delta b$ & $c \pm \delta c$ & rms$_\mathrm{\xi}$ & d.o.f.\\
\hline
\\
\hline
 \multicolumn{6}{c}{$\xi_\mathrm{CIV,1} \approx 1/(b * (a - x) * |y| + c)$} \\ 
\hline
\\
A	 &-0.3093  0.1581 &   0.3434    0.0881  &	 1.0763    0.0949 &	  0.198 & 40   \\
B	&3.9224   14.1170  &	0.0206    0.0568 	&0.9845    0.0602	  &    0.187  & 30  \\
A+B	 &   -0.1805    0.2603  &0.1978    0.0584 	&1.0117    0.0641  	  &   0.227     & 73  \\
\\
\hline
 \multicolumn{6}{c}{$\xi_\mathrm{CIV,1} \approx 1/(b * (a - x) * y + c)$} \\ 
\hline
\\
A	 &-0.4161    0.3325	  &	-0.1825    0.0693	&  1.3245    0.0919   &	 0.222 & 40 \\
B	& 28.967   496.591 &	  -0.0030   0.0486	&1.0182  0.0449 	  &    0.187 & 30  \\
A+B	 &-0.1346    0.4991  & -0.1202    0.0467	&1.1508    0.0510 	  &   0.238     & 73  \\
\\
\hline
 \multicolumn{6}{c}{$\xi_\mathrm{CIV,2} \approx  1/(a+b*x+c*y)$} \\ 
\hline
\\
A	 &0.356   0.160    	   &   -0.3480    0.0580 &    -0.351    0.0814	 	  &0.199 & 40\\
B	 &1.080    0.126   & 0.0223    0.0416         &    -0.0764    0.0515	  	  &0.186 & 30 \\
A+B		& 0.7605   0.125 	 	& -0.1460    0.0452 &  -0.2204    0.0532   	&	0.237 & 73\\

\\
\hline
 \multicolumn{6}{c}{$\xi_\mathrm{CIV,3} \approx  1/(a-b*x*y)$} \\ 
\hline
\\
A	&1.3036    0.0884 &  	0.1408    0.0481	 & \ldots  &	0.222  & 41\\ 
B	&1.0343    0.0436  & 	-0.0463    0.0255	 & \ldots &	0.187  & 31\\
A+B	&1.1470  0.0488  	&    -0.1116    0.0288     & \ldots	&    0.237 	& 74\\

\\
\hline
 \multicolumn{6}{c}{$\xi_\mathrm{CIV,A.I.} \approx 1/(b * (a - x) * z + c)$} \\ 
\hline
\\
A	 &-1.7004     0.5528 &    -1.0575     0.4703   &	 1.5515    0.1183  &	  0.240 & 40   \\
B	&-2.346    0.3153   &	0.7571    0.3026	&1.1057    0.0413 &    0.184  & 30  \\
A+B	 &  5.147   21.138 &-0.1121    0.3292	&1.2474    0.0501 	  &   0.257     & 73  \\
\\
\hline


\end{tabular}
\end{center}
\tablefoot{$x = \log \lambda L_{\lambda}(1450) - 48$, $y = $ \chm/1000, $z$ = A.I.}
\end{table*}

\section{Discussion}
\label{disc}

Recently, the problems outlined in earlier works by \citet{sulenticetal07} and \citet{netzeretal07} have been ascribed to a ``bias'' in the \civ\ \mbh\ estimates \citep{denneyetal16}. The \civ\ \mbh\ bias  is dependent on  the location in the 4DE1 quasar MS: Fig. \ref{fig:virialhb}  and \ref{fig:virialciv} clearly show the different behavior for Pop. A and B. By the same token, an \lledd\ -- dependent correction is in principle a valid approach, as \lledd\ is probably one of the main drivers of the MS \citep{borosongreen92,sulenticetal00a,sunshen15}.   Unfortunately,  several recent works still ignore  4DE1-related effects (or, in other words, MS trends). For instance,  scaling laws derived from the pairing of the virial products for {\em all} sources with reverberation mapping data should be viewed with  care \citep[as shown by the reverberation mapping results of][]{duetal18}. 

\subsection{\civ\ and \hb\ as \mbh\ estimators: input from recent works}
\label{disc:virial}

Attempts at using the \civ\ as a VBE have been renewed in the last few years, not last because \civ\ can be observed in the optical and NIR spectral ranges over which high-redshift quasars have been discovered and are expected to be discovered in the near future. The large \civ\ blueshifts indicate that part of the BLR gas is under dynamical conditions that are far from a virialized equilibrium. At high Eddington ratio  ionized gas  may   escape from the galactic bulge, and even be dispersed into the intergalactic medium, as predicted by numerical simulations \citep[e.g.,][]{debuhretal12}, and at high luminosity   ($\log L \gtrsim 47$ [\ergss]) might have a significant feedback effect on the host galaxy \citep{marzianietal16a}.

A firm premise is that the disagreement between \hb\ and \civ\ mass estimates is not  a matter of S/N \citep[][]{denneyetal13}. The \civ\ line width suffers of systematic effects which emerge more dramatically  at high S/N   i.e., when it is possible to appreciate  the complexity of the \civ\ profile. Given this basic result, recent literature can be tentatively grouped into three main strands: (1) low-$z$ studies, involving FOS and Cosmic Origin Spectrograph (COS) spectra to cover \civ; (2) high-$z$ studies, where the prevalence of large \civ\ shifts is high; (3) studies attempting to correct the \civ\ FWHM and reduce it to an equivalent of \hb, some of them employing results that are directly connected to the MS contextualization of quasar properties. 

\begin{figure}[htp!]
\centering
\includegraphics[width=0.95\columnwidth]{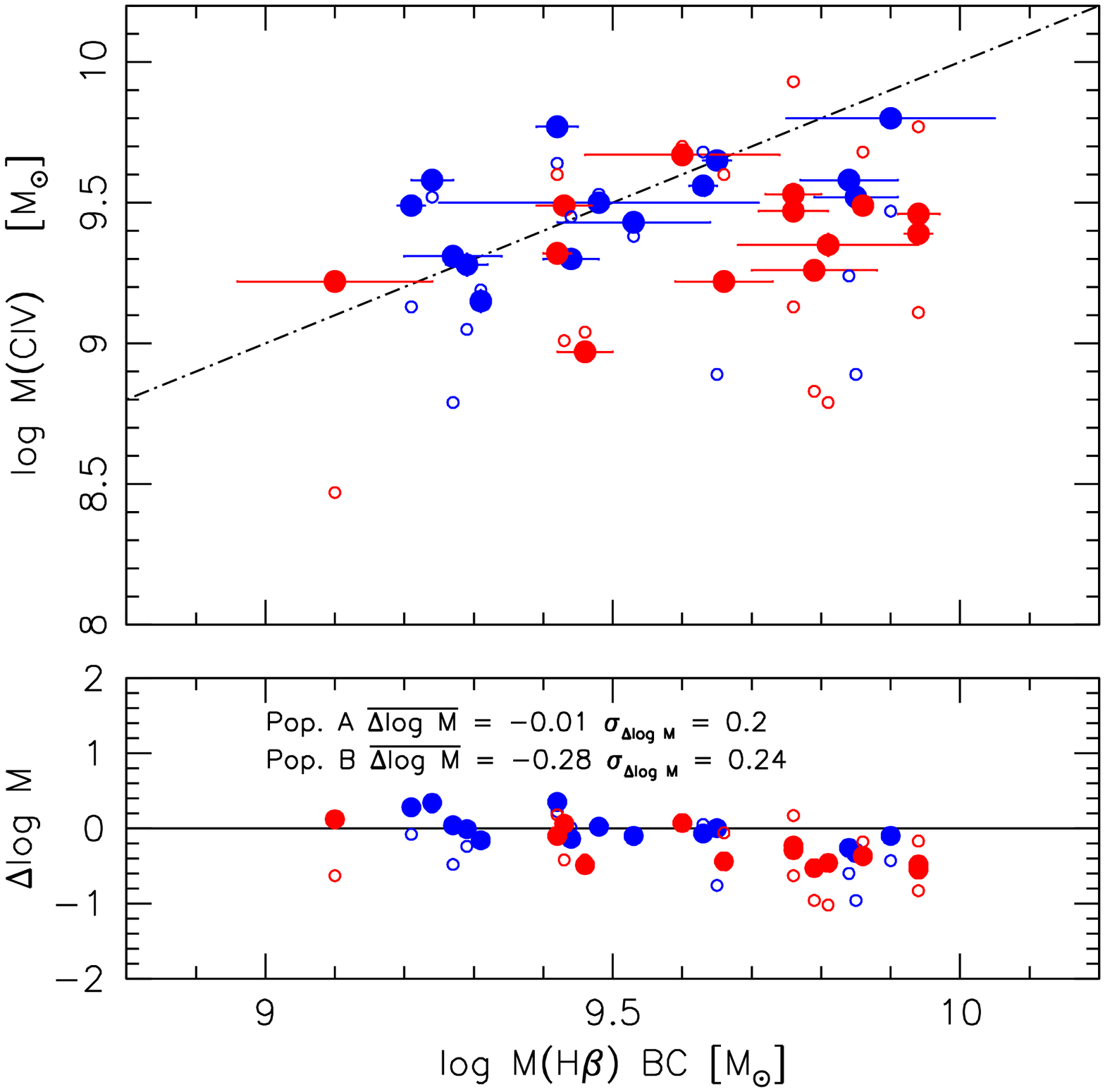}
\caption{Black hole mass computed from the fiducial relation of \citet{vestergaardpeterson06} based on FWHM \hb\ vs. the one computed from the \civ\ FWHM following \citet{parketal13}, for Pop. A (blue) and B (red) HE sources. \mbh\ values obtained from the \citet{vestergaardpeterson06} \civ\ scaling law after  the correction suggested by \citet[][small dots]{brothertonetal15}  are shown by small open circles. The lower panel shows residuals as a function of \mbh. The average and the scatter reported for Pop. A and B refer to the \citet{parketal13} scaling laws. }
\label{fig:virialcivmass}
\end{figure}

\paragraph{Low-$z$ studies} A large systematic analysis of the \civ\ profiles paired to \hb\ emission was carried out using HST/FOS (in part used for the present work) and optical observations \citep{sulenticetal07}. The results of this study emphasized the role of the \civ\ line width in the \mbh\ estimates. The Fig. 6 of \citet{sulenticetal07}  clearly shows the importance of placing sources in an E1 context:  estimates of the masses could be easily overestimated by a factor $\lesssim$ 100 for extreme Pop. A sources such as I Zw 1, while for Pop. B \civ\ and \hb\ \mbh\ estimates appeared more consistent albeit with a large scatter.   The line width (let it be the FWHM or the velocity dispersion $\sigma$)  remains a major factor in \civ\ vs \hb\ \mbh\ determinations since  broadening enters squared in the  scaling laws  \citep{kellybechtold07}. Similar warnings on using \civ\ FWHM  were issued by \citet{netzeretal07}.   Low-$z$ samples are less affected by the Eddington ratio bias that is cutting low-Eddington ratio sources at a given \mbh\ for a fixed flux limit \citep{sulenticetal14}. Therefore, it may not be  surprising to  find studies based on excellent spectra that find an overall consistency between \hb\ and \civ\ \mbh\ estimates. Intrinsic scatter is probably high if full line width without any correction are used:  \citet{tiltonshull13} find $\approx 0.5$ dex from COS observations of low-$z$\ quasars.  \citet{denneyetal13} claim to be able to reduce the disagreement between \hb\ and \civ\ derived \mbh\ to $\approx$ 0.24 dex by using the velocity dispersion of the \civ\ line.  Since the \civ\ profile in the \citet{denneyetal13} sample almost never shows large blueshifts which may be associated to a velocity shear in outflowing gas, these results appear consistent with the Pop. B properties of the FOS sample.

\paragraph{High-$z$ studies}   generally concur that the \civ\  FWHM is poorly correlated with the Balmer line FWHM.  \citet{shenliu12}  describe  the scatter between  \civ\ and \hb\ FWHM as due to an irreducible part ($\approx$0.12 dex), and a part that correlates with the blueshift of the \civ\ centroid relative to that of \hb. They propose scaling laws in which the virial assumption is abandoned i.e., with the exponent of the line FWHM significantly different from 2. For \civ, this means to correct for the overbroadening associated with the non-virial component. The  scaling law introduced by \citet{parketal13} is consistent with the \citet{shenliu12} approach and implies \mbh $\propto$FWHM$^{0.5}$ i.e., a FWHM dependence that is very different from the one expected from a virial law (\mbh $\propto$FWHM$^{2}$). As shown in Fig. \ref{fig:virialcivmass}, the scaling law suggested by \citet{parketal13} applied to the HE sample properly corrects for the overbroadening of Pop. A sources, but overcorrects the width of Pop. B, yielding a large deviation from the \hb-derived \mbh\ values (on average $\approx 0.28$\ dex).

\paragraph{Studies exploiting MS trends} 
The results reported in \S \ref{virial} and in \citetalias{sulenticetal17} indicate that any solution seeking to bring \civ\ \mbh\ estimates in agreement with the ones from \hb\ cannot exclude the strong \lledd\ dependence of the \civ\ blueshift that is in turn affecting the \civ\ FWHM (Fig. \ref{fig:virialcivbr}). The discussion in Sect. \ref{anal} and in Sect. \ref{empcorr} identifies the \civ\ blueshift as  an expedient \lledd\ proxy.  Any parameterization of the blueshifted amplitude such as \cmp, the flux bisector of \citet{coatmanetal16}  or the  ratio FWHM(\civ)/FWHM(\hb)  is taking into account the MS trends in \civ\ properties, due to  the \lledd\ and \civ\ blueshift correlation. 
Another \lledd\ proxy may involve the \siiv/\civ\ peak ratio:  at low \siiv/\civ\ the \mbh\ is underestimated with respect to \hb, at high \siiv/\civ\ the mass is overestimated  \citep{brothertonetal15}. Since the ratio \siiv/\civ\ is a known 4DE1 correlate \citep{willsetal93,bachevetal04},  these results  confirm that FWHM \civ\ leads to overestimate \mbh\ for Pop. A (as Pop. A outflows produce blue shifted emission that significantly broadens the line (Fig. \ref{fig:virialciv}; cf. \citealt{denneyetal12}).  
In our sample, however, applying the correction suggested by  \citet{brothertonetal15}, $\delta \log M \approx - 1.23 \log \frac{1400}{CIV} - 0.91$\ to the masses derived from the \citet{parketal13} would move the \mbh\ of Pop. B further down, leading to a further increase of the overcorrection, and also destroying the agreement for Pop. A sources: on top of the $\propto$FWHM$^{0.5}$\ law,  the additional correction is $\delta \log M \approx -0.91$  if $\frac{1400}{CIV} \sim 1$, as for extreme  Pop. A. The correction is lower but still  negative for most Pop. B sources where $\frac{1400}{CIV} \sim 0.3$, exacerbating the disagreement between the \hb\ and \civ\ derived masses.   {A better consistency is achieved if  the correction of \citet{brothertonetal15} is applied to the \citet{vestergaardpeterson06} scaling law for \civ. In this case (shown in Fig. \ref{fig:virialcivmass} by small open circles) the correction for Pop. A still imply non-negligible systematic residuals   $ \delta \log M = \log M_\mathrm{BH}$(\hbbc) -  $\log M_\mathrm{BH}$(\civ) $\approx 0.23$.  The average residual is higher for Pop. B \mbh, with  $ \delta \log M \approx 0.27$, and  scatter $\approx 0.39$\ dex.}

\citet{assefetal11} used a sample of  {$\approx$10 quasars} with optical spectra covering \civ\ and near-IR spectra covering \hb\ or H$\alpha$\ and show that \mbh\ estimates can be made consistent. The approach of \citet{assefetal11} may be also understood as a correction related to the MS.  \citet{assefetal11} suggest that much of the dispersion in their virial mass is caused by the poor correlation between $\lambda L_{\lambda}$ at 5100 \AA\ and at 1350  \AA\ rather than between their line widths.   Their Figs. 14 and 15 shows that the FWHM \civ\ over \hb\ ratio depends on the flux ratio at 1350 \AA\ and 5100 \AA, which is an MS correlate \citep{laoretal97b,shangetal11}. 
The \citealt{assefetal11} sample of gravitationally-lensed quasars might have lowered the Eddington ratio bias described by \citet{sulenticetal14}, leading to a preferential section of Pop. B quasars, and better agreement between \hb\ and \civ\ line width.


\subsection{A virialized  component}
\label{wind}

\begin{figure}[htp!]
\centering
\includegraphics[width=0.4\columnwidth]{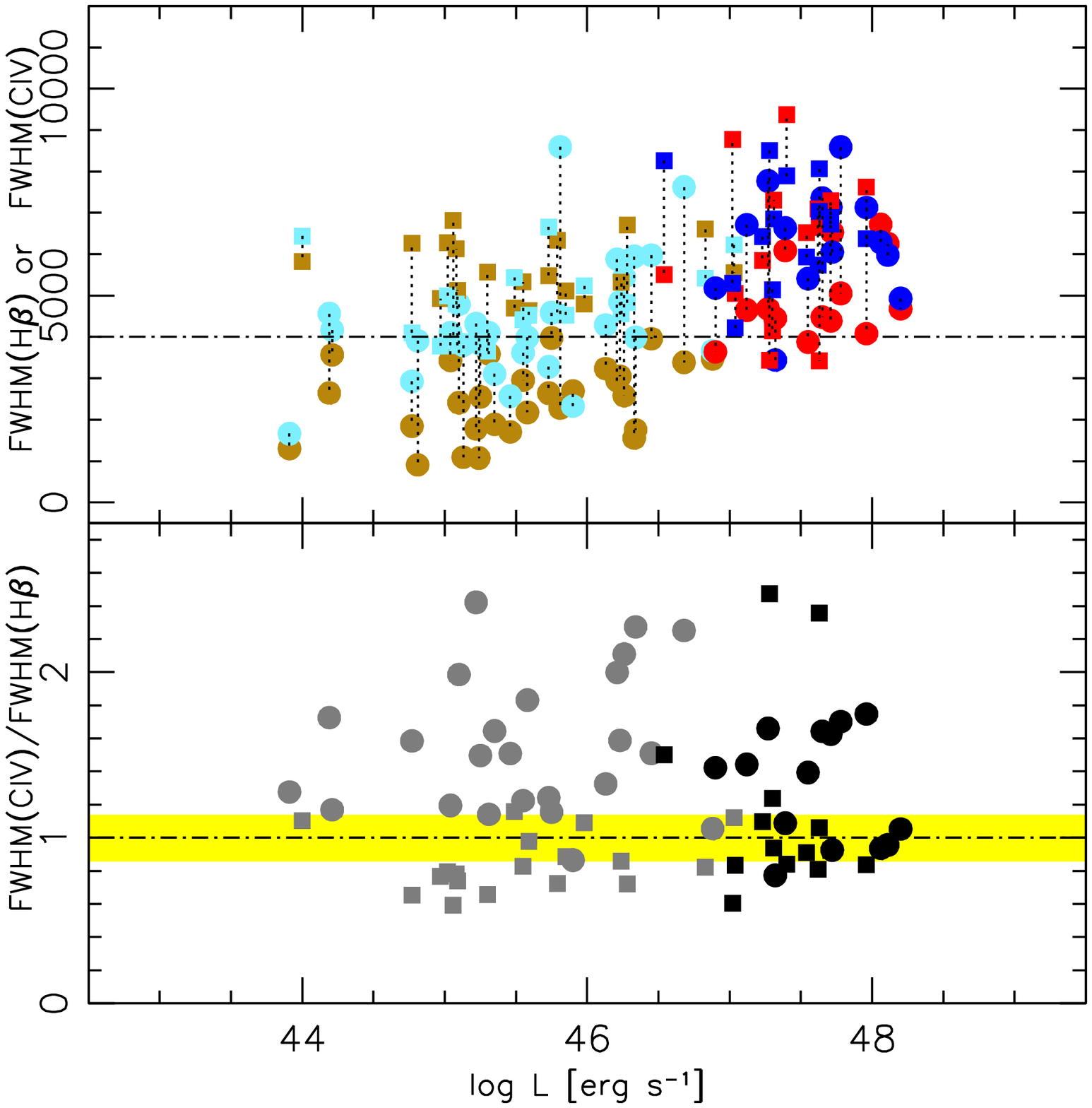}\\
\includegraphics[width=0.4\columnwidth]{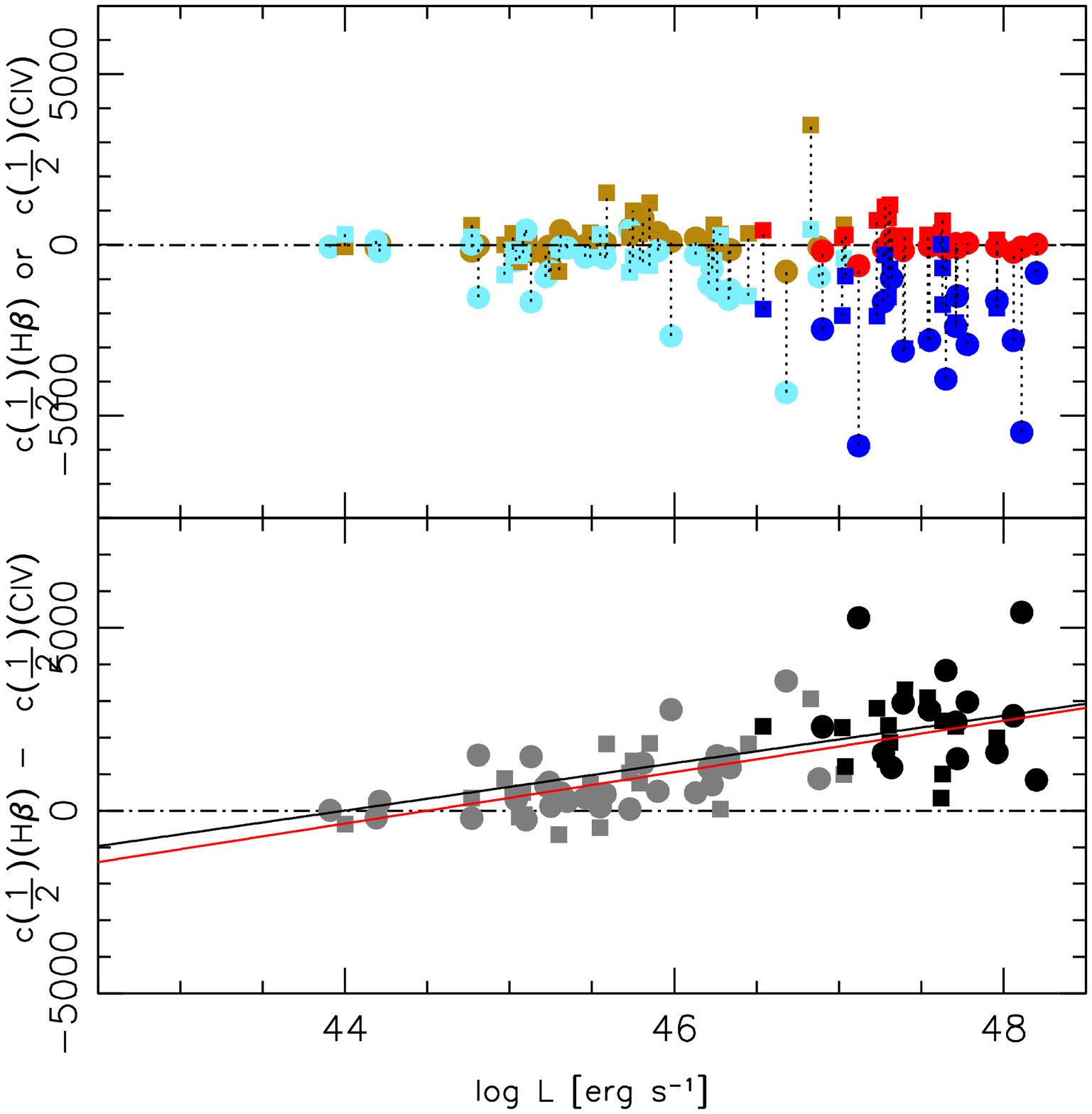}\\
\includegraphics[width=0.4\columnwidth]{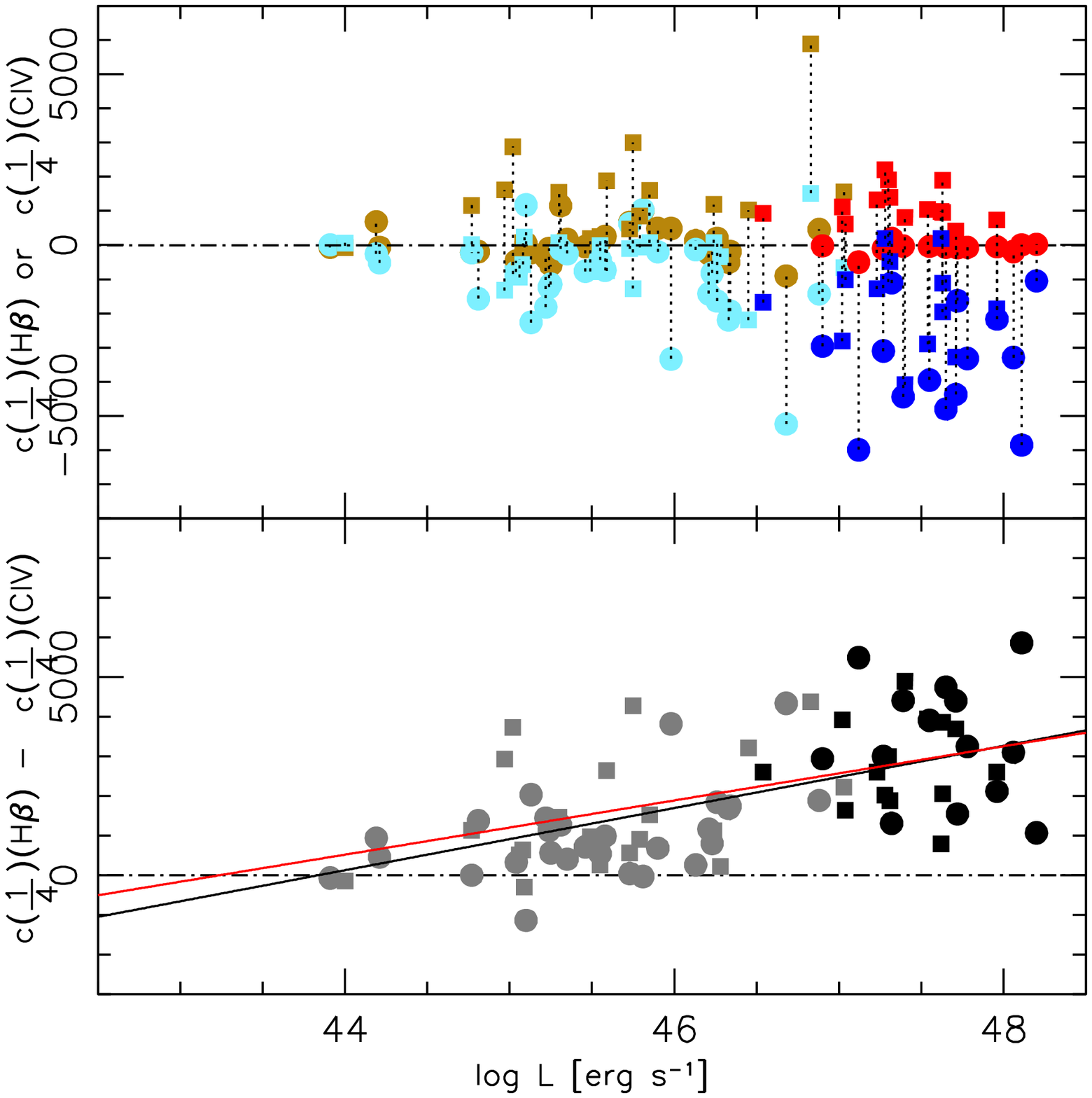}
\caption{\hb\ and \civ\ profile parameter comparison as a function of luminosity. Top panels: behavior of FWHM \civ\ and \hb\   (upper half) and of the ratio FWHM(\civ)/FWHM(\hb) as a function of $L$ \  (lower half),  for FOS (golden and pale blue) and HE sample  (red and blue). The yellow band identifies the region where FWHM(\civ)/FWHM(\hb)=1 within the errors.  Middle panels: \cmp\ of \hb\   and \civ\ (upper half), and difference $\delta(\frac{1}{2}$) as a function of $L$ (lower half). Square symbols indicate Pop. B, circles Pop. A. Lines trace an unweighted lsq fit for the Pop.  A (black) and for Pop. B (red) sources.  Bottom panels: same for \cqp\ and $\delta(\frac{1}{4}$). The vertical dotted lines join \hb\ and \civ\ parameters for the same object (e.g., they are not error bars).}
\label{fig:shiftl}
\end{figure}

A  systematic increase in line width in the HE sample is expected if the line broadening is predominantly virial: Fig. 5 of \citetalias{sulenticetal17} shows that there are no FWHM \hb$\lesssim$ 3000 \kms at $\log L \gtrsim 47$ [\ergss].   Fig. \ref{fig:shiftl} shows that a similar increase in FWHM  as a function of luminosity is occurring in the FOS+HE sample for both \hb\ and \civ. The FWHM ratio between \civ\ and \hb\  does not instead appear strongly influenced  by $L$, suggesting the interpretation that the broadening of both lines -- even if the \civ\ centroid measurements are significantly affected  by an outflowing component -- may be mostly related to the  gravitational effects of the supermassive black hole (as further discussed below). As mentioned, Balmer lines provide a  VBE up  $z \gtsim$ 2, and the results of \citepalias{sulenticetal17} extended this finding to the highest luminosities.  For \civ, Fig. \ref{fig:shiftl}  and the correlation FWHM -- \cmp\ justify the assumption of  a virial broadening component coexisting with a non virial one \citep[][]{wangetal11}.  


\subsubsection{Orientation effects on \hb}	
\label{orien}

A large part of the \civ\ -- \hb\ scatter is expected to be due to orientation effects.  The issue of orientation effects remains open for RQ sources, and orientation effects are most-likely strongly affecting the FWHM of \hb\ \citep{mejia-restrepoetal18}, even if it remains hard to distinguish them from other physical factors (such as \mbh\ and  \lledd).  A clue is  given by the 4DE1 predictions at extreme orientations:  objects observed near the disk rotation axis (i.e., nearly pole-on) have the smallest FWHM \hb, the strongest \feii\ and \caii\ intensities \citep{dultzinhacyanetal99}, the largest soft X excess, and  the largest \civ\ blue shifts/asymmetries. These predictions are motivated by the physical scenarios involving an accretion disk - wind system.  From a pole-on orientation we should see the smallest Doppler broadening of virially dominated \hb\ emitting clouds, and the strongest intensity of \feii\ and other LILs if they are emitted from clouds in the outer part of the disk \citep{martinez-aldamaetal15}. We should also observe the largest contribution of the soft X excess if it is related to disk emission \citep{wangetal96,bolleretal96,wangetal14}. And finally, if a wind is associated with an  optically thick disk, and its dynamics is dominated by radiation pressure, HILs such as \civ\ emitted in the wind would show the largest blue shifts (if the receding part of the flow is shielded from view). The case of I Zw 1 provided a prototypical case in which a flattened LIL emitting systems and a radial outflow could be seen at small inclination \citep[e.g.,][]{marzianietal96,leighly04}. 


In a more modern perspective, there are several indications that the low-ionization BLR is a flattened system \citep{mejia-restrepoetal17,mejia-restrepoetal18a,negreteetal17,negreteetal18}. We see the clearest evidence at the MS extrema: extreme Pop. B sources radiating at very-low \lledd\ frequently show LIL profiles consistent with a geometrically thin accretion disk profiles \citep[e.g.,][]{chenhalpern89,stratevaetal03,storchi-bergmannetal17}, which may be hidden in the majority of Pop. B sources \citep{bonetal07,bonetal09a}. A highly flattened LIL-BLR is also suggested in blazars, which are also Pop. B low-radiatiors \citep{decarlietal11}, by comparing the virial product to mass estimates obtained from the correlation between \mbh\ the host galaxy luminosity. At the other end of the MS, extreme Pop. A quasars show deviations from virial luminosity estimates consistent with the effect of orientation on the line width, if the emitting region is highly flattened \citep{negreteetal18}. A flattened low-ionization BLR is also suggested by comparing the virial product to mass estimates obtained from accretion disk fits to the SED \citep{mejia-restrepoetal17,mejia-restrepoetal18a}. 

The effect of orientation on the FWHM and on \mbh\ and \lledd\ estimates can be computed by assuming that we are observing randomly-oriented samples of quasars whose  line emission arises from a flattened structure -- possibly the accretion disk itself. The probability of viewing the structure with an isotropic velocity broadening $\delta v_\mathrm{iso}$ at an angle $\theta$  between line-of-sight and the symmetry axis of a flattened structure is $P(\theta)=\sin(\theta)$. The radial velocity spread (in the following we use the FWHM as a measure, $\delta v_\mathrm{obs} =$FWHM) can be written as 

\begin{equation}
\frac{\mathrm{FWHM}^{2}}{4} = \delta v^{2}_\mathrm{iso} + \delta v^{2}_\mathrm{K}\sin^{2}\theta, 
\label{eq:flat}
\end{equation} 

which implies that

\begin{equation}
\frac{M_\mathrm{BH,obs}}{M_\mathrm{BH,K}}  = \frac{\delta v^{2}_\mathrm{obs}}{\delta v^{2}_\mathrm{K}} = 4 \cdot (\kappa^{2} + \sin^{2} \theta), \label{eq:xihb}
\end{equation}


%
%
%

where $\kappa =  \delta v_\mathrm{iso} / \delta v_\mathrm{K}$. From Eq. \ref{eq:flat} one can   estimate the ratio $\delta v_\mathrm{obs}$ $\over$ intrinsic velocity $\delta v_\mathrm{K}$ either by computing a most  probable value of $\theta$ or by deconvolving the observed velocity distribution from $P(\theta)$. The calculations are described in Appendix \ref{incl}.  The average ratio is $ < \frac{M_\mathrm{BH,obs}}{M_\mathrm{BH,K}} >\approx  1.1$ if $\kappa=0.1$. If the FWHM of the \hb\ line is used, the \mbh\ suffers of a small bias, if the LIL emitting region is highly flattened. If $\kappa=0.5$\ (a ``fat'' emitting region), then the bias is much larger $ < \frac{M_\mathrm{BH,obs}}{M_\mathrm{BH,K}} >\approx  2.1$. The \mbh\ dispersion in the case of $\kappa =0.1$ was  estimated for  large samples ($10^{6}$ replications) with $\theta$\ distributed according to $P(\theta)$\ (Appendix \ref{origin}), and was found to be $\sigma_{M_\mathrm{BH}} \approx 0.33$ dex. Therefore, even if  $P(\theta)$\ strongly disfavor cases with $\theta \rightarrow 0$, the viewing angle can account for a large fraction of the dispersion in the \mbh\ scaling laws with line width and luminosity.   

\subsection{A wind component}
\label{toy}

The interpretation of the \civ\ profile  (and \hb\ profile differences) rests on the main results of \citetalias{sulenticetal17}: the \civ\ shifts are dependent on \lledd\ and, to a lesser extent on $L$; the \civ\ broadening is due to a blueshifted component whose strength with respect to a virialized component increases with \lledd\ and $L$.

At $\frac{1}{4}$\ and $\frac{1}{2}$\ fractional intensity the difference in the line centroid radial velocity of \hb\ and \civ\ i.e. \chm(\hb) -- \chm(\civ) and \cqm (\hb) - \cqm (\civ) are  almost always positive, and can reach  7000 \kms\  and 4000 \kms\ in the HE sample and FOS respectively mainly because of the large \civ\ blueshifts (Fig. \ref{fig:shiftl}).   A luminosity dependence  of $\delta(\frac{1}{2})$ =  \chm(\hb) -- \chm(\civ) and $\delta(\frac{1}{4})$ =  \cqm  (\hb) - \cqm (\civ)  is illustrated in Fig. \ref{fig:shiftl}.  The centroid separations are correlated with $L$, with a similar slope at both $\frac{1}{2}$\  and $\frac{1}{4}$\  fractional intensity (Fig. \ref{fig:shiftl}):  for \chm\ of Pop. A, 
\begin{equation}
\delta(\frac{1}{2}) \approx (648 \pm121) \log L -  (28530\ \pm\ 5600)\, \mathrm{km s}^{-1},
\end{equation}
in the range $44 \lesssim \log L \lesssim 48.5$.

The trends of Fig. \ref{fig:shiftl} suggest that the \civ\ broadening is however affected by \mbh, as both the \hb\ and \civ\ widths steadily increase   with luminosity,  and their ratio shows no strong dependence on luminosity. This may  be the case if the outflow velocity is a factor $k$\ of the virial velocity ($k = \sqrt{2}$ \ would correspond to the escape velocity). The correlation between shift and FWHM of \citetalias{sulenticetal17} indicates that we are seeing an outflow component ``emerging'' on the blue side of the BC. If we assume that   line emission arises from a flattened structure with velocity dispersion $v_\mathrm{iso}$\ (i.e., as in Eq. \ref{eq:flat}), and that the outflowing component from the accretion disk contributes to an additional broadening term  proportional to $\cos \theta$\ (the projection along the line of sight of the  outflow velocity), then the observed \civ\ broadening can be written as 


\begin{equation}
\mathrm{FWHM}^{2}_\mathrm{CIV} = 4(\delta v^{2}_\mathrm{iso} + \delta v^{2}_\mathrm{K}\sin^{2}\theta) + \gimel^{2} \delta v^{2}_\mathrm{K} \mathcal{M} \frac{L}{L_\mathrm{Edd}}\cos^{2}\theta, 
\label{eq:fciv}
\end{equation}

where 
$\gimel$ is a proportionality constant, and $\cal{M}$\ the force multiplier. It follows that the total broadening can easily exceed $\delta v_\mathrm{K}$ for a typical viewing angle $\theta=\pi/6$, provided that the factor $\mathcal{Q} =\gimel^{2}\mathcal{M}\frac{L}{L_\mathrm{Edd}}$ is larger than 1. The factors $\gimel$\  and $\cal{M}$\  depend  on physical properties (density, ionization level) and should be calculated in a real physical model linking ionization condition and dynamics.  The factor $\mathcal{Q}$\ encloses the dependence of wind properties on radiation forces, opacity, etc. \citep{stevenskallman90} along with the dependence on ionization. For example, in the case of  optically thick gas being accelerated by the full absorption of the ionizing continuum, the force multiplier is ${\cal M} = \frac{\alpha}{\sigma_\mathrm{T} N_\mathrm{c}} \approx 7.5$\ for column density $N_\mathrm{c} = 10^{23}$cm$^{-2}$, and $\alpha = 0.5$\ ($\alpha$ is the fraction between the ionizing and bolometric luminosity, \citealt{netzermarziani10}).  If \lledd $\rightarrow 1$, and $\gimel \sim 1$, implying $\mathcal{Q} \sim O(10)$,  the {FWHM}$_\mathrm{CIV}$ can exceed by up to a factor of several the virial broadening, as indeed observed in the most extreme radiators from the comparison between \civ\ and \hb.  

Eqs. \ref{eq:flat} and \ref{eq:fciv}   account for the consistent increase in broadening of \civ\ and \hb\ (Fig. \ref{fig:shiftl}). In the context of the present sample  covering a wide range in luminosity, $L$\ can be considered a proxy for the increase  in \mbh\  ($L \propto $ \mbh, with a scatter set by the \lledd\ distribution) and therefore in Keplerian velocity.  The top and middle panels of Fig. \ref{fig:shiftl} show a consistent increase of the centroid, and of the centroid difference $\delta(\frac{1}{2})$  and $\delta(\frac{1}{4})$ with $L$. This result motivated the introduction of a luminosity-dependent correction to the line width. The centroid difference can be written as: 

\begin{equation}
\delta(\frac{i}{4}) \sim \frac{1}{2}\delta v_\mathrm{K}(\frac{i}{4}) \cos\theta \left(-\gimel  \left(\mathcal{M} \frac{L}{L_\mathrm{Edd}} \right)^{\frac{1}{2}} + {f} \right),   i=1,2, \\ \label{eq:dc4}
\end{equation} 

with $f \equiv 0$\ for Pop. A, and $f$\ defined by the infall velocity $v_\mathrm{inf} = f \delta v_\mathrm{K}$\ as a fraction of the radial free-fall velocity for Pop. B. Eq. \ref{eq:dc4} and \ref{eq:fciv} imply that FWHM$^{2}_\mathrm{CIV}$ = FWHM$_\mathrm{H\beta}^{2} + 4\delta^{2}(\frac{1}{2})$, if $f = 0$. 

If we ascribe the redward displacement of the \hb\ wing in Pop. B sources to gravitational and transverse redshift \citep[e.g.,][]{corbin90,bonetal15}, 

\begin{eqnarray}
\delta(\frac{i}{4}) &=& \frac{1}{2}\left(- \delta v_\mathrm{K}(\frac{i}{4}) \gimel \left( \mathcal{M} \frac{L}{L_\mathrm{Edd}}\right)^{\frac{1}{2}} \cos\theta +\frac{3}{2}c z_\mathrm{g}(\frac{i}{4})\right)\\
 &  =& \frac{1}{2} \delta v_\mathrm{K}(\frac{i}{4}) \left(- \gimel \left( \mathcal{M} \frac{L}{L_\mathrm{Edd}}\right)^{\frac{1}{2}}\cos\theta + \frac{3}{2} \frac{\delta v_\mathrm{K}(\frac{i}{4})}{c} \right)\nonumber
 \label{eq:dc4zg}
\end{eqnarray}


where $cz_\mathrm{g} \sim c {GM_\mathrm{BH}}/{c^{2}r} $ is the \cmp\ or \cqp\ of \hb, which usually can be 0 (Pop. A) or $\ge$0 (Pop. B), and where we have used the weak field approximation for the gravitational redshift. 

Eqs. \ref{eq:dc4} and \ref{eq:dc4zg} account for the steady increase of the centroid difference $\delta$\ with luminosity. The \hb\ centroid displacement in Pop. B may be associated with free fall or gravitational redshift. The amplitude of blueshift  depends on luminosity in Pop. A. The point is that both \hb\ redward displacement and blueshift of \civ\  (and hence their differences)  are proportional to the $\delta v_\mathrm{K}$\ and hence to the \mbh. 


The $\xi_\mathrm{CIV}$ factor can be written as:

\begin{eqnarray}
\xi_\mathrm{CIV}   & = &   \left(\frac{\delta v^{2}_\mathrm{iso} + \delta v^{2}_\mathrm{K}\sin^{2}\theta}{4(\delta v^{2}_\mathrm{iso} + \delta v^{2}_\mathrm{K}\sin^{2}\theta) + \gimel^{2} \delta v^{2}_\mathrm{K} \mathcal{M} \frac{L}{L_\mathrm{Edd}}\cos^{2}\theta} \right)^{\frac{1}{2}} \\ \nonumber
 & = & \left( \frac{1}{1+ \frac{{\mathcal Q} \cos^{2}\theta }{4(\kappa^{2} +\sin^{2}\theta)}} \right)^{\frac{1}{2}}\\ \nonumber
\end{eqnarray}

\begin{figure}[htp!]
\centering
\includegraphics[width=0.8\columnwidth]{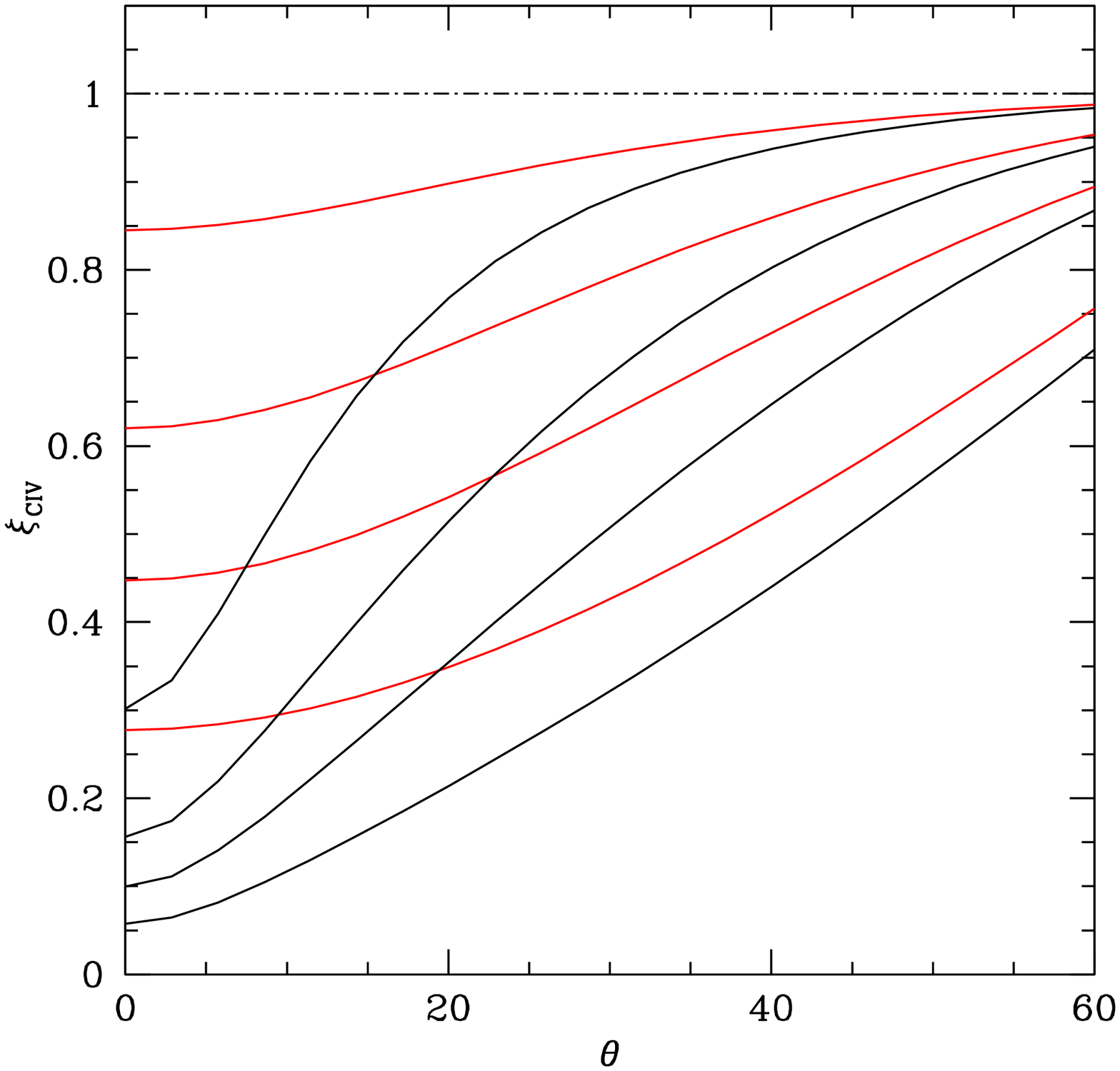}
\includegraphics[width=0.8\columnwidth]{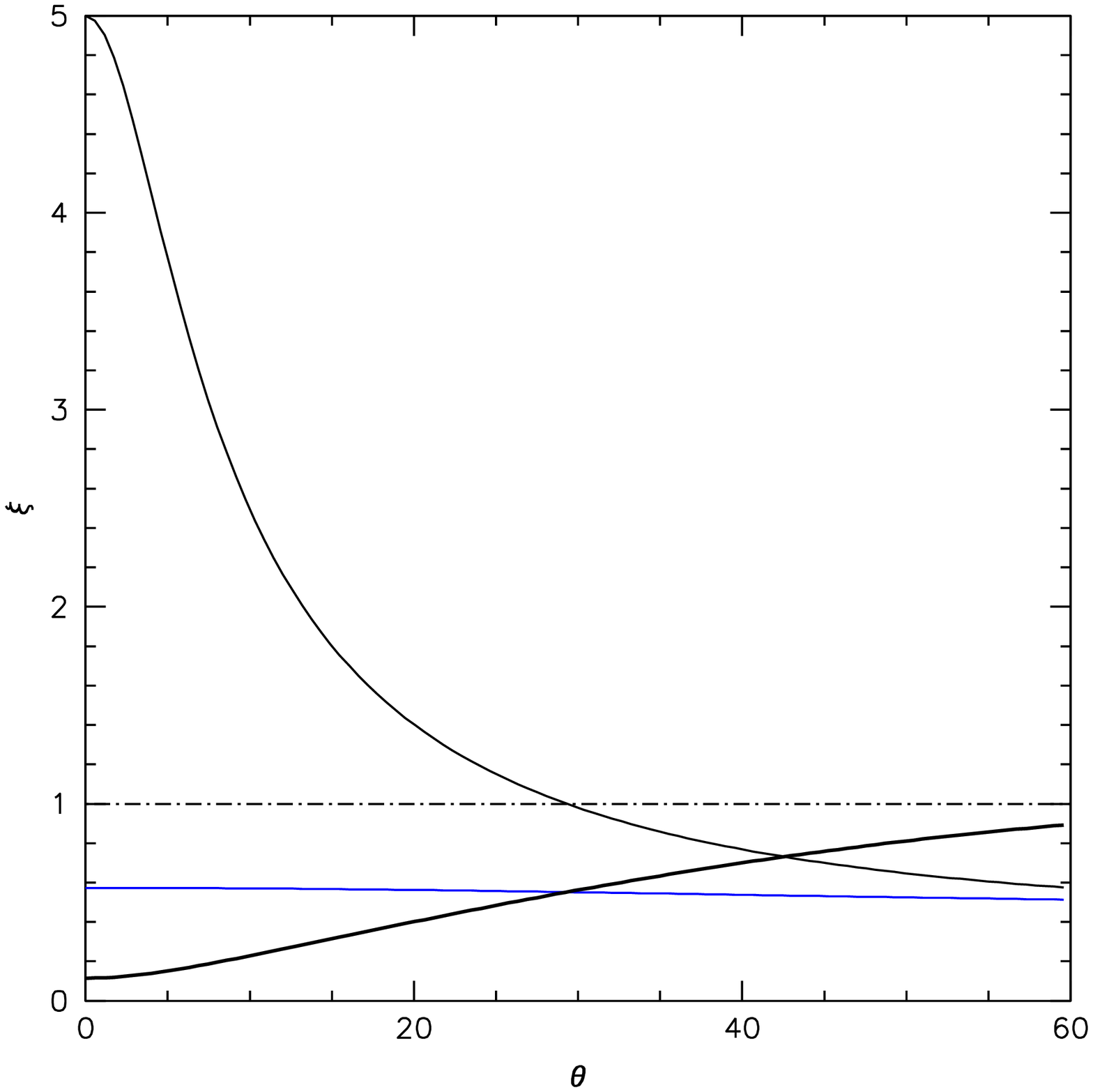}
\caption{Top: parameter $\xi_\mathrm{CIV}$  \ behavior as a function of viewing angle as a function of $\theta$\  for a ``thin'' emitting region with $\kappa = 0.1$, for different $\mathcal{Q}$\ values (0.4, 1.6, 4.0, 12.0; black lines). Red line: $\xi_\mathrm{CIV}$ behavior for a thick emitting region   $\kappa =0.5 $, for the same $\mathcal{Q}$\ values. { Bottom: same as above for $\kappa = 0.1$, with $\mathcal{Q} = 12$\ . The thin lines  are the $\tilde{\xi}$ values for \hb\ (black) and \civ\ (blue). See text for more details. The thick line is their ratio (also shown in the top panel). }}
\label{fig:xibehav}
\end{figure} 

\begin{figure}[htp!]
\centering
\includegraphics[width=0.42\columnwidth]{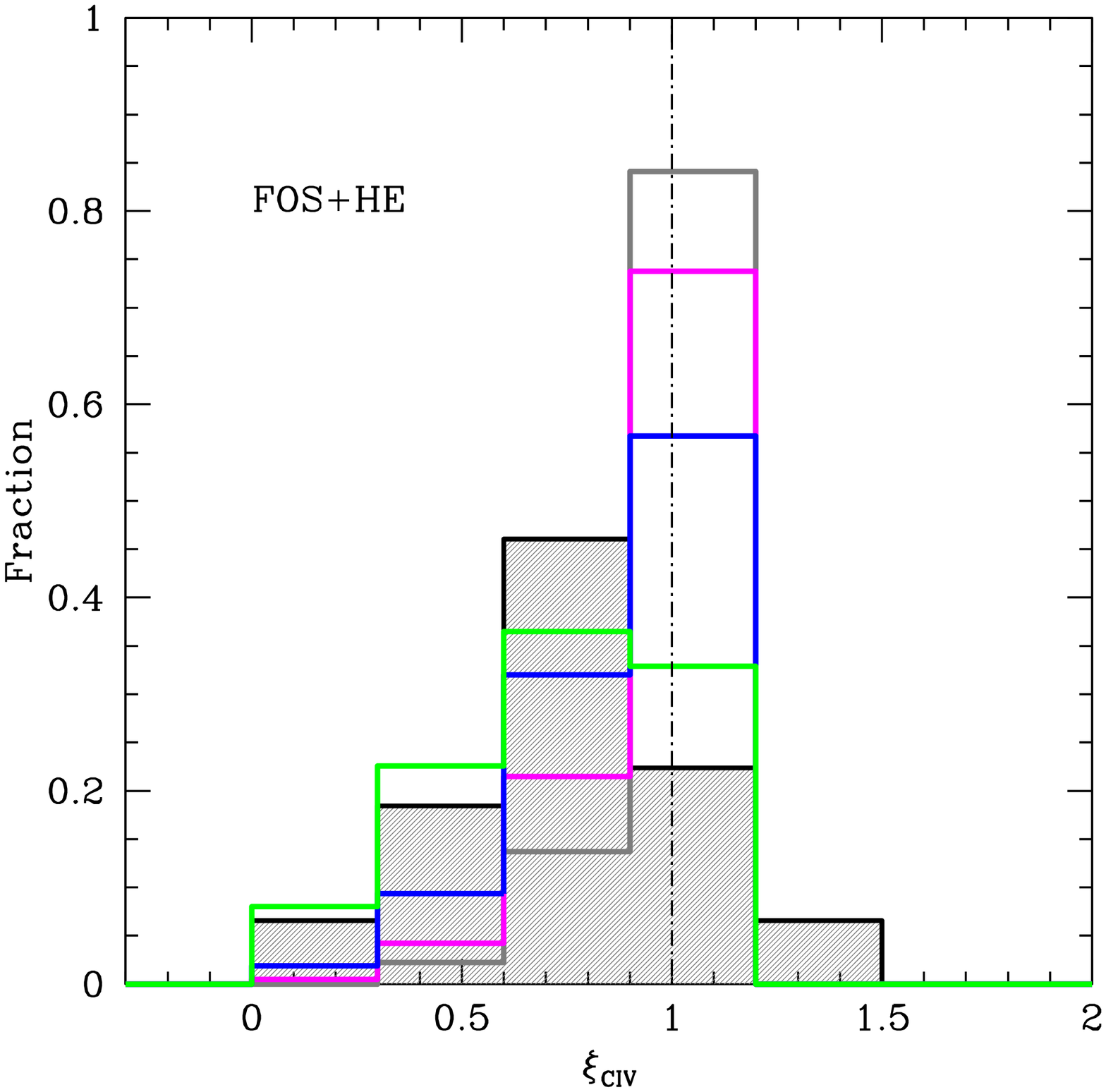}
\includegraphics[width=0.42\columnwidth]{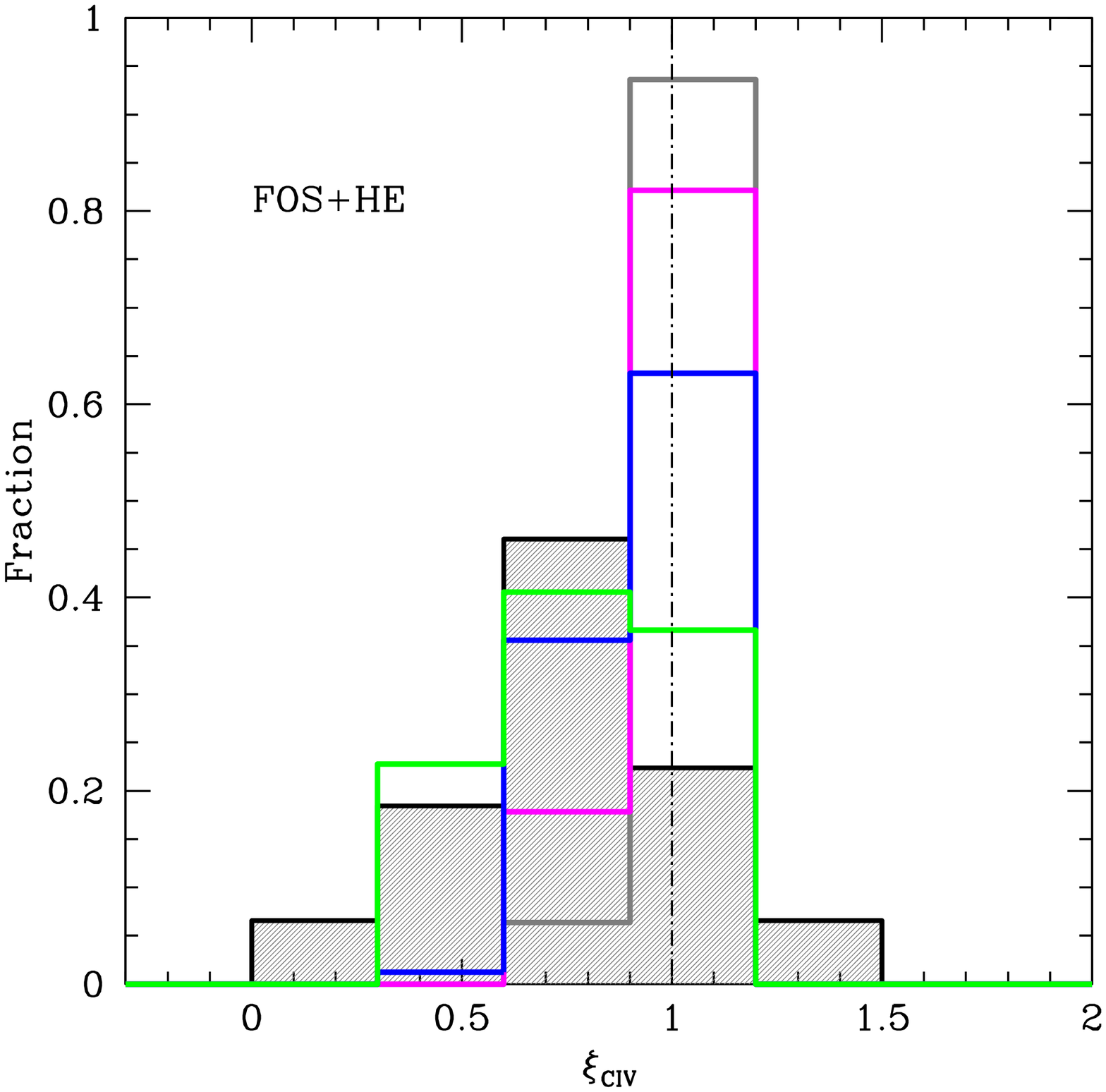}\\\includegraphics[width=0.42\columnwidth]{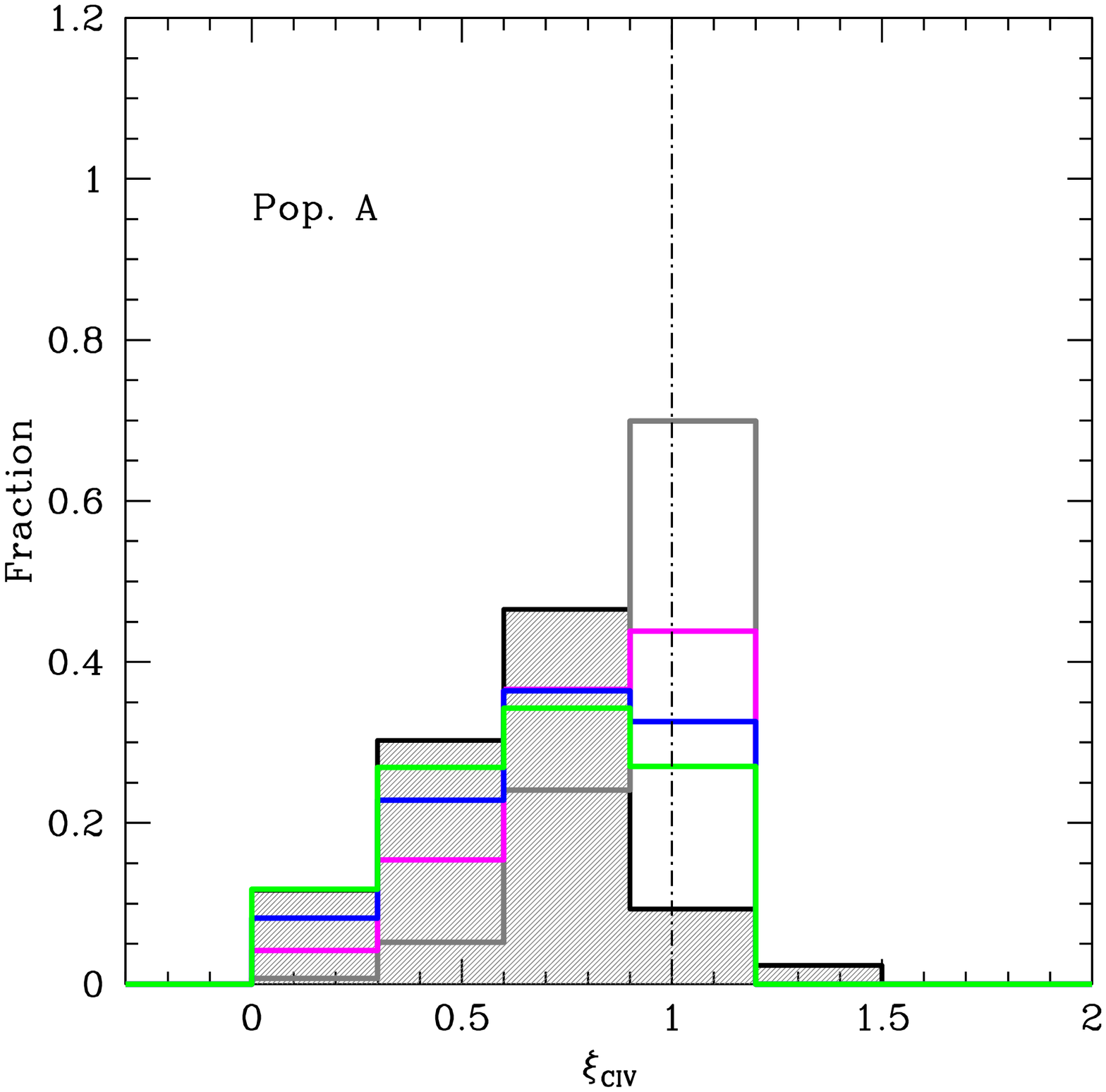}
\includegraphics[width=0.42\columnwidth]{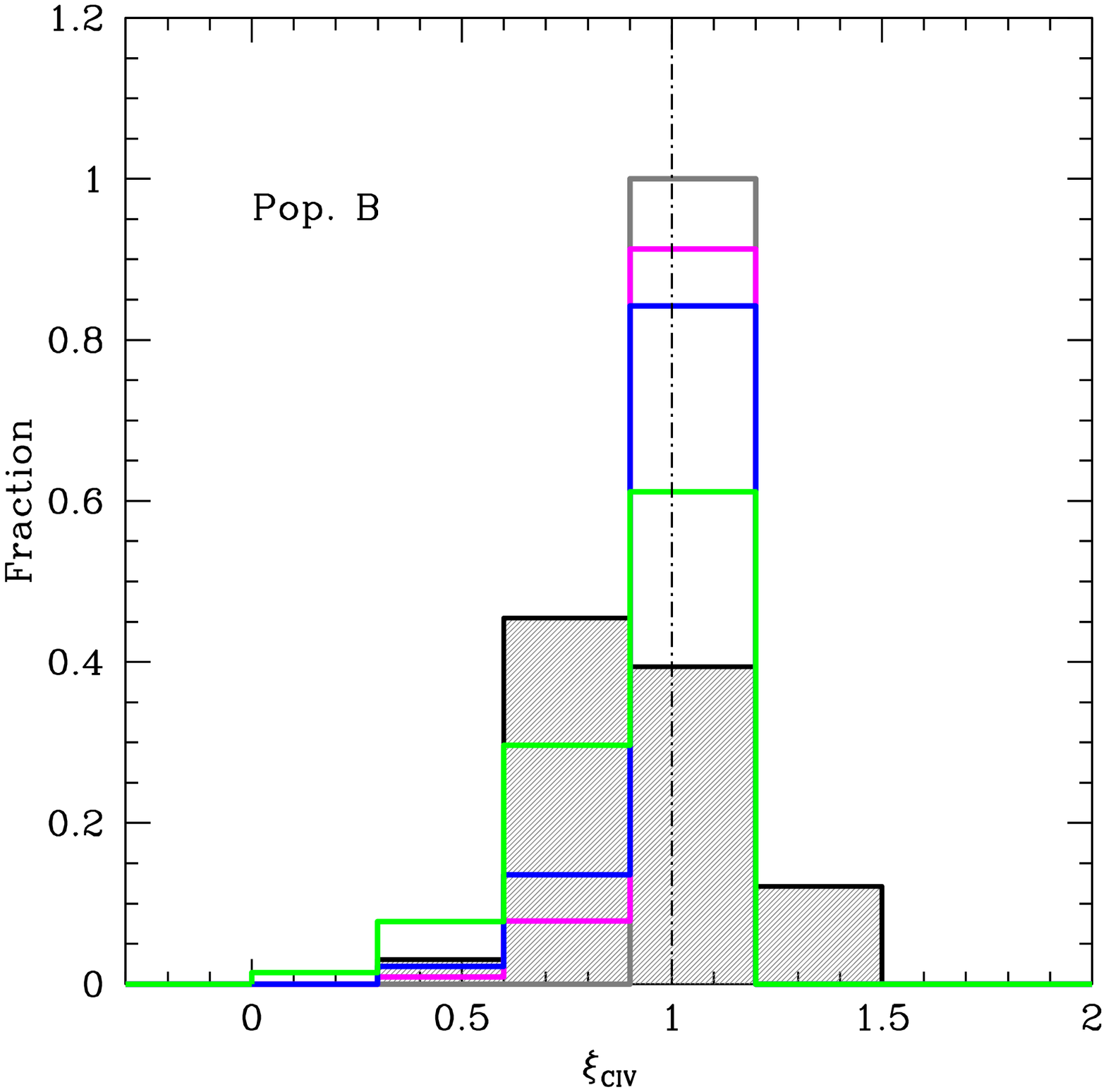}
\caption{Top left: Observed distribution of $\xi_\mathrm{CIV}$\ for the full FOS+HE sample (shaded histogram), and distribution of $\xi_\mathrm{CIV}$ for $\mathcal{Q}$=0.4 (grey),0.8 (magenta), 2.0 (blue), 7.8 (green), assuming $\kappa=0.1$, for randomly-oriented synthetic samples. Top right: same as in top left  panel, for $\kappa = 0.5$. Bottom left: distribution of $\xi_\mathrm{CIV}$ restricted to Pop. A sources  for $\mathcal{Q}$=1.0 (grey), 2.0 (magenta), 4.0 (blue),  12.0 (green), for $\kappa=0.1$. Bottom right: same for Pop. B sources, with $\mathcal{Q}$=0. (grey),0.2 (magenta), 0.4 (blue), 1.6 (green), for $\kappa=0.1$. See text for more details.}
\label{fig:xidistr}
\end{figure} 

The $\xi_\mathrm{CIV}$\ behavior as a function of the viewing angle $\theta$ is described in of Fig. \ref{fig:xibehav}. Fig. \ref{fig:xibehav} shows the dependence in case of a flat $\kappa =0.1$\ (black) or fat $\kappa = 0.5$ (red) for four values of ${\cal Q}$. { The bottom panel of Fig. \ref{fig:xibehav} shows the behaviour of the ratios $\tilde{\xi}_\mathrm{H\beta} = 1 / [4(\kappa^2 + \sin^2 \theta)]^{1/2}$  and $\tilde{\xi}_\mathrm{CIV} = 1/(4\kappa^2 + 4\sin^2 \theta + \mathcal{Q}\cos^2\theta)^{1/2}$. The $\tilde{\xi}$\ are the ratios between the $\delta v_\mathrm{K}$\ and the observed FWHM. At low $\theta$, the FWHM(\hb) underestimates the $\delta v_\mathrm{K}$\ by a large factor, while the overestimation of $\delta v_\mathrm{K}$\ by the FWHM(\civ) is almost independent of $\theta$\ and a factor $\approx 2$.}

The panels of Fig. \ref{fig:xidistr} compare the observed distribution of  $\xi_\mathrm{CIV,1}$\ (shaded histogram) with the prediction of  synthetic samples randomly oriented, at different ${\cal Q}$. We are not seeking a fit of the observed distribution especially around $\xi_\mathrm{CIV} \approx 1$\  because of the many biases affecting our sample and of the problem raised by $\xi_\mathrm{CIV} > 1$ (see below), but a qualitative consistency in the distribution of $\xi_\mathrm{CIV} < 1$.  \  

If we focus the analysis of Fig. \ref{fig:xidistr} mainly on large shifts, the presence of low $\xi_\mathrm{CIV}$ values and their higher frequency favors a highly flattened low-ionization BLR, as well as high ${\cal Q}$ for the full sample. A fat $\kappa = 0.5$\  BLR is unable to reproduce the largest shift amplitudes.   The scatter in $\xi_\mathrm{CIV}$ linear values at ${\cal Q}\gtrsim 2$ is $\approx 0.2$, implying a dispersion in the \mbh\ of $\approx$\ 0.15 dex.  If we separate Pop. A and B, more extreme values of ${\cal Q}\gtrsim 2$\ are required to fit the large shift distribution in Pop. A, with ${\cal Q}\sim 10$.  The distribution of    $\xi_\mathrm{CIV}$ for Pop. B is more peaked around $\xi_\mathrm{CIV} \approx 1$, and the $\xi_\mathrm{CIV}  $\ distribution can be qualitatively accounted for if ${\cal Q}\lesssim 2$.

The $\xi_\mathrm{CIV}$\ observed distribution includes values $> 1$.  These values are not possible following our model: the FWHM(\civ) should be always in excess or comparable to FWHM(\hbbc).
In the case  ${\cal Q} \gg 1$, the \civ\ line is broadened by an outflowing component; if ${\cal Q} \rightarrow 0$, $\xi_\mathrm{CIV} \lesssim 1$. In the latter case, the excess broadening may come from the smaller emissivity-weighted distance expected for \civ\ in a virial velocity field. The existence of cases with FWHM \civ $<$ FWHM \hb\ was already noted by \citet{mejia-restrepoetal18}, so it is not unique to the FOS+HE sample. The bottom right panel of Fig. \ref{fig:xidistr} shows that such cases are relatively frequent among Pop. B.  Inspection of the HE spectra in \citetalias{sulenticetal17} reveals that the \civ\ profile is significantly affected by semi-broad absorptions such as the ones often found in mini-BAL quasars \citep{vestergaard03,sulenticetal06a}. Since mini-BALs cluster around the line core, it is most likely that these Population B sources would satisfy the condition FWHM \civ $\gtrsim$ FWHM \hb\ if the effect of the absorptions could be removed.


\begin{figure}[htp!]
\centering
\includegraphics[width=0.6\columnwidth]{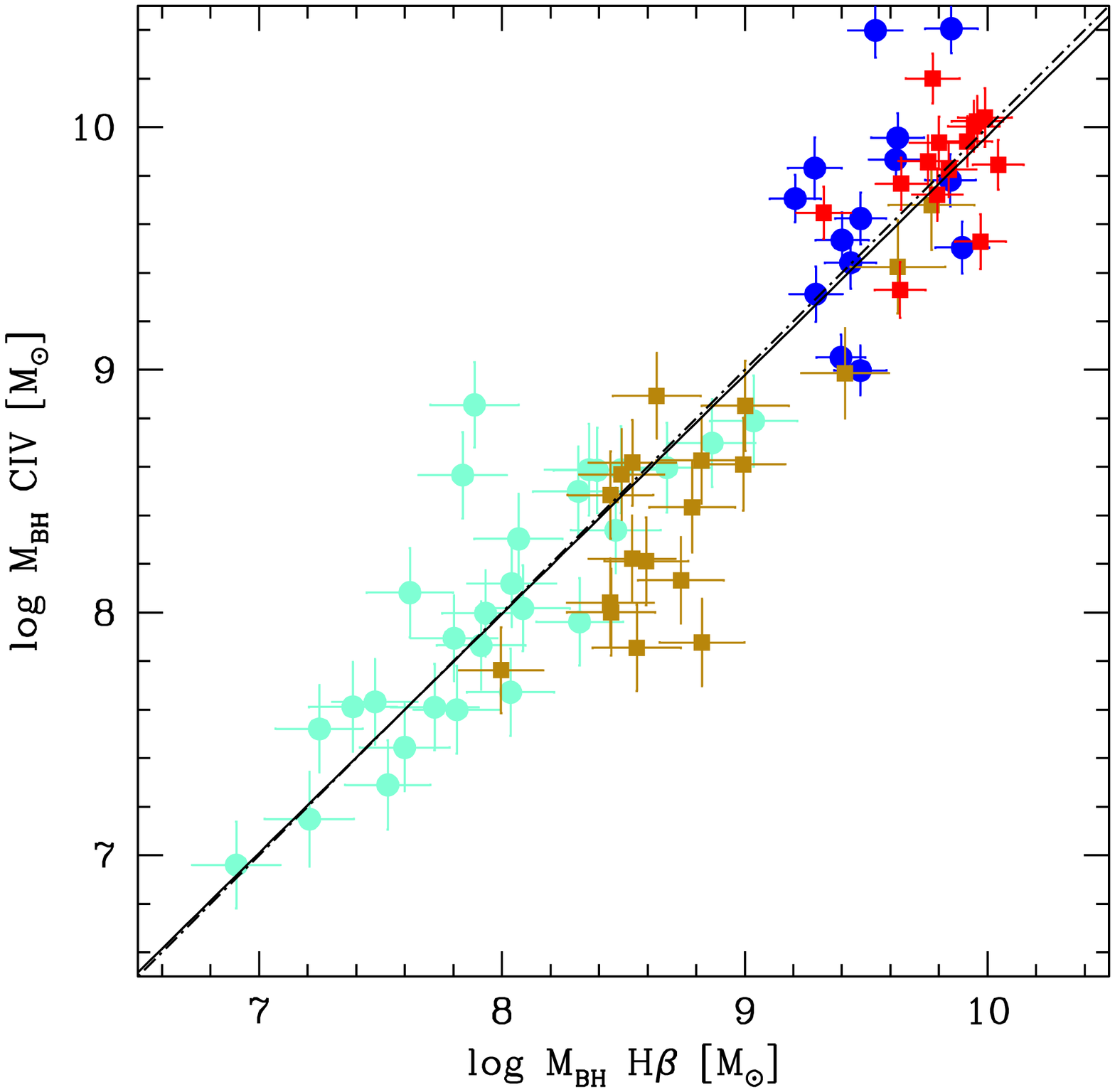}\\
\vspace{-0.25cm}
\includegraphics[width=0.6\columnwidth]{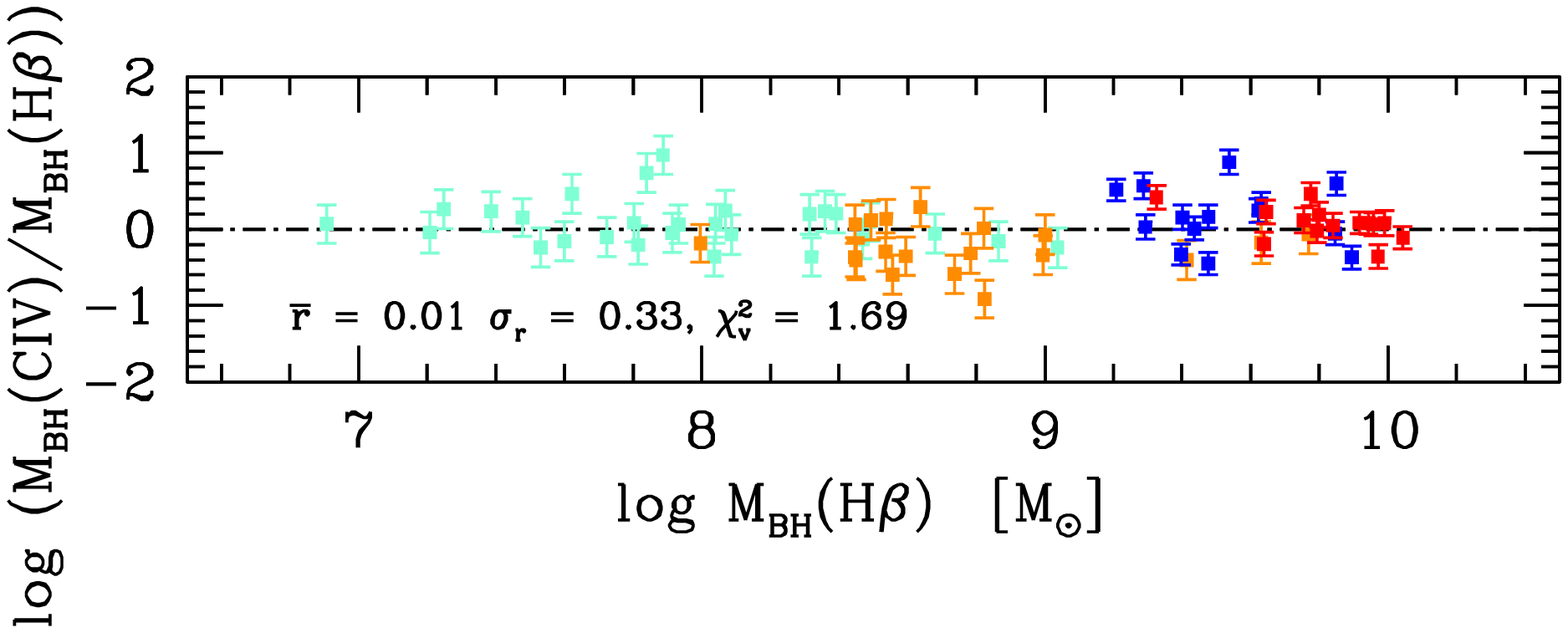}\\
\vspace{-3.cm}
\caption{\mbh\ computed from the fiducial relation of \citet{vestergaardpeterson06} based on FWHM \hb\ vs. the one computed from the \civ\ FWHM corrected following Eq. \ref{eq:xic1}. Error bars  include luminosity uncertainties estimated by the scatter in $L$\ derived from the UV and the visual spectral ranges along with errors on FWHM propagated quadratically. The lower panel shows the residuals for the three cases. Meaning of color code is the same as in the previous Figures. }
\label{fig:virialcivcorrmass}
\end{figure}

\subsection{\mbh\ scaling laws dependent on \lledd\ and $L$}
\label{mass}

The goal is to obtain an \mbh\ estimator based on \civ\ that is consistent with  the scaling law derived for \hb. In this context, a second-order dependence on luminosity of FWHM \civ\ cannot be ignored especially if samples encompass a broad range in luminosity. This will be  the case in deep, forthcoming surveys. Considering the corrections to FWHM \civ\ of \S \ref{virial}, the \mbh\ scaling law is derived in the form $\log M_\mathrm{BH} = \alpha \cdot \log L + 2 \cdot \log {\mathrm FWHM} + \gamma$ by minimizing the scatter and any systematic deviation of   \mbh\   estimated from \civ\ with respect to the \hb-derived masses:  the unweighted least square fit of Fig. \ref{fig:virialcivcorrmass} yields 

\begin{equation}
M_\mathrm{BH}(\mathrm{CIV})  \approx (0.99 \pm 0.04) M_\mathrm{BH}(\mathrm{H\beta}) + (0.11\pm 0.39). \label{eq:eqmass}
\end{equation}

 The \civ\ scaling law takes the form:
\begin{eqnarray}\label{eq:masscorr}
\log M_\mathrm{BH,1}\, \mathrm{CIV} & \approx & (0.64^{+0.045}_{-0.025}) \log L_{1450} + \\ \nonumber 
 && 2 \log \left(\xi_\mathrm{CIV,1}\mathrm{FWHM}(\mathrm{ CIV}) \right) + (0.525^{+0.22}_{-0.18})
 \end{eqnarray}

for the FWHM correction using Eq. \ref{eq:xic1}. Applying  Eq. \ref{eq:xilinear}, the scaling law does not change appreciably, and uncertainties in the coefficients are only  slightly different.  

\begin{eqnarray} \label{eq:masscorr1}
\log M_\mathrm{BH,2}\, \mathrm{CIV} & \approx & (0.63^{+0.045}_{-0.035}) \log L_{1450} + \\ \nonumber 
 && 2 \log \left(\xi_\mathrm{CIV,2}\mathrm{FWHM}(\mathrm{ CIV}) \right) + (0.525^{+0.275}_{-0.19})
\end{eqnarray}

The scaling law parameter uncertainties have been estimated following the standard approach in \citet[][p. 210ff]{bevingtonrobinson03}, with the constrain that unbiased consistency  between \mbh\ from \hb\ and \civ\ (Eq. \ref{eq:eqmass}) is satisfied within the $1 \sigma$ uncertainties.  The rms scatter is  $\sigma \approx 0.33$\ for the Eq. \ref{eq:xic1} and to $\sigma \approx 0.35$\ for Eq. \ref{eq:xilinear}. Assuming a single correction for both Pop. A and B significantly worsens the fit quality, and no scaling law is reported. 

{ An application of the bisector fitting technique using {\tt SLOPES}  \citep{feigelsonbabu92} yields :}

\begin{eqnarray}\label{eq:masscorrb}
\log M_\mathrm{BH,1}\, \mathrm{CIV} & \approx & (0.5925^{+0.0275}_{-0.030}) \log L_{1450} + \\ \nonumber 
 && 2 \log \left(\xi_\mathrm{CIV,1}\mathrm{FWHM}(\mathrm{ CIV}) \right) + (0.62\pm032)
 \end{eqnarray}

\begin{eqnarray}\label{eq:masscorrb}
\log M_\mathrm{BH,2}\, \mathrm{CIV} & \approx & (0.572^{+0.0285}_{-0.032}) \log L_{1450} + \\ \nonumber 
 && 2 \log \left(\xi_\mathrm{CIV,2}\mathrm{FWHM}(\mathrm{ CIV}) \right) + (0.64\pm035)
 \end{eqnarray}


The \citet{vestergaardpeterson06} \hb\ scaling laws suffer from a significant scatter (see the discussion in their paper) that can be explained on the basis of the scatter induced by orientation (0.33 dex at 1$\sigma$) according to the results of Appendix \ref{incl}.   The \civ\ and \hb\ relation should be considered equivalent. The luminosity exponent ($\approx 0.64$)\ is { in agreement with previous observations \citep{petersonetal05}. It is slightly above } the exponent of the \civ\ radius dependence on luminosity found in more recent reverberation mapping studies \citep[$\approx 0.52 - 0.55$][]{kaspietal07,liraetal17,liraetal18}.  
 
Fig. \ref{fig:virialcivcorrmass} suggests the presence of a well-behaved distribution with a few outlying points. It is possible to reduce the scatter to $\sigma \approx 0.25$\ applying a $\sigma$\ clipping algorithm (i.e., eliminating all sources deviating more than  $\pm$ 2$\sigma$), with no significant change in the best fitting parameters. This selective procedure is however unwarranted: as shown in Appendix \ref{origin}, outlying points are expected right because of the possible occurrence of low-probability viewing angles. The residual rms  can be largely accounted for by orientation effects if $\cal{Q}$ is small, and by the combination of orientation effects and outflow prominence if $\cal{Q}$\ is much larger than 1 (Appendix \ref{origin}).

\subsection{Application to a large sample with \hb\ and \civ\ data}
\label{large}

\begin{figure}[htp!]
\centering
\includegraphics[width=0.9\columnwidth]{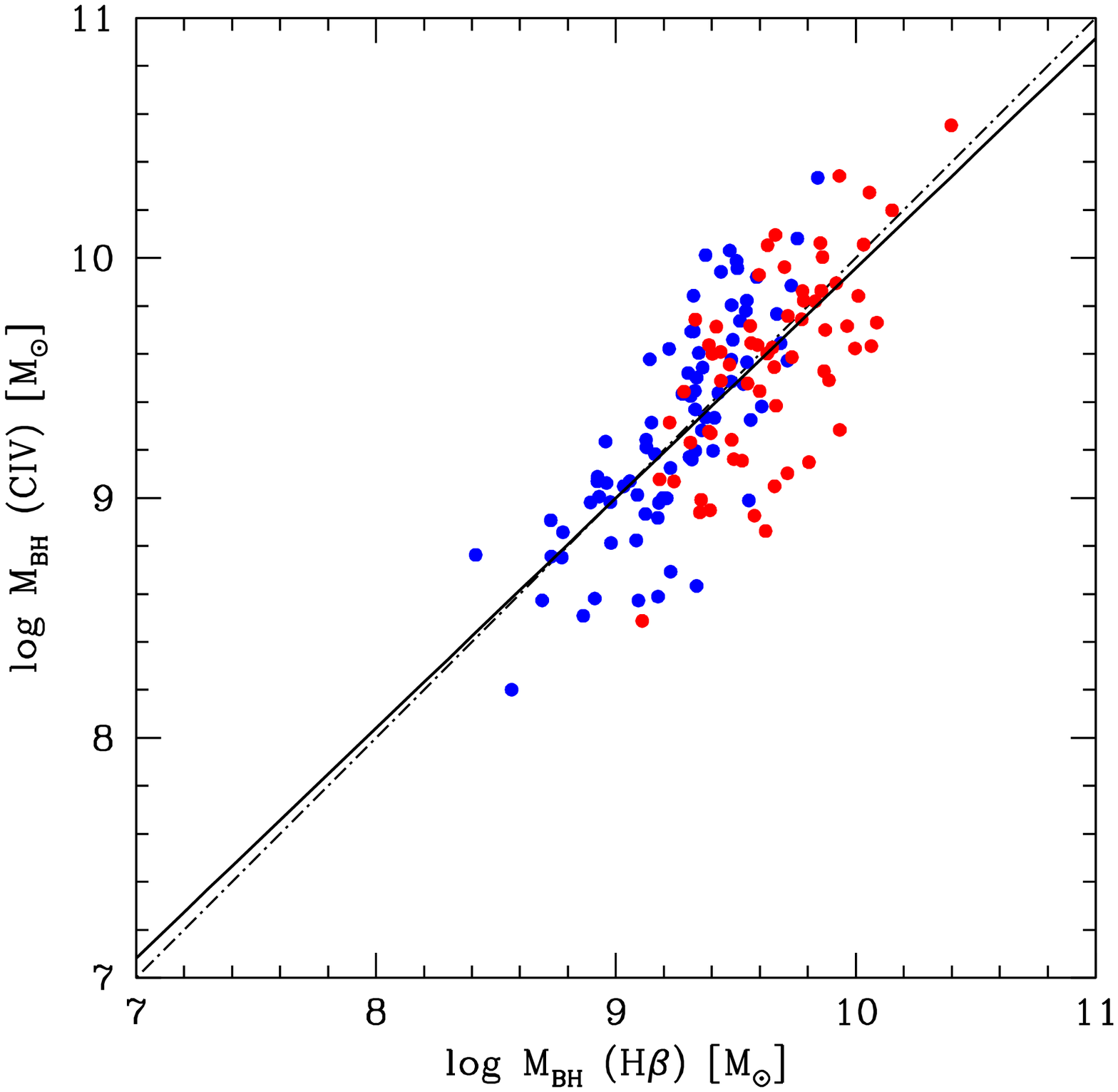}\\
\caption{Estimates of \mbh\ using corrected \civ\ FWHM as a VBE  versus \mbh\ computed from \hb\ FWHM using the scaling-law of \citet{vestergaardpeterson06}. The luminosity and FWHM data from \citet{coatmanetal17} with Pop. A (blue) and B sources kept separated (see Sect. \ref{large} for more details), and with \mbh\ computed according to Eq. \ref{eq:masscorr}. The dot-dashed line traces the 1:1 relation between \mbh\ from \civ\ and \hb; the black line is an unweighted lsq fit  for the A+B \mbh\ estimates.}
\label{fig:c17}
\end{figure}

The \citet{coatmanetal17} data provides a different sample for the testing of the 
 scaling law of Eq. \ref{eq:masscorr}.  The flux bisector can be converted into \cqp\  (Sect. \ref{empcorr}).  An application of Eq. \ref{eq:masscorr} to Pop. A and B (applying the luminosity-dependent separation as in \citetalias{sulenticetal17}) is yielding agreement with the expectation of an unbiased \mbh\ estimator with respect to the \hb\ \mbh\ estimates using the scaling law of \citet{vestergaardpeterson06}: { the slope of an unweighted lsq fit is $\approx 0.958  \pm  0.065$\ (using {\tt SLOPES}, \citealt{feigelsonbabu92}, Fig. \ref{fig:c17}). } 
 

\section{Conclusion}

The present investigation was focused on the \civ\ \  relations to  \hb\  over a broad range of luminosity ($\log L \sim 43 - 48$, including very high luminosities $\log L \gtrsim$ 47)	 in the Eigenvector 1 context, with the goal of testing the \civ\  suitability as  virial broadening estimators when  low-ionization lines observations are not available. The  Eigenvector 1 context means that the quasar main sequence is considered to properly interpret first-order Eddington ratio effects and luminosity effects that appear to be second order in low-$z$\ samples. 

The main conclusions reached in this paper are as follows:

\begin{enumerate}
\item  Within the limits of our sample size, and of our UV spectral coverage,  it does not appear that the \civ\ FWHM  can be used as a reliable virial broadening estimator without significant corrections. There is a large scatter between FWHM measurements on \hb\ and \civ\  that seems to defy the definition of a meaningful  trend. 
\item 
The \civnc\ removal improves the agreement between \civ\ and \hb\ measures in the HE sample, but   \civnc\ is not a major factor hampering the definition of a \civ\ VBE consistent with \hb. 
\item Corrections to FWHM \civ and \civ-based \mbh\ estimates that  vary systematically along the 4DE1 sequence and are strongly dependent on Eddington ratio are promising and should be further explored. 
\item Following the results of \citetalias{sulenticetal17}, we define a correction to the FWHM(\civ) based on the full-profile \civ\ \cmp\   (a proxy for \lledd) and on the luminosity at 1450 \AA.  Given the intrinsic differences between Pop. A and B, and their ``threshold'' separation dependent on a critical \lledd, two different correction laws were considered for the two populations. We remark that the correction for Pop. B as derived from the FOS+HE sample is highly uncertain. 

\item  {The \mbh\ scaling law (Eq. \ref{eq:masscorr}) associated with the corrected FWHM(\civ) following Eq. \ref{eq:xic1}, as explained in Sect. \ref{mass}, allows for the preservation of the virial dependence on line broadening. Its practical usefulness rests on the ability to distinguish Pop. A and B quasars. This can be achieved in a large fraction of quasars following the guideline set forth by \citet{negreteetal14}.} 
\item We constructed a toy model that helped the interpretation of the scatter. Orientation effects induce scatter $\approx 0.3 - 0.4$ dex in mass estimates that account for a large fraction of the dispersion in the landmark scaling law of \citet{vestergaardpeterson06}. { A physical model of the disk + wind system might allow to recover the viewing angle $\theta$\ for individual quasars.} 
\end{enumerate}

\begin{acknowledgements}

 The authors acknowledge the contribution of Jack Sulentic to the development of this paper which  is based  on observations made with ESO Telescopes at the Paranal Observatory under programme 082.B-0572(A), and  with the Italian Telescopio Nazionale Galileo (TNG) { operated on the island of La Palma by the Fundaci\' on Galileo Galilei of the INAF at the Spanish Observatorio del Roque de los Muchachos of the Instituto de Astrof\'{\i}sica de Canarias.}. PM acknowledges the Programa de Estancias de Investigaci\'on (PREI) No. DGAP/DFA/2192/2018 of UNAM, where this paper was completed.  The relevant research is part of the projects 176001 ``Astrophysical spectroscopy of extragalactic objects'' and 176003 ''Gravitation and structure of universe on large scales'' supported by the Ministry of Education, Science and Technological Development of the Republic of Serbia. MLMA acknowledges a CONACyT postdoctoral fellowship, and wishes to thank the IAA for support during her postdoctoral stay from February 2016 to July 2018.  MLMA acknowledges a CONACyT  postdoctoral fellowship and National Science Centre,  Poland,  grant  No.2017/26/A/ST9/00756  (Maestro 9). AdO and MLMA acknowledge financial support from the Spanish Ministry of Economy and Competitiveness through grant AYA2016-76682-C3-1-P. { AdO acknowledges financial support from the State Agency for Research of the Spanish MCIU through the ``Center of Excellence Severo Ochoa" award for the Instituto de Astrof\'{\i}sica de Andaluc\'{\i}a (SEV-2017-0709).}  MLMA, PM and MDO acknowledge funding from  the INAF PRIN-SKA 2017 program 1.05.01.88.04. DD and AN acknowledge support from CONACyT through  grant CB221398. DD  also thanks for support from grants IN108715 and IN113719  PAPIIT, DGAPA, UNAM.  MAMC was partially supported by the Spanish Research project MTM2015-64095-P and by Diputaci\'on General de Arag\'on, Group E24-17R.  {  The scientific results reported in this article are also
based on publicly available HST spectra.}
\end{acknowledgements}

\vfill

\begin{appendix}

\section{ The MS / 4DE1 formalism: a glossary }
\label{glossary}

\begin{figure}[htp!]
\centering
\includegraphics[width=0.95\columnwidth]{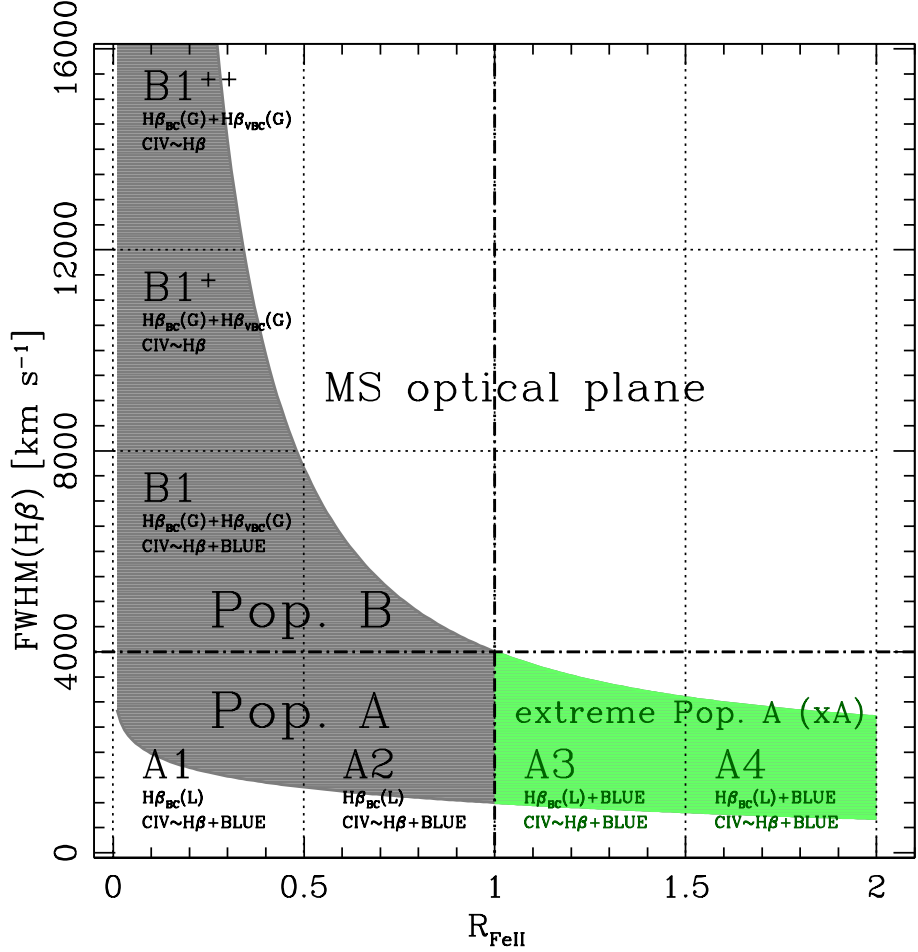}\\
\caption{ Schematic representation of the optical plane of the quasar MS, with the subdivisions identifying spectral types along the sequence. The main components that are blended in the \hb\ and  \civ\ profile are listed in each spectral bin. The shaded area shows the approximate occupation of low-$z$ quasars sample in the plane. }
\label{fig:sketch}
\end{figure}

{ In the optical plane of the quasar main sequence,  spectral types are isolated following \citet{sulenticetal02}. Fig. \ref{fig:sketch} provides a sketch with the spectral types identification in the optical plane of the MS.  Here we provide a glossary of the MS-related terms  and acronyms employed in the paper, along the order of the quasar main sequence. A more thorough description can be found in \citet{sulenticetal11} and \citet{marzianietal18}. The rational for two type-1 quasar populations (A and B) was originally given by \citet{sulenticetal00a}.
}

\begin{center}\scriptsize
\begin{tabular}{p{1.75cm}p{6cm}}\hline
ST & \hspace{2.5cm}Definition \\
\hline
\\
\hline
\multicolumn{2}{p{8cm}}{ Population B: FWHM(\hb) $\gtrsim$  4000 \kms. Virial-dominated with  redward-asymmetric profiles in \hb\ and \mgii. Also described a disk-dominated \citep{richardsetal11}.} \\
\hline
B1++ & 12000 \kms $\le$ FWHM(\hb) $<$ 16000 \kms,\rfe $<$0.5 \\
B1+   & 8000 \kms $\le$ FWHM(\hb) $<$ 12000 \kms \rfe $<$0.5 \\
B1     & 4000 \kms $\le$ FWHM(\hb) $<$ 8000 \kms\ \rfe $<$0.5\\
\hline
\\
\hline
\multicolumn{2}{p{8cm}}{Population A:  FWHM(\hb) $\lesssim$  4000 \kms. Sources frequently show \civ\ blueshifts, and \hb\ Lorentzian-like profiles \citep{duetal16}. Includes a range of FeII emission.  }\\
\hline
A1    & FWHM(\hb) $<$ 4000 \kms; \rfe $<$0.5 \\
A2    & FWHM(\hb) $<$ 4000 \kms; $\le$ 0.5 \rfe $<$1\\
A3 [xA]    & FWHM(\hb) $<$ 4000 \kms; $\le$ 1 \rfe $<$1.5\\
A4  [xA] & FWHM(\hb) $<$ 4000 \kms; $\le$ 1.5 \rfe $<$2\\
\hline
\end{tabular}
\end{center}

{Spectral types A3 and A4 are are also indicated as extreme Population A, with \rfe $\gtrsim 1$. They are the highest radiators per unit mass and possibly super-Eddington accretors \citep{wangetal14}. }

{ The 4D Eigenvector-1 formalism was introduced to limit the set of MS-correlated parameters to the four ones that are most relevant for the MS physical interpretation. In addition to FWHM \hb, \rfe, and \cmp\ \civ, the soft X-ray photon index $\Gamma_\mathrm{soft}$ is also considered \citep{sulenticetal00b}. The four parameters were meant to represent the velocity dispersion of the LIL-emitting part of the BLR, the physical condition within the LIL-BLR (\rfe), the dynamical condition of the HIL emitting gas, and the accretion state of the black hole. $\Gamma_\mathrm{soft} > 2$\ implies a soft X-ray excess that is exclusive of Population A \citep{wangetal96,bolleretal96,shenho14,benschetal15}. }

{  Line components assumed in the decomposition along the MS are defined as follows (see also the sketch of Fig. \ref{fig:sketch}).}
\vfill
\begin{center}\scriptsize
\begin{tabular}{p{1.75cm}p{7cm}}\hline
Line component & Definition \\
\hline
\\
\hline
\multicolumn{2}{p{7cm}}{{\hspace{3cm}Low-ionization lines (LILs): HI \hb}}\\
\hline
\hbvbc\ & \multicolumn{1}{p{7cm}}{Gaussian FWHM $\sim 10000$ \kms, redshifted by $1-2000$\kms\ (defining property of Pop. B; absent in Pop. A)}\\
\hbbc\ & Lorentzian, FWHM $\sim 1-4000$ \kms\ (Pop. A); Gaussian FWHM \hb $\gtrsim$ 4000 \kms\ (Pop. B)\\
\hb\ BLUE & Asymmetric Gaussian which models an excess of emission on the blue side of \hb; usually weak save in ST A3 and A4\\
\hb\  & Sum of \hbbc, \hbvbc, and BLUE (when applicable); full broad \hb\ profile \\
\hbnc\ & \hb\ narrow component \\
\\
\hline
\multicolumn{2}{p{7cm}}{\hspace{3cm} High-ionization lines (HILs): \civ}\\
\hline
\civvbc\ & Gaussian FWHM $\sim 10000$ \kms, redshifted by $1-2000$\kms \\
\civbc\ &Lorentzian, if FWHM(\hb)$\lesssim $ 4000 \kms (Pop. A); Gaussian FWHM $\gtrsim 4000$ \kms (Pop. B) \\
\civ\ BLUE & Asymmetric Gaussian which models an excess of emission on the blue side of \civ; detected in most quasar and most prominent in spectral types A3 and A4\\
\civ & Sum of \civbc, \civvbc, and \civ\ BLUE; full broad \civ\ profile \\
\civnc\ & \civ\ narrow component, prominent in Pop. B at low-$z$; almost absent in most Pop. A sources\\
\hline
\end{tabular}
\end{center}
\vfill

{ BLUE becomes detectable as a blueward excess in the \hb\ profile mainly in A3 and A4 (more infrequently in A2). For \civ, BLUE increases in prominence along the sequence from B1$^{++}$, to A4. In   B1$^{++}$ and  B1$^{+}$ is weak and often undetectable, while in A3 and A4 it may dominate \civ\ emission.  }
\vfill
\onecolumn

\section{Effect of orientation on \hb\ \mbh\ estimates}
\label{incl}

From the inversion of Eq. \ref{eq:flat}, we obtain an expression for the viewing angle
$\theta = \arcsin{\sqrt{x^2/4 - \kappa^2}}$,
where $x = v_\mathrm{obs}/\delta v_\mathrm{K}$ and $\kappa = \delta v_\mathrm{iso}/\delta v_\mathrm{K}$.\ The probability to observe  $v_\mathrm{obs}$ for a given  $ \delta v_\mathrm{K}$\ is then 

\begin{equation}
\centering
P(x) = \sqrt{x^2/4-\kappa^{2}} \frac{d\theta}{dx}  =  \frac{x/4}{\sqrt{\kappa^2-x^2/4+1}} 
\end{equation}

The black hole mass is $\propto x^2$. Therefore the average effect can be written as

\begin{equation}
< \frac{M_\mathrm{BH,obs}}{M_\mathrm{BH,K}} > = \int_{x(v_\mathrm{iso})}^{x(\mathrm{edge})} x^2 P(x) dx{  /}  \int_{x(v_\mathrm{iso})}^{x(\mathrm{edge})}   P(x) dx\label{eq:mratio}
\end{equation}

The integrals  of Eq. \ref{eq:mratio} can be computed analytically:

\begin{equation}
 \int x^2 P(x) dx =  (-\frac{4}{3}\sqrt{4\kappa^2-x^2+4})-\frac{4}{3}\kappa^2 \sqrt{4\kappa^2-x^2+4} -\frac{1}{6}(x^2 \sqrt{4\kappa^2-x^2+4}) 
\end{equation}

\begin{equation}
 \int  P(x) dx =  -\sqrt{1+ \kappa^2 - x^2/4}
\end{equation}

Note that the integration limits in $x$ (which correspond to $\theta =0 $ and $\theta=45$) are different for $\kappa=0.1$ and 0.5. For $\kappa =0.1$, $x_\mathrm{iso} =0.2$ and $x_\mathrm{edge} \approx1.43$ In the latter case, $\theta=0$ corresponds to $x_{min}$ = 1 and $\theta = 45$ to $x_\mathrm{edge} \approx1.73$: \mbh\ will be always overestimated, for every possible $\theta$ value larger than 0.  

If we consider the Eddington ratio, we obtain: 

\begin{equation}
 \int  P(x)/x^2 dx=   -\arctan{\mathrm h} \left(\frac{\sqrt{\kappa^2+1}}{\sqrt{\kappa^2-x^2/4+1}}\right)/4{\sqrt{\kappa^2+1}}
\label{eq:lledd} \end{equation}
 
Eq. \ref{eq:lledd} implies a significant effect on \lledd\ for both $\kappa = 0.1 $ \ and $\kappa = 0.5$. In the first case, the \lledd\ will be overestimated by a factor $\approx 1.8$.  In the second case, the \lledd\ will be underestimated by a factor $\approx$2, due to the systematic overestimation in \mbh.  


\twocolumn
\section{Origin of scatter in scaling laws}
\label{origin}

\begin{figure}[htp!]
\centering
\includegraphics[width=0.4\columnwidth]{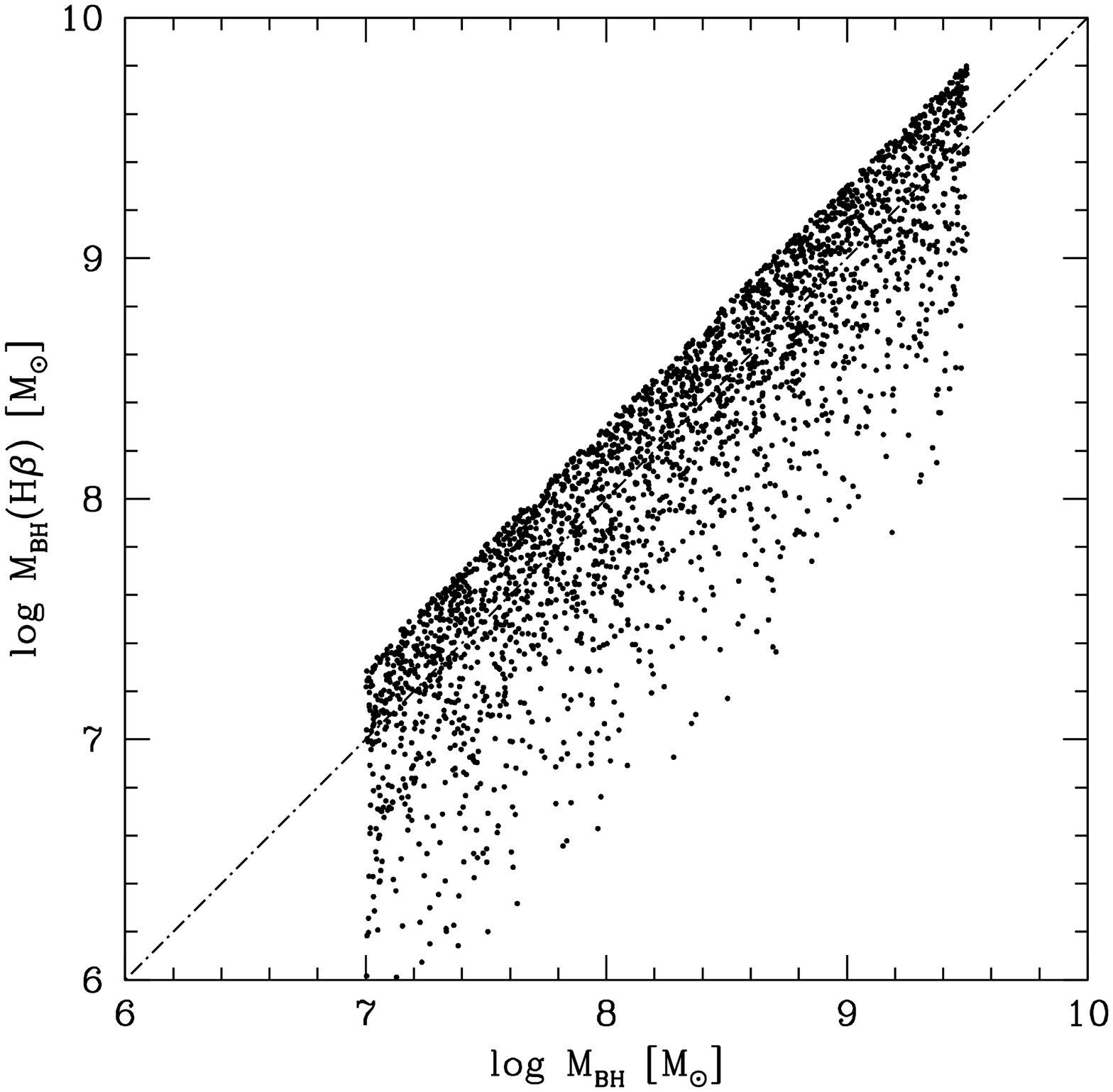}\\
\vspace{-0.25cm}
\includegraphics[width=0.4\columnwidth]{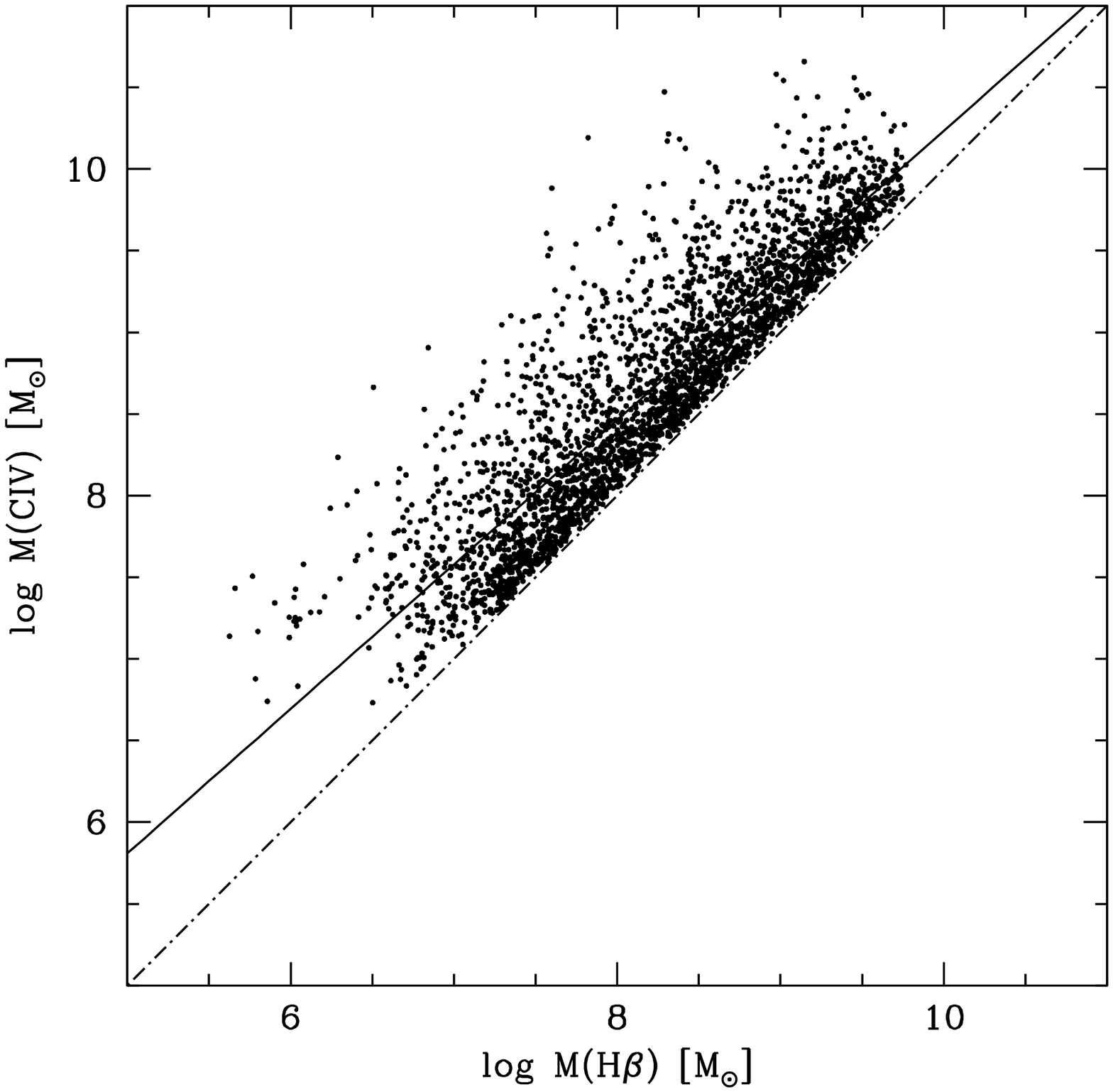}\\
\vspace{-0.25cm}
\includegraphics[width=0.4\columnwidth]{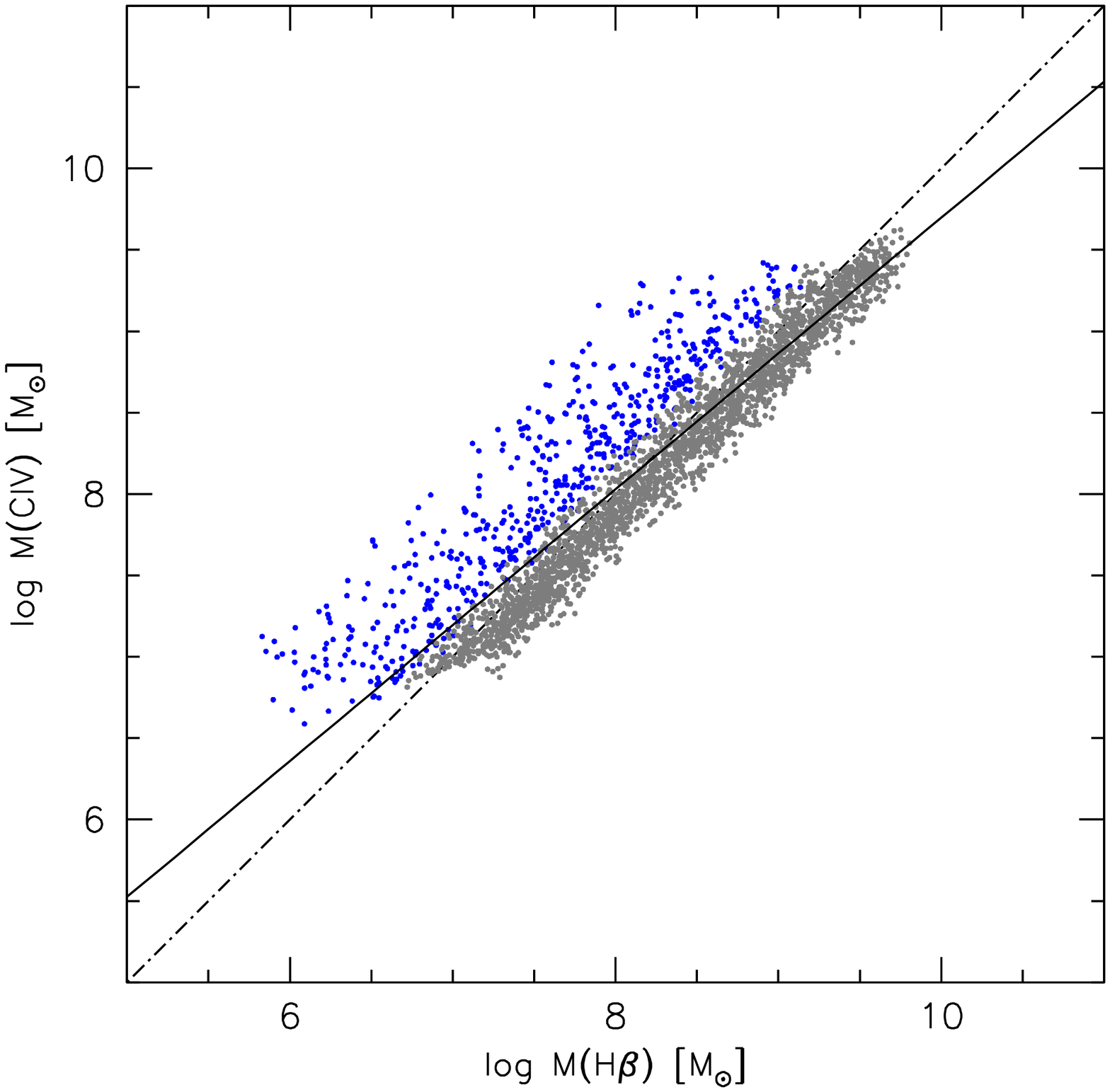}\\
\vspace{-.25cm}
\includegraphics[width=0.4\columnwidth]{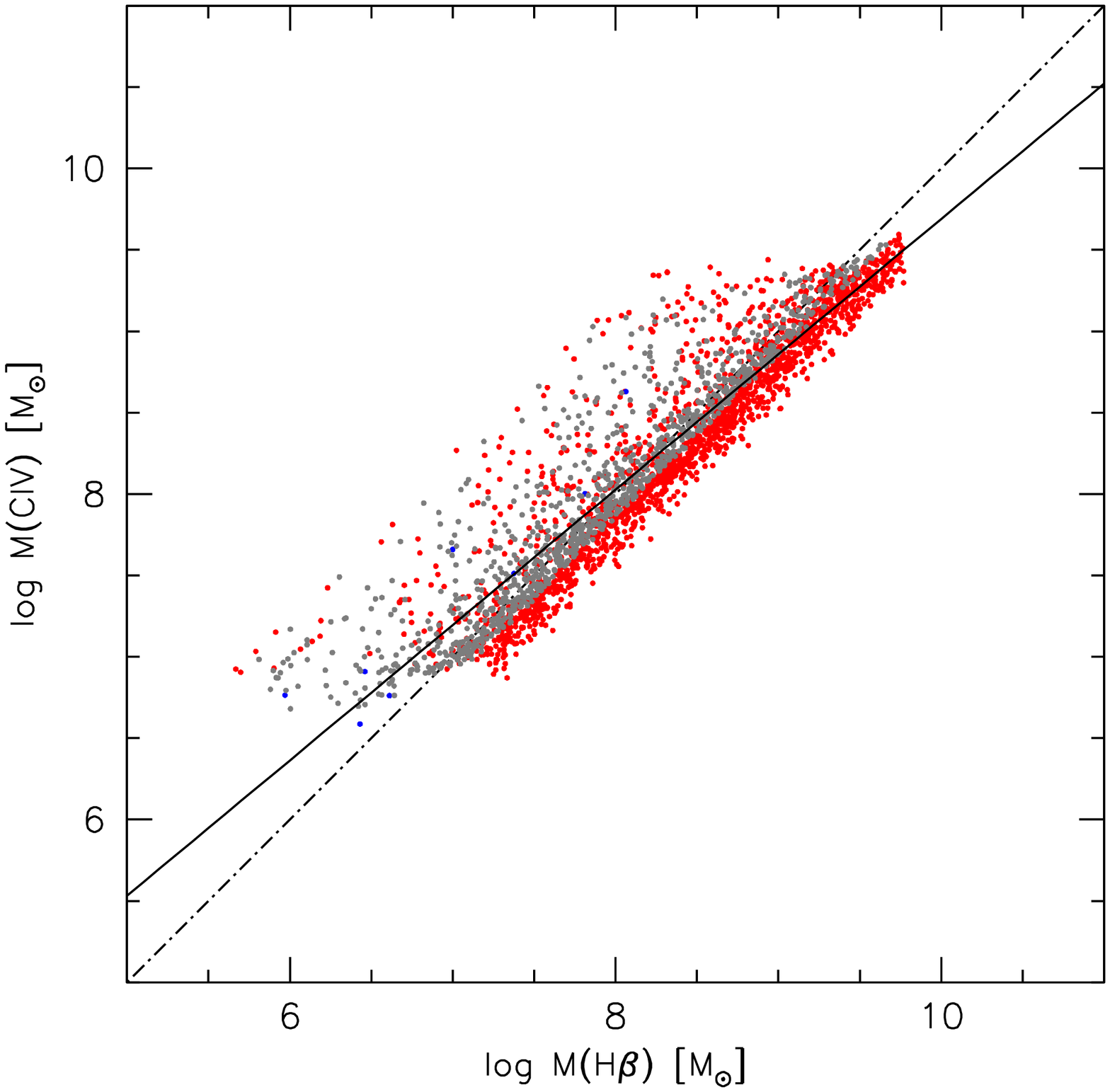}\\
\vspace{-.25cm}
\includegraphics[width=0.4\columnwidth]{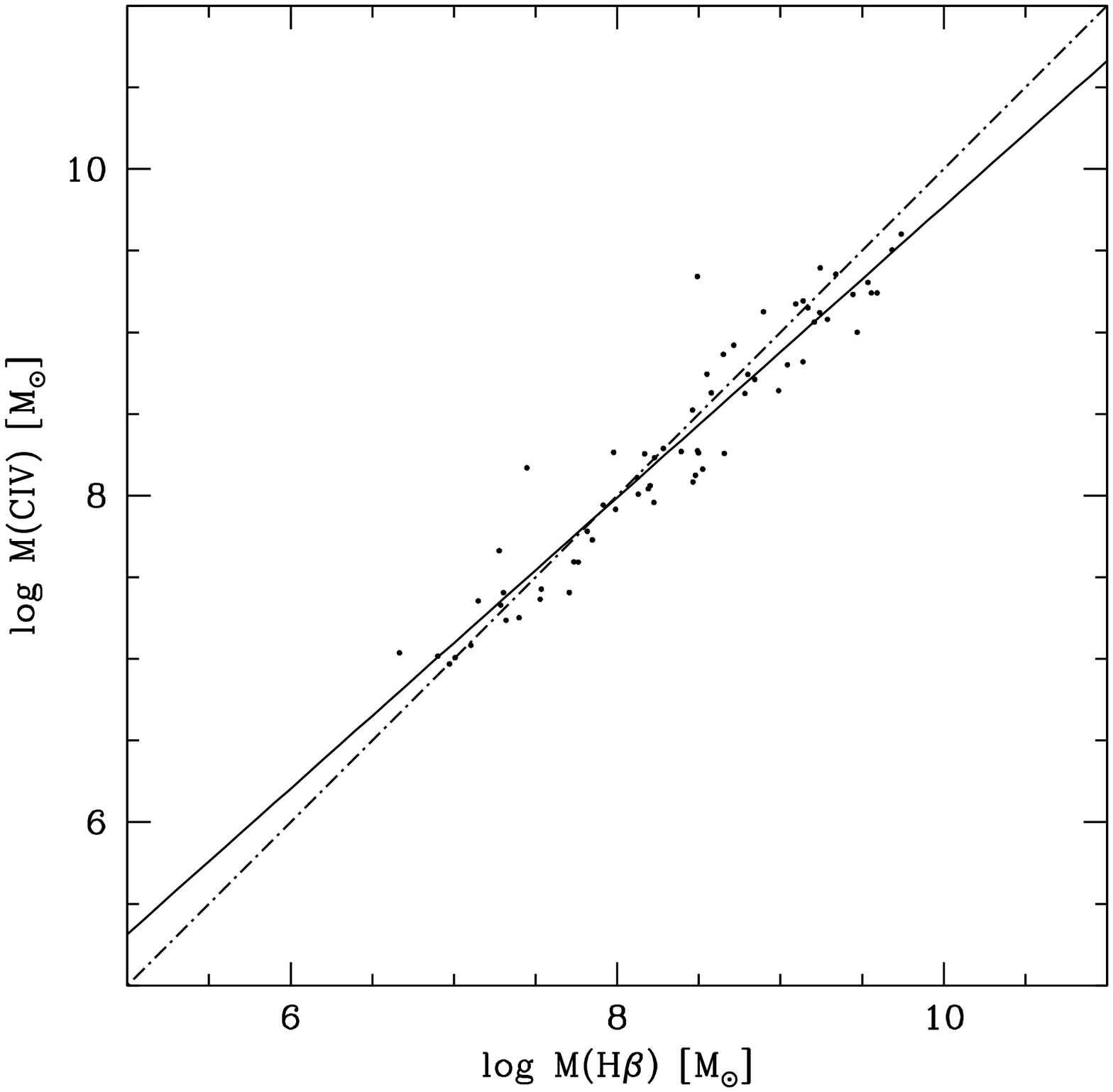}\\
\vspace{-.25cm}
\caption{Top panel: \mbh\ with effect of orientation via Eq. \ref{eq:xihb} vs ``true''  \mbh, for a synthetic sample of 10000 sources.  Second from top: \mbh\ \civ\ vs \mbh\ \hb\ estimated for an Eddington ratio distribution as described in the text and no correction.
{ Third from top: The FWHM \mbh\ \civ\ has been corrected because of outflow broadening using a correction factor $\xi$. The blue dots identify the \mbh\ \hb\ estimates that are under 0.33 dex the true \mbh; the red ones are for overestimates by more than 0.33 dex. The grey dots represent mass estimates within $-$0.33 and +0.33 dex from the true value. Fourth from top: same, with color coding referring to \civ\ \mbh. }  
Bottom: a synthetic sample with $n \lesssim 80$ sources, as in the FOS+HE sample. See text for more details. In all panels, the dot-dashed line is the equality line; the filled line traces an unweighted least square fit. }
\label{fig:synthmass}
\end{figure}

We considered  synthetic samples of $\sim 10000$\ objects obtained from random variates with distribution $P(\theta) \propto \sin \theta$\ ($0 \le \theta \le \pi/4$), and ``true'' \mbh\ uniformly distributed in the range $10^{7} M_{\odot} \le M_\mathrm{BH} \le 10^{9} M_{\odot}$.  The dependence on orientation of the \hb\ FWHM is assumed to follow Eq. \ref{eq:flat}, with $\kappa = 0.1$. The effect of orientation on FWHM is such that, in a randomly oriented synthetic sample, the \mbh\ estimated from \hb\ deviates from the true \mbh\ as in Fig. \ref{fig:synthmass} (top panel).  The dispersion is $\approx 0.35$, which is comparable to the uncertainty in the scaling-law \mbh\ estimate following \citet[][$\approx$ 0.5 at 1$\sigma$\ confidence level]{vestergaardpeterson06}. 

The estimates of \mbh\ \civ\ show a significant scatter if plotted against \mbh\ \hb\ (Fig. \ref{fig:virialcivcorrmass}). The origin of the scatter is in part related to orientation, in part to the outflow component.  The second panel from top of Fig. \ref{fig:synthmass} shows the \mbh\ \civ\ vs \hb\ for a synthetic sample to which no correction has been applied. FWHM of \civ\ and \hb\ are expected to be related to the Keplerian velocity by Eqs. \ref{eq:fciv} and \ref{eq:xihb}, respectively.  The distribution of \lledd\ has been assumed Gaussian, peaking at $\log$\lledd$=-0.3$, and $\sigma \approx 0.5$.  Typical values of $\xi_\mathrm{CIV}$ and ${\cal Q}$\ are appropriate for Pop. A sources. There is a strong bias (0.5 dex)  and a standard deviation of the mass ratios  of $\approx 0.6$ dex. In some rare instances (a combination of face-on orientation and large outflow velocity) the ratio between \mbh\ from \hb\ and \civ\ can reach a factor $\sim 10^{2}$, as actually found by \citet{sulenticetal07}.

The third and fourth panels from top shows the same configuration, but after applying a correction factor  $\xi_\mathrm{CIV}$  in the form $1/\zeta(L, $ \lledd) $= 1/(1+ k L^{a}$(\lledd)$^{b})$, with $a\approx 0.1$, $b \approx 1$\ and no dependence on orientation, intended to mimic the correction actually applied to the data in Sect. \ref{virial}.   The orientation effect is changing the \civ\ FWHM following Eq. \ref{eq:fciv}, and therefore displacing the \mbh\ from the true mass also after correcting the FWHM. { Note that the outlying blue points are due to \hb\ underestimates of the mass by more than 0.33 dex because of low values of the viewing angle $\theta$. On the converse, the relatively high value of ${\mathcal Q}$ in the simulation produces a significant fraction of \mbh\ \civ\ with are overestimating the \mbh\ by more than 0.33 dex (red points). } The dispersion is however reduced with respect to the case with no correction, with an rms $\approx 0.3$.  The bottom panel is a realization of the synthetic sample for a number of sources $\lesssim 100$, comparable to the size of the FOS+HE sample, and Gaussian distribution of \lledd\ as for case shown and correction in the second panel from top.

In all of these cases save the one of the second panel from top the dispersion remains $\sim 0.3 $, comparable to the one measured for the scaling laws of Eq. \ref{eq:masscorr} and Eq. \ref{eq:masscorr1}. It is interesting to note that, in the framework of the toy model, if ${\cal Q}=0$\ the orientation-induced scatter is the same for \hb\ and \civ; if ${\cal Q} \approx 4$, \civ\ becomes an almost perfect VBE, with all the scatter being due to \hb, in a plot \mbh\ \civ\ vs \mbh\ \hb. 
 This results may be consistent with  no strong dependence on orientation of the \civ\ line shift in RL quasars \citep{runnoeetal14}.\footnote{However, it is not clear whether the results of \citet{runnoeetal14} are applicable to radio quiet quasars: RL sources show no strong evidence of large blueshifts \citep{sulenticetal07,richardsetal11} as the disk outflow properties may be strongly affected by the powerful radio ejecta \citep[e.g.][ and references therein]{punsly10,punslyzhang11,sulenticetal15}.}

A uniform distribution of  ${\cal Q}$\  between 0 and 1.6, a situation more appropriate for Pop. B, was also considered. Results are similar with smaller dispersion and biases.   

\end{appendix}

\clearpage


\bibliographystyle{aa} 

\end{document}